\documentclass[lettersize,journal]{IEEEtran}
\usepackage{amsmath,amsfonts}
\usepackage{algorithmic}
\usepackage{array}
\usepackage{textcomp}
\usepackage{stfloats}
\usepackage{url}
\usepackage{verbatim}
\usepackage{graphicx}
\usepackage{subcaption}
\usepackage{xcolor}
\def\BibTeX{{\rm B\kern-.05em{\sc i\kern-.025em b}\kern-.08em
    T\kern-.1667em\lower.7ex\hbox{E}\kern-.125emX}}
\usepackage{balance}

\usepackage{makecell}
\begin{document}
\title{Subjective Evaluation of Low Distortion Coded Light Fields\\ with View Synthesis}
\author{Daniela~Saraiva,~\IEEEmembership{Student Member,~IEEE}, Joao~Prazeres,~\IEEEmembership{Student Member,~IEEE},
Manuela~Pereira,~Antonio~M.~G.~Pinheiro,~\IEEEmembership{Senior Member,~IEEE}
\thanks{Daniela Saraiva, Joao Prazeres and Antonio M. G. Pinheiro are with Instituto de Telecomunicacoes \& Universidade da Beira Interior, Portugal.}
\thanks{Manuela Pereira is with NOVA LINCS \& Universidade da Beira Interior, Portugal.}
\thanks{This work is supported by UID/04516/NOVA Laboratory for Computer Science and Informatics (NOVA LINCS) with the financial support of FCT.IP}
\thanks{This work is funded by FCT/MECI through national funds and when applicable co-funded EU funds under UID/50008: Instituto de Telecomunicações}}


\maketitle

\begin{abstract}
 Light field technology is a powerful imaging method that captures both the intensity and direction of light rays in a scene, enabling the reconstruction of 3D information and supporting a range of unique applications. However, light fields produce vast amounts of data, making efficient compression essential for their practical use. View synthesis plays a key role in light field technology by enabling the generation of new views, yet its interaction with compression has not been fully explored.

In this work, a subjective analysis of the effect of view synthesis on light field compression is conducted. To achieve this, a sparsely sampled light field is created by dropping views from an original light field. Both light fields are then encoded using JPEG Pleno and VVC. View synthesis is then applied to the compressed sampled light field to reconstruct the same number of views as the original. The subjective evaluation follows the proposed JPEG AIC-3 test methodology designed to assess the quality of high-fidelity compressed images. This test consists of two test stimuli displayed side-by-side, each alternating between an original and a coded view, creating a flicker effect on both sides. The user must choose which side has the stronger flicker and, therefore, the lower quality. Using these subjective results, a selection of metrics is validated.
\end{abstract}


\section{Introduction}
Light Fields stand out from standard imaging technologies due to their inherent ability to capture multiple views. This enables unique applications such as refocusing after capture or enhancing image resolution using super-resolution methods. The biggest challenge to the success of this technology is the vast amount of data it generates, making storage and transmission difficult. 
In this context, light field compression emerges as an area of interest. Research in this field ranges from adaptations of standard video codecs like HEVC and VVC, where coding is applied to a pseudo-temporal sequence composed of light field views. Another well-known approach is the use of specialized codecs, such as the plenoptic coding standards developed by the Joint Photographic Experts Group (JPEG).

View synthesis is another well-researched area in light fields. It is a powerful tool that enables the reconstruction or prediction of new views from a limited set of existing views, a technique commonly used in light field super-resolution tasks.~\cite{mahmoudpour_learning-based_2024}
Compression models using view synthesis have been considered in previous works~\cite{VSC-9105980,VSC-8902614}.

Currently, the typical subjective evaluation protocol compares two pseudo  videos composed of the  views of the reference and distorted light field running side by side \cite{viola2016objective,viola2017comparison}. Some variants of this initial model have been considered recently by the JPEG Committee~\cite{JPEG-WD}, but all of them suffer from the problem that in high quality, it is very difficult for a subject to identify distortions.
Furthermore, there is a lack of quality models for the quality evaluation of the angular consistency that might result of the compression or view synthesis.

In this work, a subjective quality evaluation study is conducted on light field compression methods using view synthesis. In previous work, this same data was analyzed with objective metrics. Even though some of the objective metrics used like MS-SSIM try to take into account the human visual system, subjective quality evaluation cannot be replaced when it comes to image quality assessment.
Since the images/views used in this work are very high quality, the methodology used needs to match its requirements. The methodology chosen for this work was proposed by JPEG AIC-3~\cite{testolina2023assessment}, where two stimuli are shown side-by-side. Each shows the source images in-place with the test images, creating a flickering effect that will allow the subject to observe the otherwise barely noticeable distortions. 

The data was obtained by creating a sparsely sampled light field from an original one. Both complete and sampled light fields are then compressed using JPEG Pleno and VVC. A chosen view synthesis method is then applied to the coded sampled light field. By doing so the sparsely sampled light field becomes a reconstructed light field with the same amount of views as the original.

A correlation will be established between the newly obtained subjective results and the objective results.

The remainder of this paper is structured as follows. Section \ref{sec:relatedwork} presents relevant works regarding subjective quality evaluation of light fields, compression of light fields, and view synthesis methods. Section~\ref{sec:methodology} describes the experimental setup, as well as the objective quality metrics considered in this study. Section~\ref{dataanalysis} analyses the results obtained from both the subjective and objective quality studies. Finally, Section~\ref{sec:conclusions} presents the conclusions drawn from this work.

\section{Related Work}
\label{sec:relatedwork}

\subsection{Subjective Quality Evaluation}

Subjective quality evaluation is usually conducted using either single-stimulus or double-stimulus methods, with the latter being the most commonly used in light field subjective evaluations. Double-stimulus methods involve showing two stimuli simultaneously, allowing the subject to compare and assess their quality. Although generally more time consuming, they are more accurate in some types of artifacts like shifts in colors~\cite{jpeg2024low,5412098}.

One of the most widely used methods is the Double Stimulus Impairment Scale (DSIS) \cite{ascenso2020learning}. In this approach, both the reference and coded stimulus are shown. The subject is then asked to rate the impairment between them using the following scale: very annoying, annoying, slightly annoying, perceptible but not annoying, and imperceptible.

Another commonly used method is the Double-Stimulus Continuous Quality-Scale (DSCQS) \cite{9465445,viola2019depth,shan2019no}. In this method, participants are shown both the reference and coded images, without knowing which is which, and are asked to rate the quality of each using a continuous scale. This method is slow but reliable, especially for cases where learning-based compression methods are used.

Advancements in image capture devices, compression, storage, and display technologies have raised the standard for expected image quality to a very high level. As a result, new subjective quality evaluation methodologies are required to address this demand. With the previously mentioned methods, the differences between stimuli can be extremely subtle, making it difficult to assess quality accurately.


Recently, JPEG AIC-3 proposed two test methodologies for evaluating the visual quality of high-fidelity contents, boosted triplet comparison (BTC) and Plain Triplet Comparison (PTC) \cite{testolina2024fine}.

Both methods show two stimuli that alternate between an original image and a coded image, creating a flicker effect and therefore enhancing the observers sensitivity in visual quality evaluation, particularly in the high-quality range. For each triplet (original image and two coded versions of it), observers are asked to identify the stimulus with the strongest flicker effect, answering by choosing either “Left”, “Right”, or “Not Sure”.

The BTC method consists in boosting techniques so the artifacts produced are more 
noticeable. In contrast, the PTC method presents the decoded images without any alterations. For this work, the latest method, PTC, was chosen.

\subsection{Compression}
\label{sec:compression}

Extensive research into light field compression methods has surged over recent years.
Those methods range from adaptations of standard video codecs, like H.264, HEVC \cite{H.264/HEVC}, and VVC \cite{avramelos2019light} to specialized methods tailored for light field data, including the plenoptic coding standard developed by the Joint Photographic Experts Group (JPEG), considered for this work.
JPEG Pleno provides a standard framework for representing new imaging modalities, such as light field, point cloud, and holographic imaging \cite{astola2020jpeg,
1400-1700_isoiec_nodate}. 
It also provides a low-complexity alternative to  other codecs~\cite{amirpour_performance_2021}.

Versatile Video Coding (VVC), was also considered for this work~\cite{VVC}. It uses light field views to define a sequence and then encodes the light field as a pseudo-video. 
This model is particularly effective for compressing light fields, as it explores the higher similarity between different views. 
VVC is a codec developed by the Joint Video Exploration Team (JVET) and 
MPEG. It incorporates innovative transformation and quantization methods, optimizing data representation while minimizing perceptual losses \cite{9328514}. It also presents a promising framework for light field compression, by leveraging its advanced coding capabilities, allowing to maintain high fidelity while achieving substantial bitrate reductions.

Although briefly, some works have explored the integration of view synthesis into light field compression. Mukati \textit{et al}. \cite{VSC-9105980} proposed Distributed Source Coding (DSC) and applied learning-based view synthesis to generate high-quality side information at the decoder, significantly reducing the number of key views that need to be transmitted while achieving similar performance. 
Another study by Bakir \textit{et al}. \cite{VSC-8902614} reveals that encoding a sparse set of views and synthesizing the rest at the decoder yields higher subjective visual quality than conventional light field coding, highlighting view synthesis’s potential for improved compression efficiency.

\subsection{View Synthesis}
\label{sec:viewsynthesis}
When capturing light fields, there is an inherent trade-off when it comes to the spatial and angular resolution that can be obtained, due to hardware limitations. 
To overcome the problem of having a sparse set of views (therefore low angular resolution), intermediate views are synthesized between the views
to obtain dense light fields. View synthesis has been used in this context to improve the quality of light fields, though its impact on compression still needs to be further researched.

SepConv++ \cite{niklaus_revisiting_2020} was selected as the view synthesis method for this work, which is an improved version of SepConv \cite{niklaus_video_2017}. 
SepConv has shown strong performance in light field view synthesis, outperforming other methods like Shearlet and LFEPI considering metrics such as PSNR and SSIM \cite{chen_study_2020}. 
Moreover, it was successfully integrated as part of a layered light field coding strategy~\cite{amirpour2021slfc}. 
More recent view synthesis models have been proposed, that usually present slightly better results when compared with SepConv as it is the case of Chen et al.~\cite{chen2020self}.
However, these models have no available implementation. 
Furthermore, as these methods are learning-based, they strongly depend of the training proccess and data, which makes very unlikely to reproduce the claimed performance.


SepConv++ extends the original SepConv neural network architecture, where given input frames, an encoder-decoder network extracts features that are given to four sub-networks that each estimate one of the four 1D kernels for each output pixel in a dense pixel-wise manner. 
The estimated pixel-dependent kernels are then convolved with the input frames to produce the interpolated frame \cite{niklaus_video_2017}.

One of the main enhancements in the updated model is the inclusion of residual blocks, which take advantage of the significant advancements in deep learning architectures developed after the original release of SepConv. Along with other network improvements. These updates contribute to enhanced interpolation quality.

The kernel normalization strategy was also changed in SepConv++. The updated approach applies adaptive separable convolution to both the input and a mask, then normalizes by dividing the filtered input by the filtered mask. This modification significantly improves synthesis quality and model convergence.

\section{Methodology}\label{sec:methodology}

\begin{figure*}
\centering
    \subfloat[\textit{Bikes}]{\includegraphics[width=0.23\linewidth]{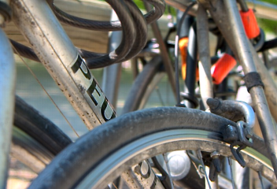}} \quad
    \subfloat[\textit{Fountain}]{\includegraphics[width=0.23\linewidth]{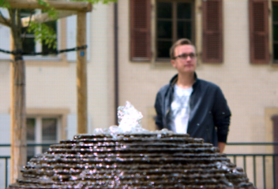}} \quad
    \subfloat[\textit{Bicycle}]{\includegraphics[width=0.23\linewidth]{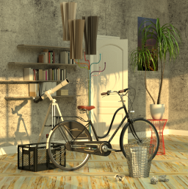}} \quad
    \subfloat[\textit{Sideboard}]{\includegraphics[width=0.23\linewidth]{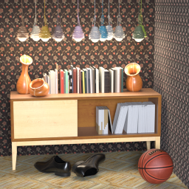}} \quad\\
    \caption{Center view of the selected light fields.}
\label{fig:res}
\end{figure*}

For this work, four light fields were used, namely \textit{Bikes}, \textit{Fountain and Vincent 2}~\cite{amirpour2022efficient}, \textit{Bicycle} and \textit{Sideboard}~\cite{amirpour_performance_2021}. Their respective central views can be seen in Fig.~\ref{fig:res}.
The first two were captured by a Lytro Illum camera. They present natural and outdoors content, and consist of 15$\times$15 views, with a resolution of 625$\times$434 with a 10 bit-depth. \textit{Bicycle} and \textit{Sideboard} are synthetically generated light fields. They  consist of 9$\times$9 views with a resolution of 512$\times$512 with a 8-bit depth.

\begin{figure}
\centering
    \subfloat[\textit{Original}]{\includegraphics[width=0.47\linewidth]{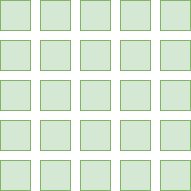}} \quad
    \subfloat[\textit{Sparsely Sampled}]{\includegraphics[width=0.47\linewidth]{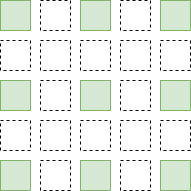}} \quad
    \caption{Light field view selection process.}
\label{fig:ViewSelection}
\end{figure}

For simplicity purposes, only the inner 5$\times$5 views were used for the original light field set. Then, a sparsely sampled light field is created by selecting a 3$\times$3 set from the original. The selection process is described in Fig.~\ref{fig:ViewSelection}.

Both light fields (original and sparsely sampled), are encoded using the chosen Codecs, namely JPEG Pleno 4D-TM \cite{plenoConformanceAndReferenceSW2020} and VVC~\cite{avramelos2019light} using the Random Access configuration. The target bitrates used in this process were defined using JPEG Pleno and the Bikes light field, and are the following: 0.118, 0.236, 0.472 and 1.003.

\begin{figure}
    \centering
    \includegraphics[width=0.7\linewidth]{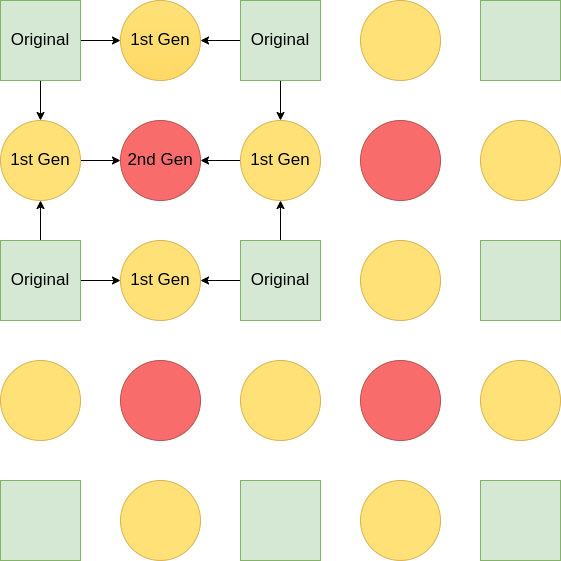}
    \caption{View synthesis process}
    \label{fig:viewsynthesis}
\end{figure}

The light field decoding was followed by a view synthesis process applied to the sparsely sampled 3$\times$3 light field. The view synthesis process reconstructs the missing views, creating a 5$\times$5 light field consisting of both coded and synthesized views. View synthesis is applied in a two-step process shown in Fig.~\ref{fig:viewsynthesis}. The first stage views (yellow circles) were synthesized from adjacent original compressed views (green squares). The second stage views (red circles) were then synthesized from the first stage synthesized adjacent views (yellow circles).

To reduce the amount of data used in the subjective evaluation, 3 views were selected from each light field. The selection was based on the sparsely sampled light field encoded with VVC that underwent view synthesis. One view of each type (original compressed, first-generation synthesized, and second-generation synthesized) was chosen. For each type, the selected view was the one with the lowest MS-SSIM at the highest bitrate.

\subsection{Subjective Test}

\subsubsection{Test Methodology}

The JPEG AIC standardization project is currently defining new subjective testing methodologies to address the need for appropriate visual quality evaluation methods 
for the near visually lossless quality  range.

In this work, an adaptation of the Plain Triplet Comparison (PTC) methodology proposed by JPEG AIC-3 was used \cite{testolina2024fine}. Triplets consist of two coded versions of a reference view, and the reference view itself. For each triplet evaluation, two stimuli are shown side-by-side, each alternating between one of the coded views and the original view. For each triplet the subject is asked to choose the stimulus with the strongest flicker effect, by selecting one of the three button options ``Left", ``Right" and ``Not Sure". As for the flicker, the coded and reference views were temporally interleaved at a change rate of 2 Hz, with each image displayed for 500 ms per switch.


\subsubsection{Triplet Question Types}
Five types of triplet questions were employed in the subjective quality evaluation. In all triplet comparisons, the lowest included bitrate was never directly compared against the highest bitrate to avoid trivial quality judgments. The different resultant light fields were grouped by codec (Pleno or VVC) and by coding method (complete light field encoding or sparse-view encoding with view synthesis). The question types included:

\begin{itemize}
    \item \textbf{Intra-Method Comparisons}: 
    Triplets comparing decoded images with different bitrates within the same codec and encoding method (full or sparse-view encoding with view synthesis).
    
    \item \textbf{Cross-Codec Comparisons}: 
    Triplets comparing the same encoding method (complete or sparse-view) across different codecs at different distortion levels. Specifically, the two lowest JPEG Pleno bitrates were never compared against VVC, as preliminary tests showed these comparisons yielded obvious results.
    
    \item \textbf{Encoding-Method Comparisons}: 
    Triplets comparing complete light field encoding versus sparse-view encoding with synthesis, using the same codec at distinct distortion levels.
    
    \item \textbf{Bias-Control Comparisons}: 
    Triplets containing two identical renderings of the original uncompressed light field to detect any systematic response biases.
    
    \item \textbf{Attention-Check Questions}: 
    Triplets containing extreme quality variations to verify observer attentiveness. More specifically one of the stimulus contains a decoded image with the strongest distortion level and the other side displays the original image.
\end{itemize}

Considering all the mentioned triplet types, across all light fields and the three selected view types for each LF, a total of 776 unique triplets were included in the subjective test. Limiting the test to four light fields was a necessary constraint, as including more would have substantially increased the number of unique triplets, making the experiment too time-consuming and demanding for a controlled laboratory setting. To evaluate a more diverse and extensive light field dataset, the subjective test would require a crowdsourcing  approach.

\subsubsection{Environment Setup}


The subjective quality study was conducted following the ITU-T BT.500-15~\cite{BT500} recommendations for subjective quality evaluations, in a controlled lighting situation, with the color of all background walls and curtains being mid-gray. The test was conducted on an EIZO CG318-4K monitor. However, since the stimuli had a low resolution (e.g., 512$\times$512), the display was set to Full HD (1920$\times$1080) instead of 4K, to ensure proper visibility.
The distance of the subjects
from the monitor was approximately equal to 7 times the
image height, as recommended in ITU-R BT.2022~\cite{bt2012general}.

\subsubsection{Test procedure and Participants}

Due to the large scale of the experiment, it was not feasible for each participant to evaluate all the stimuli. Instead, each stimulus was evaluated 16 times to ensure reliable results, which was achieved by involving 32 subjects. Each subject performed half of the total evaluations, allowing the test duration to remain manageable while still meeting the required number of evaluations per stimulus. Additionally, to reduce visual fatigue and maintain response quality, a mandatory break was taken halfway through each evaluation session. 

The order in which the stimuli were displayed was randomized, ensuring that distortions of the same content were never presented in consecutive comparisons. Each triplet was shown inverted for half of the evaluations, where the left and right stimuli trade their placement to avoid any additional biases.

Before the test session, a training session was conducted using additional light field content to allow the participants to familiarize themselves with the evaluation procedure.


An informed consent form was also previously handed to the participants for signature. All the subjects were tested to ensure normal or corrected-to-normal vision using the Snellen\footnote{https://visionscreening.zeiss.com/en-INT} visual acuity test and absence of color blindness using the  Ishihara\footnote{https://www.blindnesstest.com/ishihara-test/} test.

A total of 32 subjects took part in these subjective evaluations, 11 female and 21 male. The subjects ages ranged from 19 to 54 with an average of 27.7.

\subsubsection{Subjective score screening and processing}
To analyze the results of the subjective quality evaluation, the number of times one condition/stimulus is selected over another is computed. A comparison matrix \( V \) is then formed, where each element \( v_{ij} \) reveals the number of times condition \( i \) is selected over condition \( j \). If an observer indicates ``Not Sure", the score is evenly split, assigning half of the scores to each condition. 
To transform the raw comparison scores to a quality scale, the Thurstone Case V~\cite{thurstone2017law}, was  employed, based on a previous study Testolina \textit{et al.}~\cite{testolina2023performance}. 
The implementation of Thurstone Case V provided by Perez \textit{et al.} was used~\cite{perez2017practical} to convert the scores into a continuous quality scale.

The removal of outliers for this subjective quality evaluation was also performed using the method proposed by Perez \textit{et al.}~\cite{perez2017practical}, as well as the method proposed for the 95\% confidence intervals. The implementation of the software used in this work is publicly available\footnote{https://github.com/mantiuk/pwcmp}.



\section{Data Analysis}\label{dataanalysis}

\begin{figure*}
\centering
    \subfloat[\textit{Bikes}]{\includegraphics[width=0.25\linewidth]{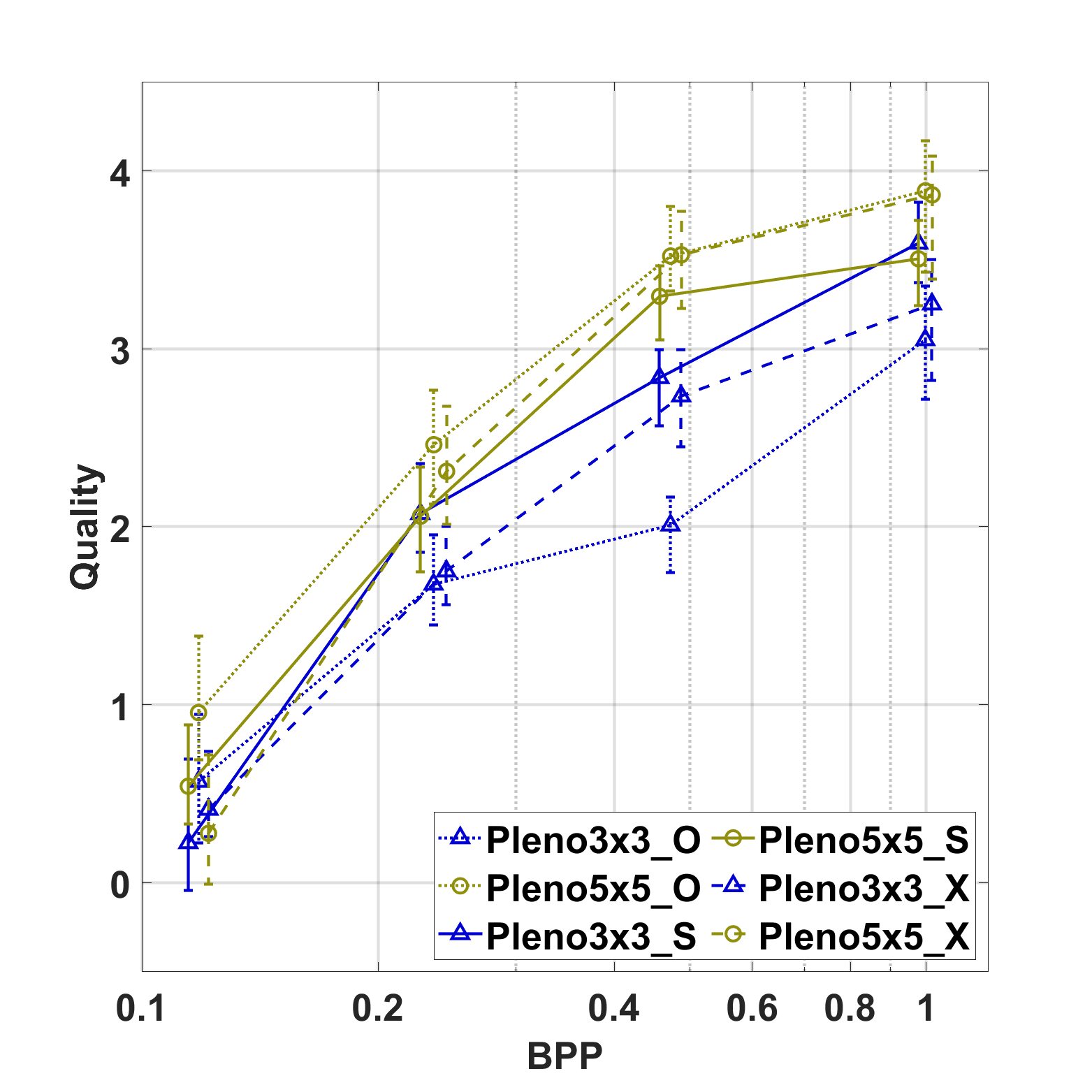}}
    \subfloat[\textit{Fountain}]{\includegraphics[width=0.25\linewidth]{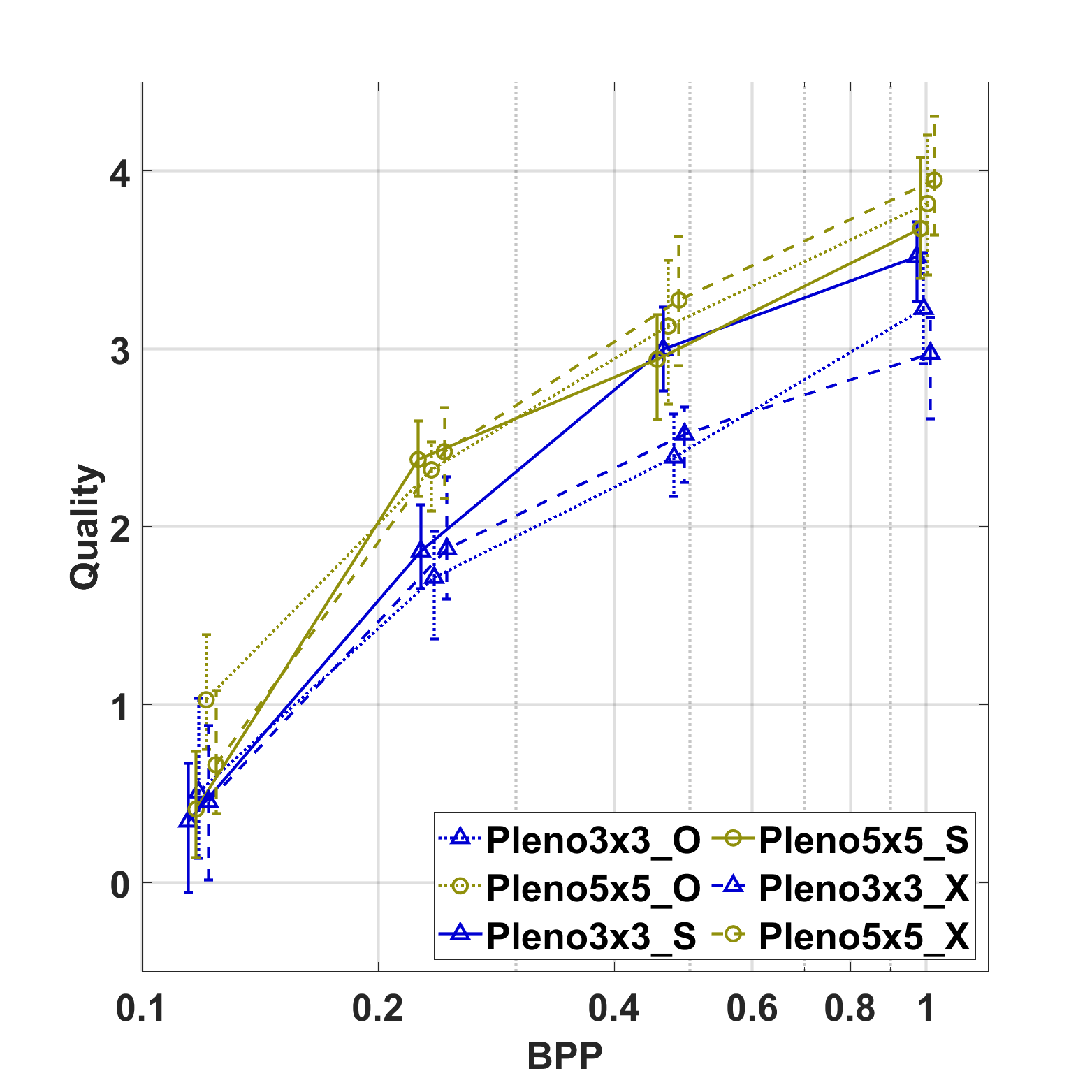}}
    \subfloat[\textit{Bicycle}]{\includegraphics[width=0.25\linewidth]{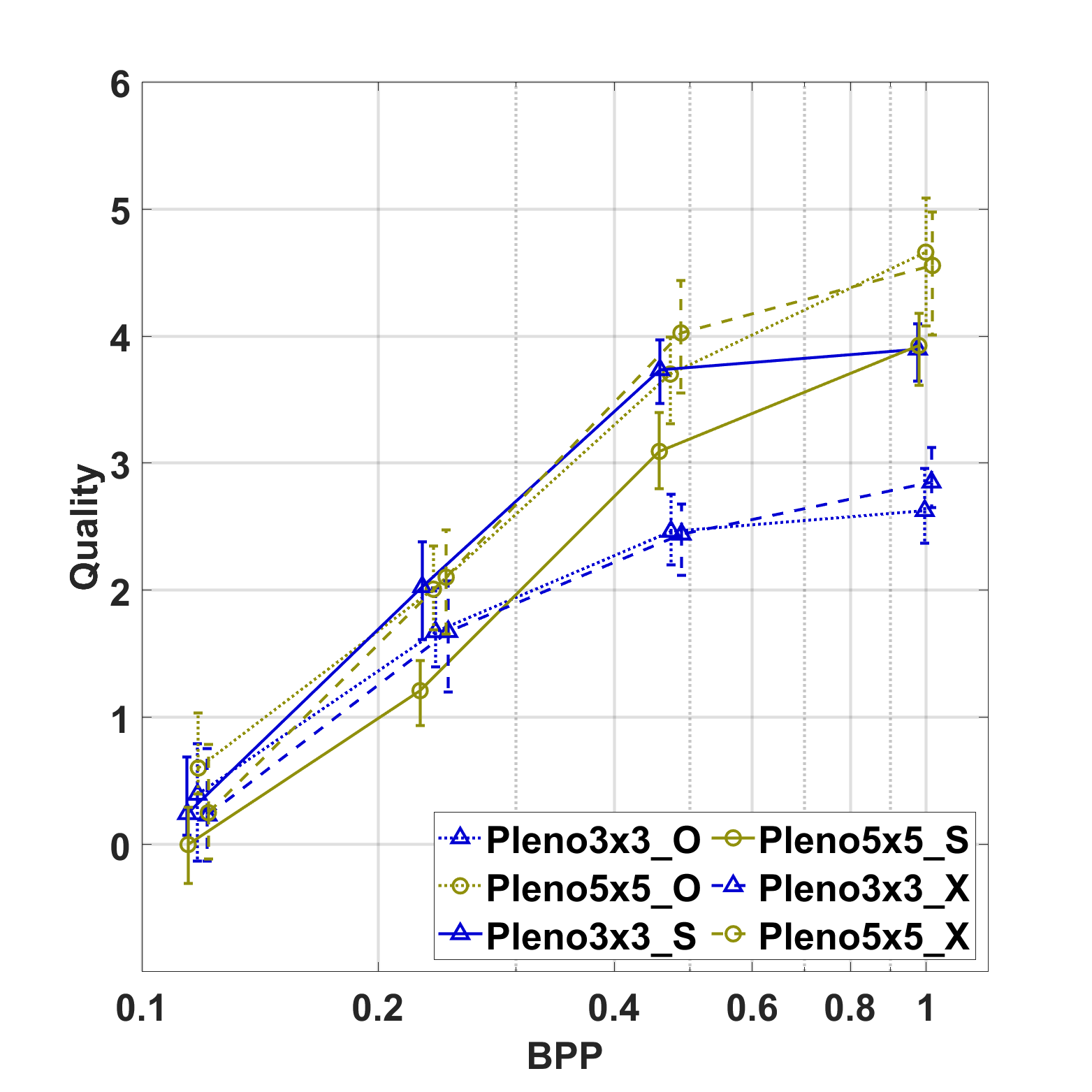}}
    \subfloat[\textit{Sideboard}]{\includegraphics[width=0.25\linewidth]{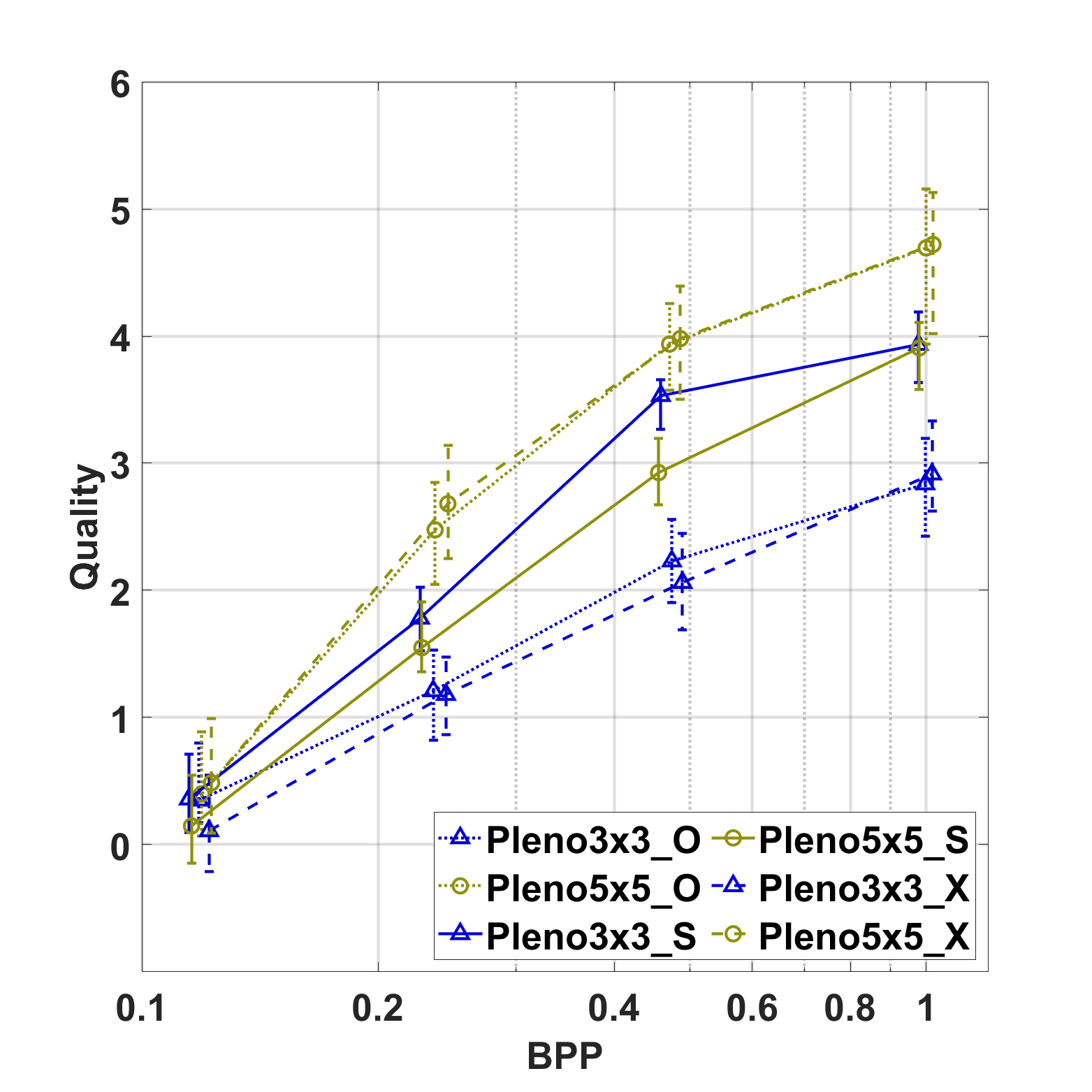}}\\
    \caption{Subjective Quality Scale with 95\% confidence interval 
    vs bpp, for 
    JPEG Pleno.}
\label{fig:Subjective_CM_Pleno}
\end{figure*}

\begin{figure*}
\centering
    \subfloat[\textit{Bikes}]{\includegraphics[width=0.25\linewidth]{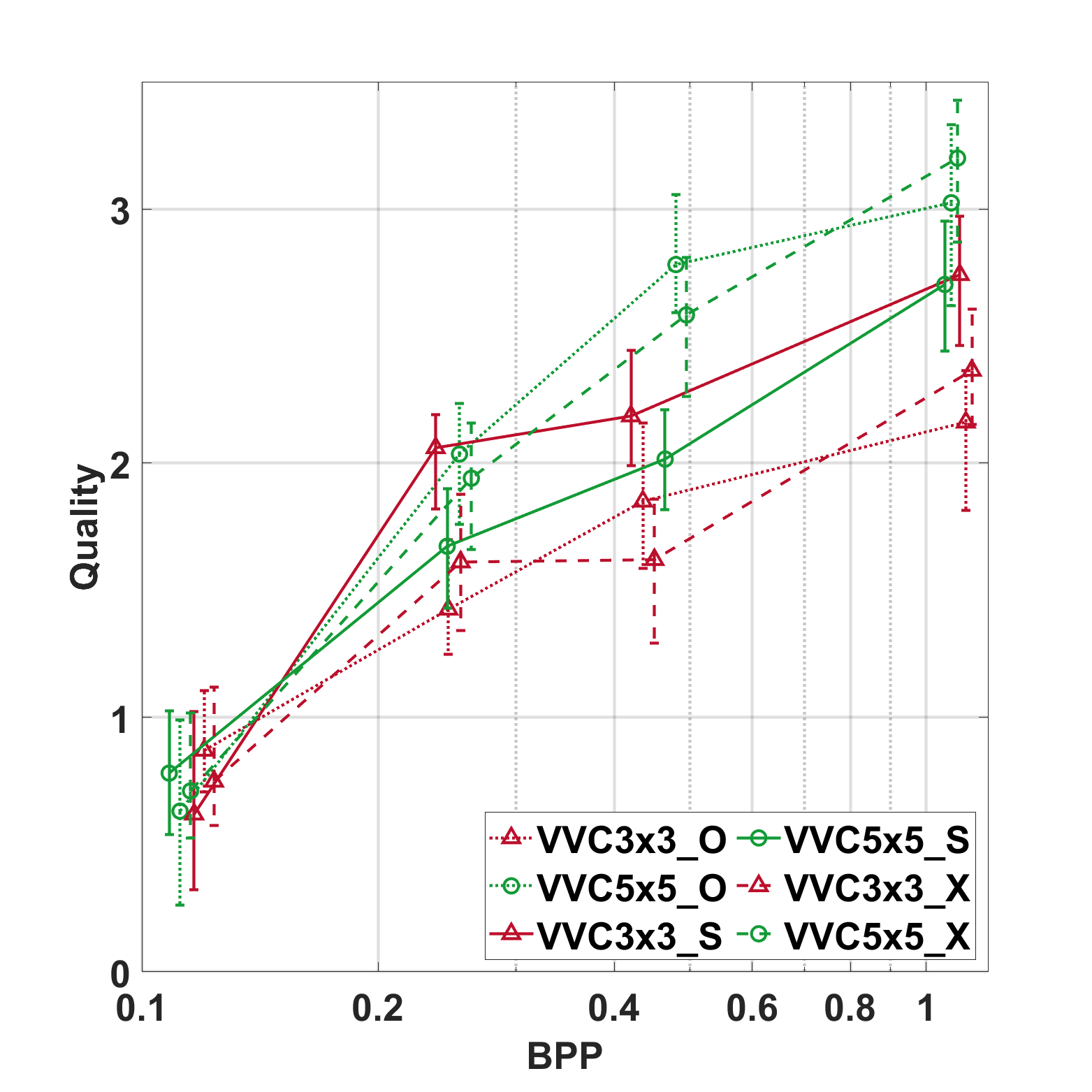}}
    \subfloat[\textit{Fountain}]{\includegraphics[width=0.25\linewidth]{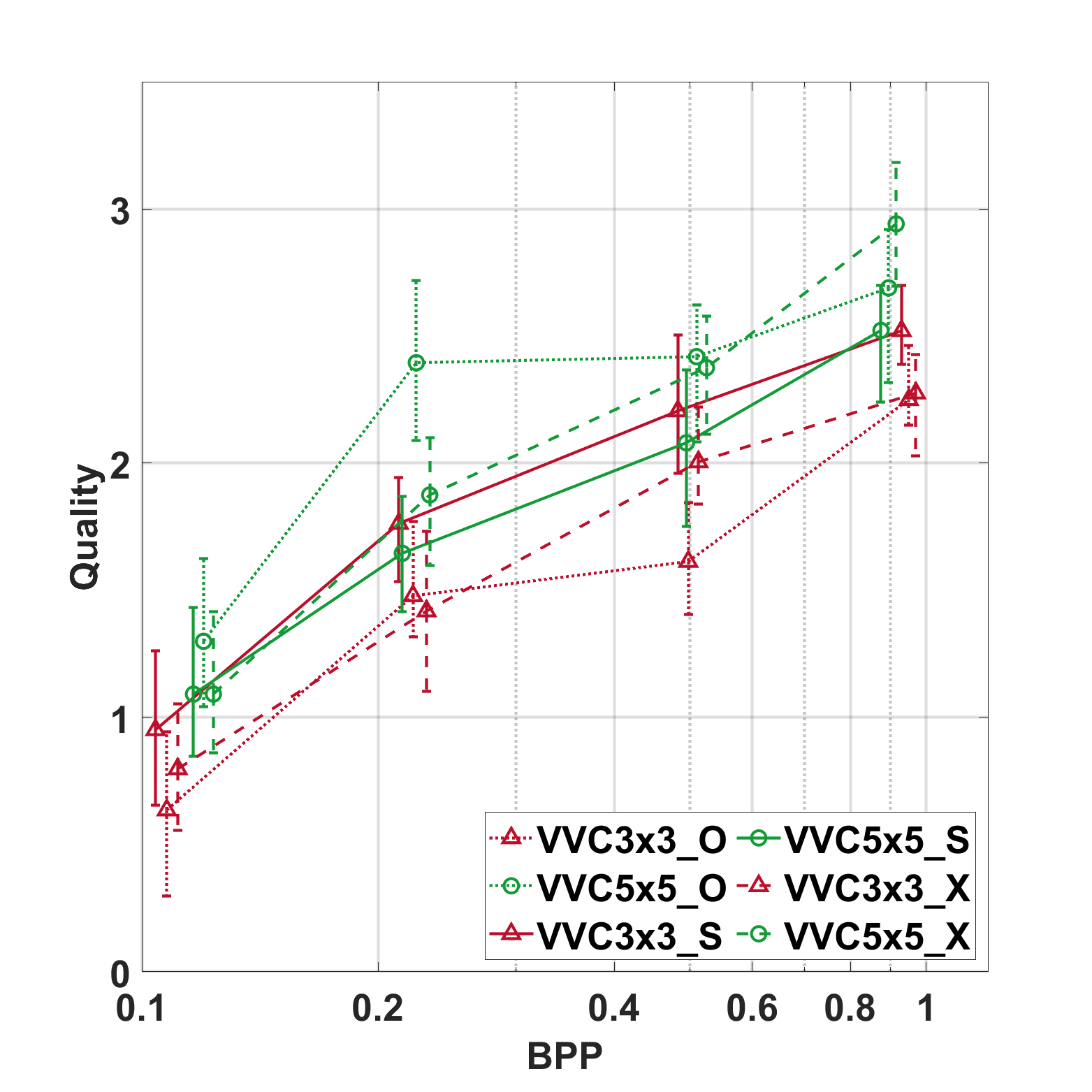}}
    \subfloat[\textit{Bicycle}]{\includegraphics[width=0.25\linewidth]{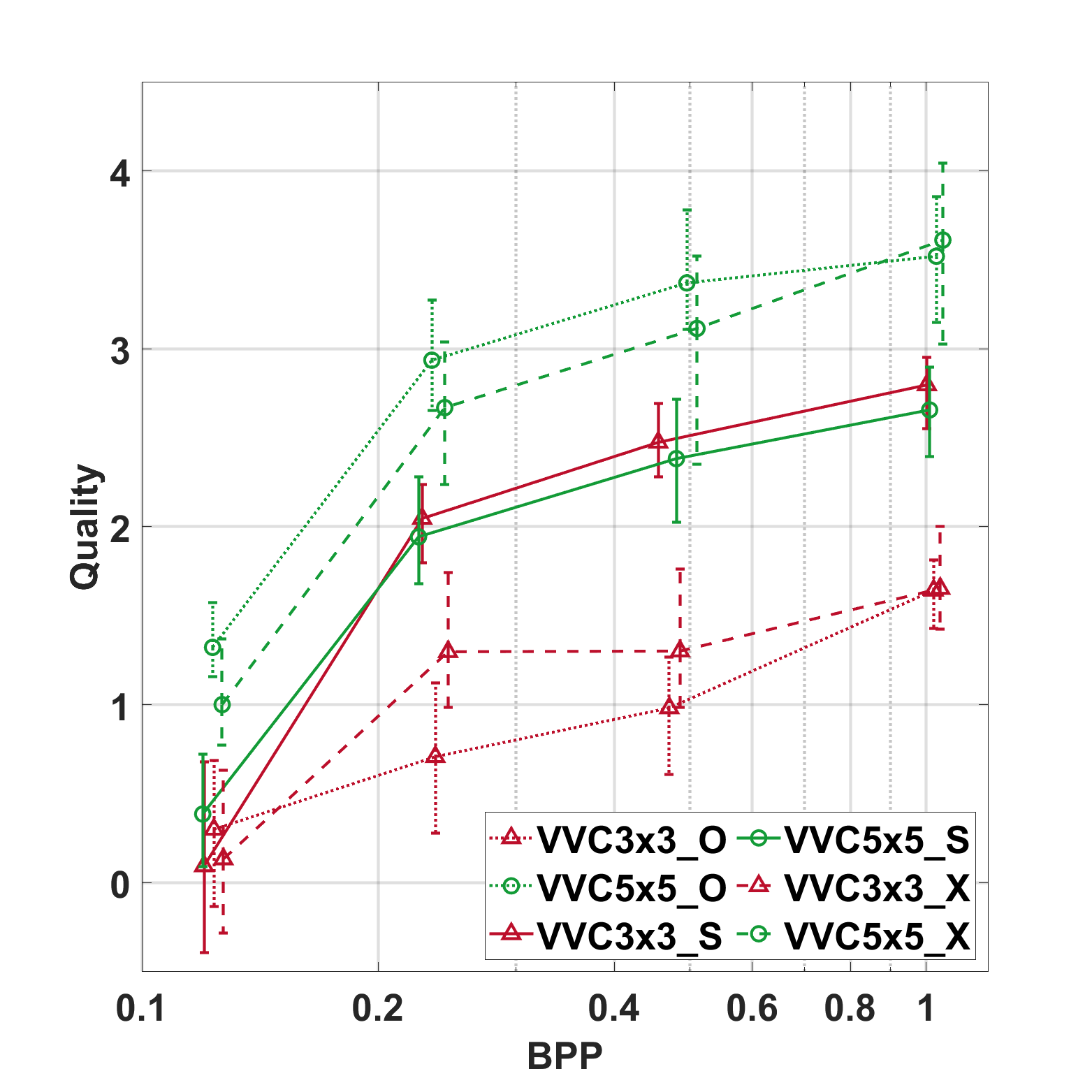}}
    \subfloat[\textit{Sideboard}]{\includegraphics[width=0.25\linewidth]{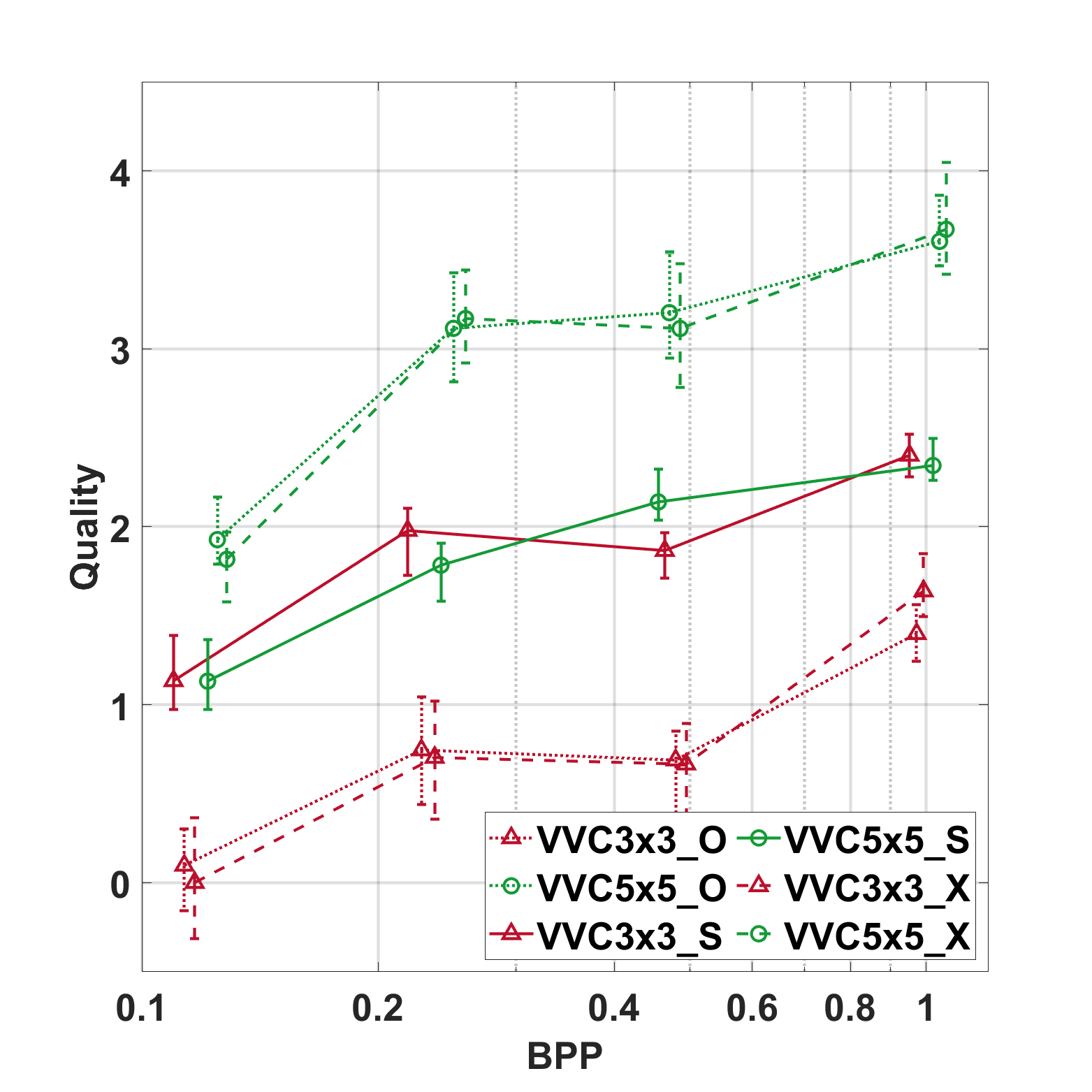}}\\
    \caption{Subjective Quality Scale with 95\% confidence interval 
    vs bpp, for 
    VVC.}
\label{fig:Subjective_CM_VVC}
\end{figure*}

\begin{figure*}
\centering
    \subfloat[\textit{Bikes}]{\includegraphics[width=0.25\linewidth]{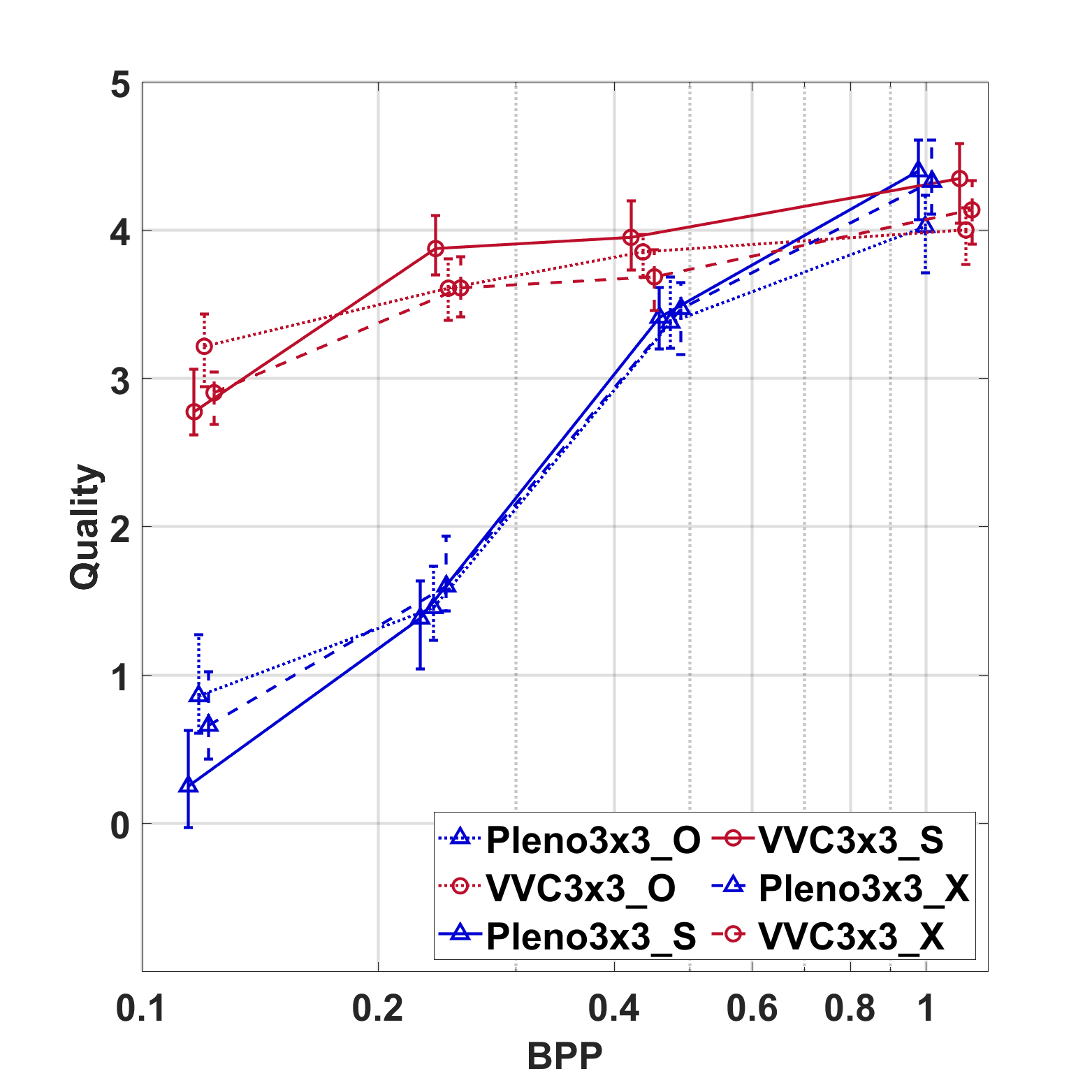}}
    \subfloat[\textit{Fountain}]{\includegraphics[width=0.25\linewidth]{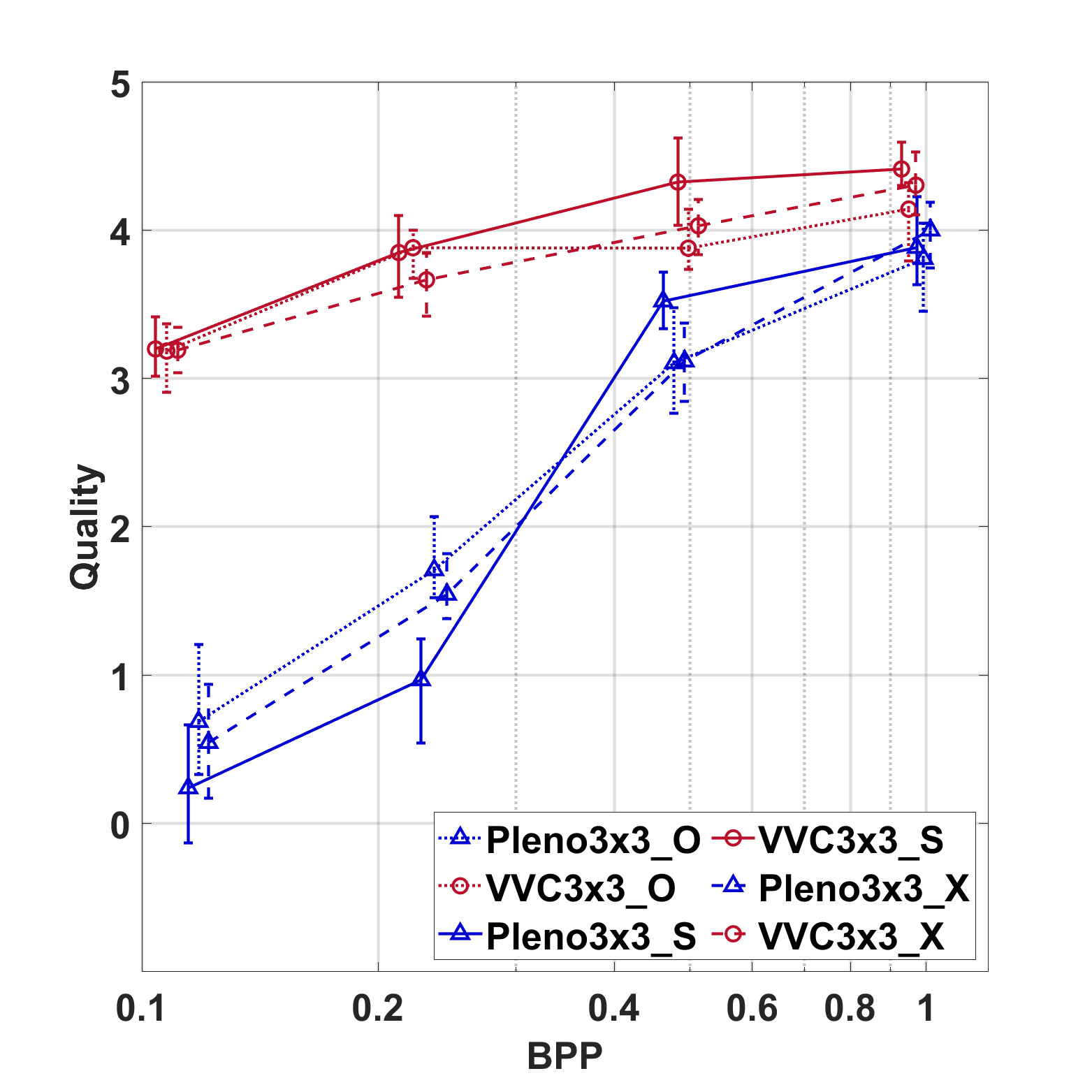}}
    \subfloat[\textit{Bicycle}]{\includegraphics[width=0.25\linewidth]{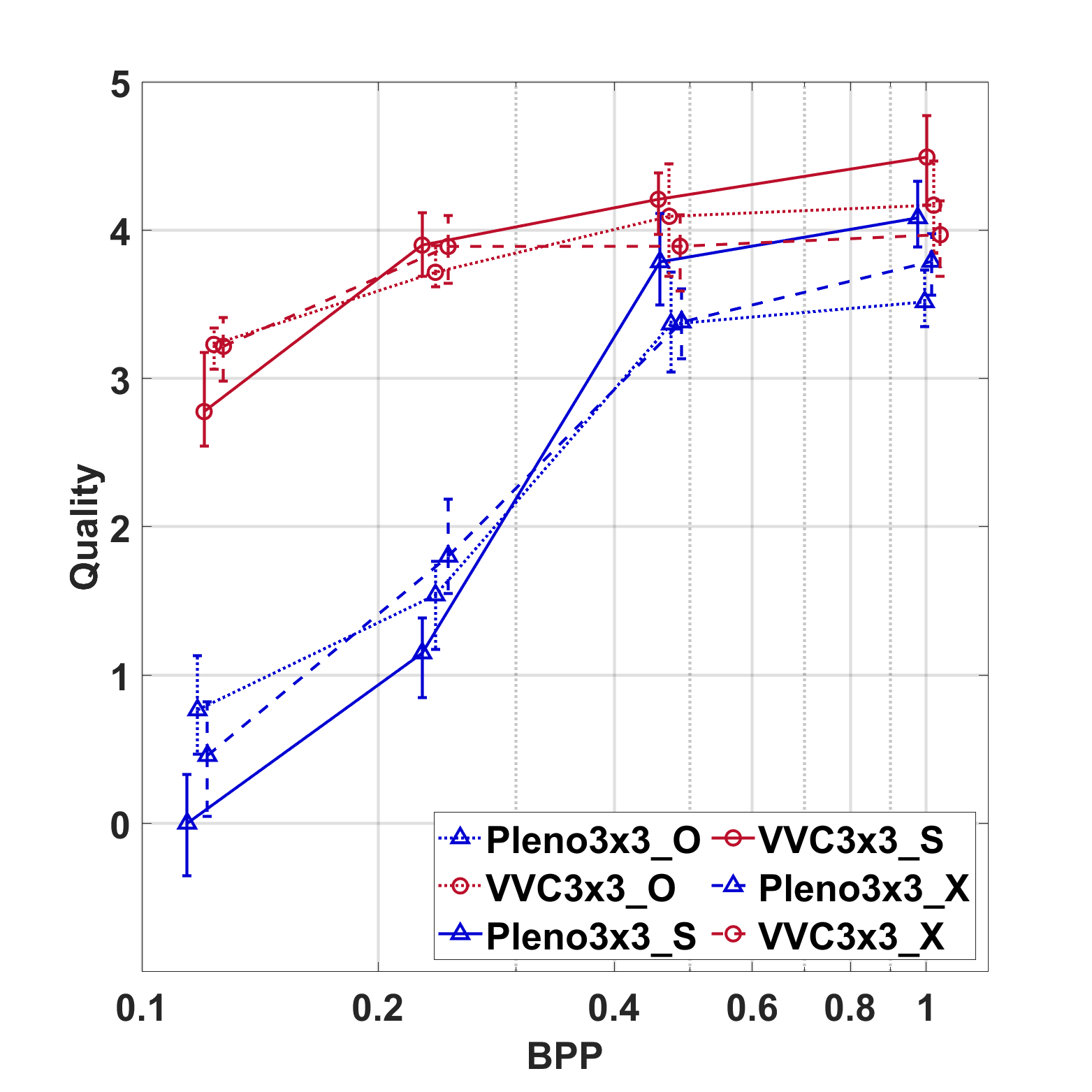}}
    \subfloat[\textit{Sideboard}]{\includegraphics[width=0.25\linewidth]{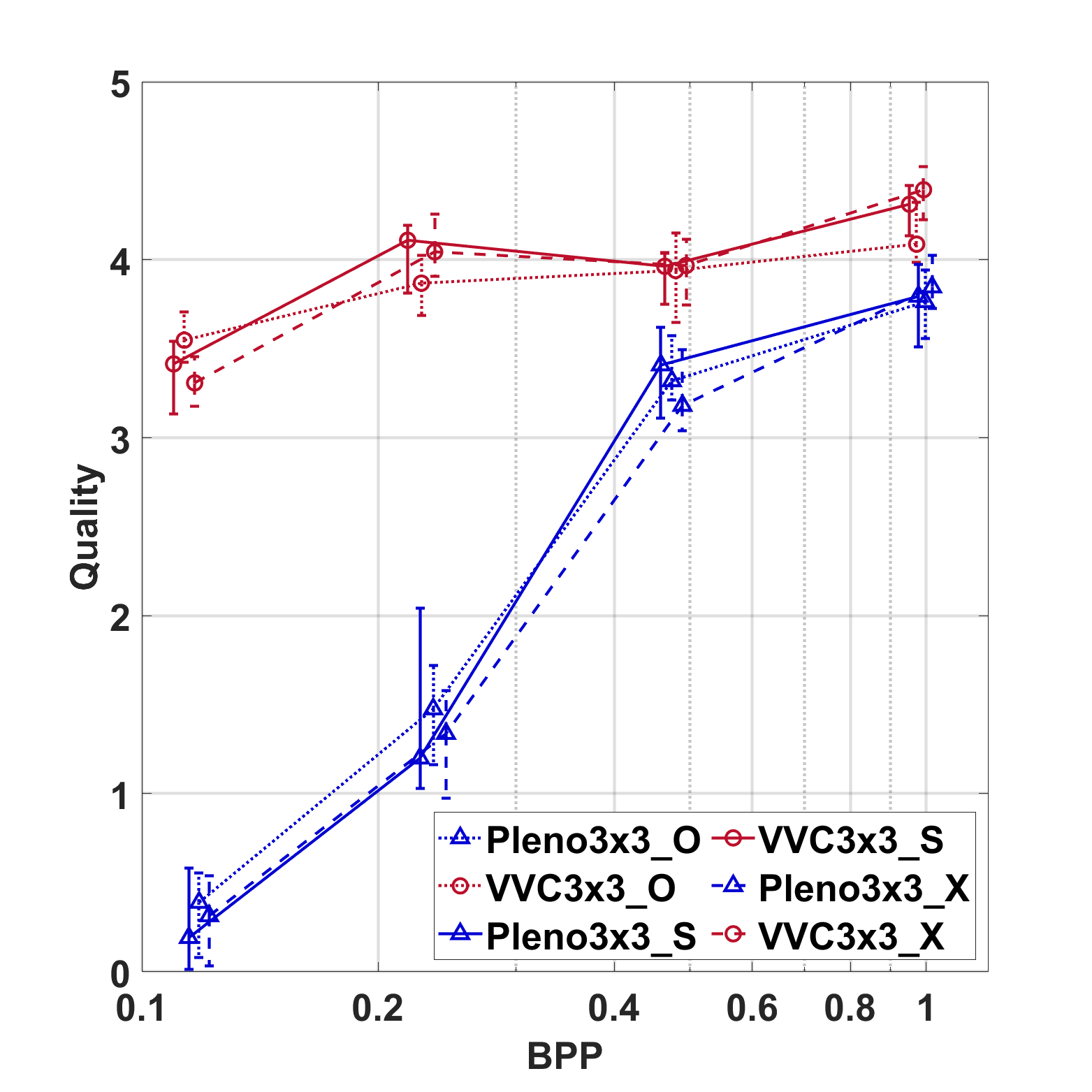}}\\
    \caption{Subjective Quality Scale with 95\% confidence interval 
    vs bpp, 
    comparing JPEG Pleno 3$\times$3 and VVC 3$\times$3.}
\label{fig:Subjective_CC_3x3}
\end{figure*}

\begin{figure*}
\centering
    \subfloat[\textit{Bikes}]{\includegraphics[width=0.25\linewidth]{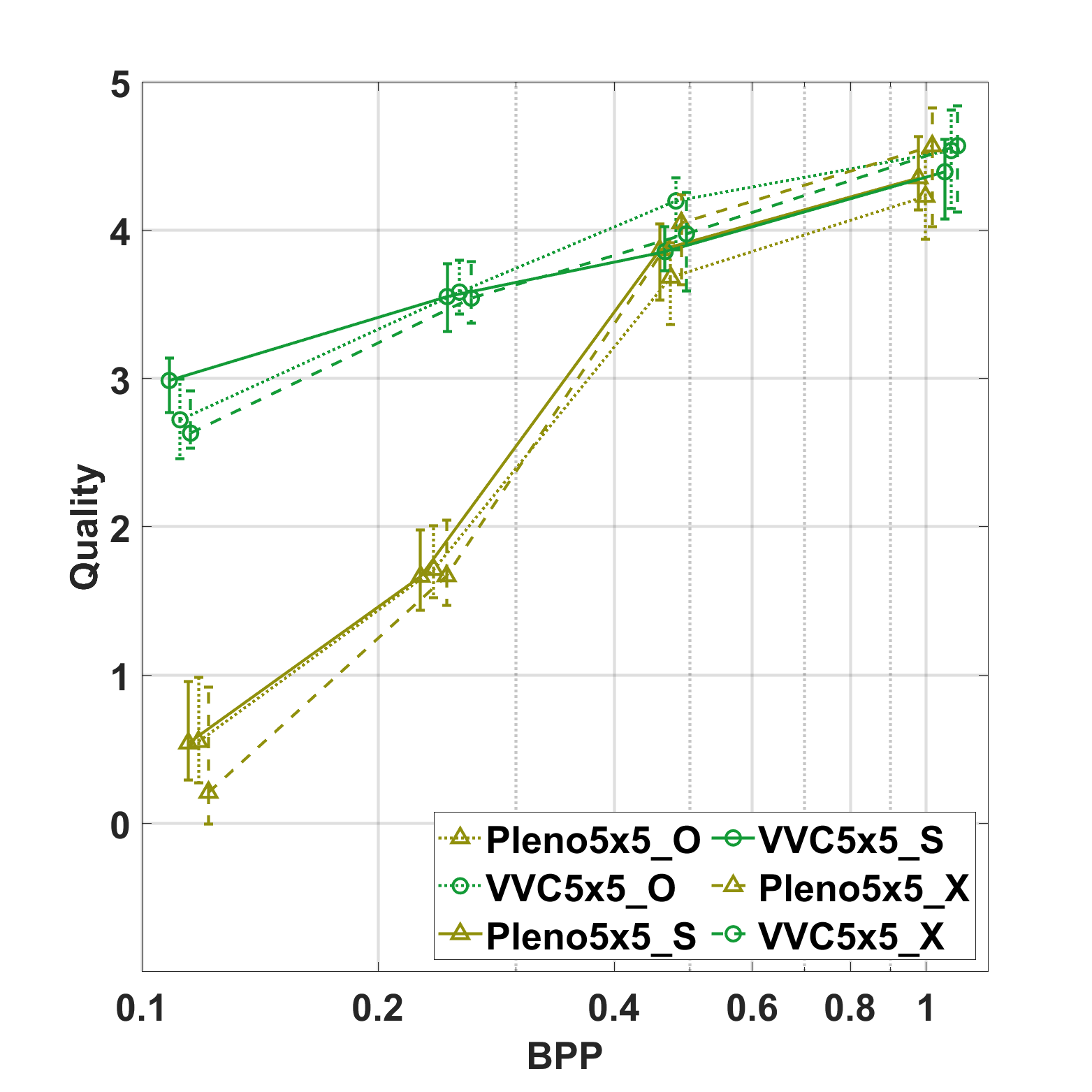}}
    \subfloat[\textit{Fountain}]{\includegraphics[width=0.25\linewidth]{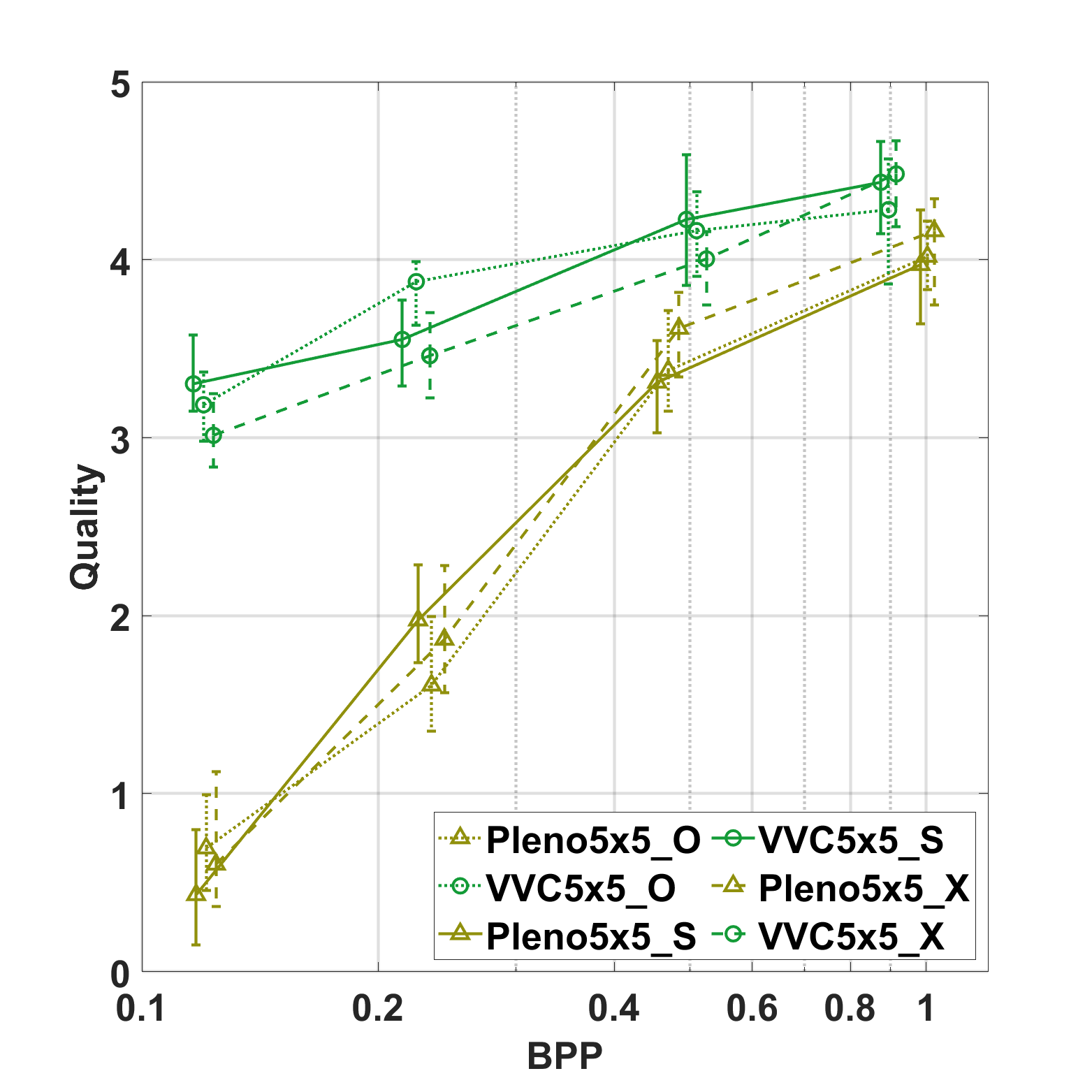}}
    \subfloat[\textit{Bicycle}]{\includegraphics[width=0.25\linewidth]{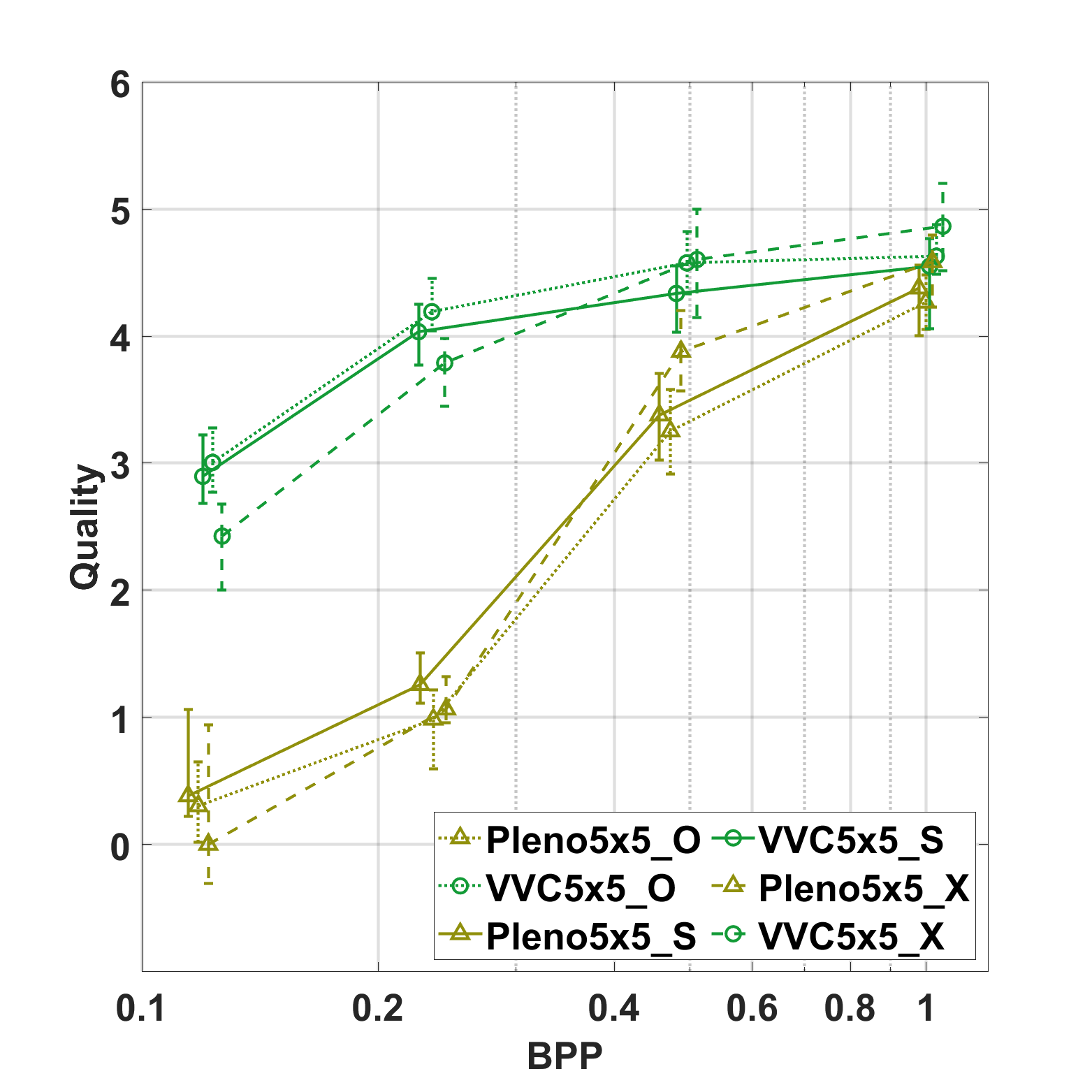}}
    \subfloat[\textit{Sideboard}]{\includegraphics[width=0.25\linewidth]{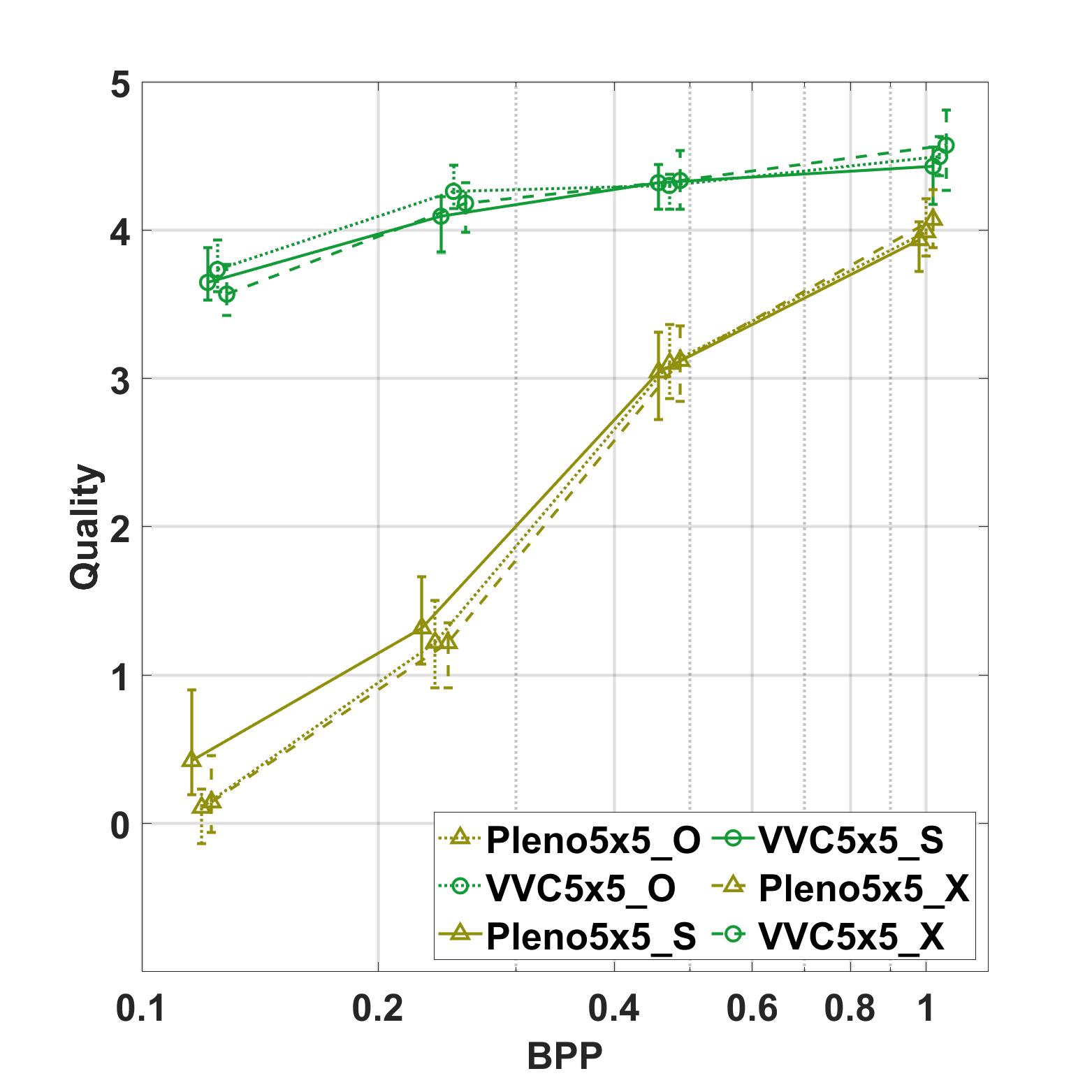}}\\
    \caption{Subjective Quality Scale with 95\% confidence interval 
    vs bpp, for cross-codec comparison between JPEG Pleno 5$\times$5 and VVC 5$\times$5.}
    \label{fig:Subjective_CC_5x5}
\end{figure*}

During the subjective quality evaluation, participants were instructed to select the stimulus with the most noticeable flicker. 
The results in Fig.~\ref{fig:Subjective_CM_Pleno} to \ref{fig:Subjective_CC_5x5} were obtained using the Thurstone Case V model, which estimates the probability of each stimulus being chosen over the others. The values on the quality scale were normalized between 0 and 1. An adjustment was made so that higher values indicate higher perceived quality, allowing for a more intuitive analysis.

\begin{figure*}
\centering
    \subfloat[\textit{PSNR HSV}]{\includegraphics[width=0.25\linewidth]{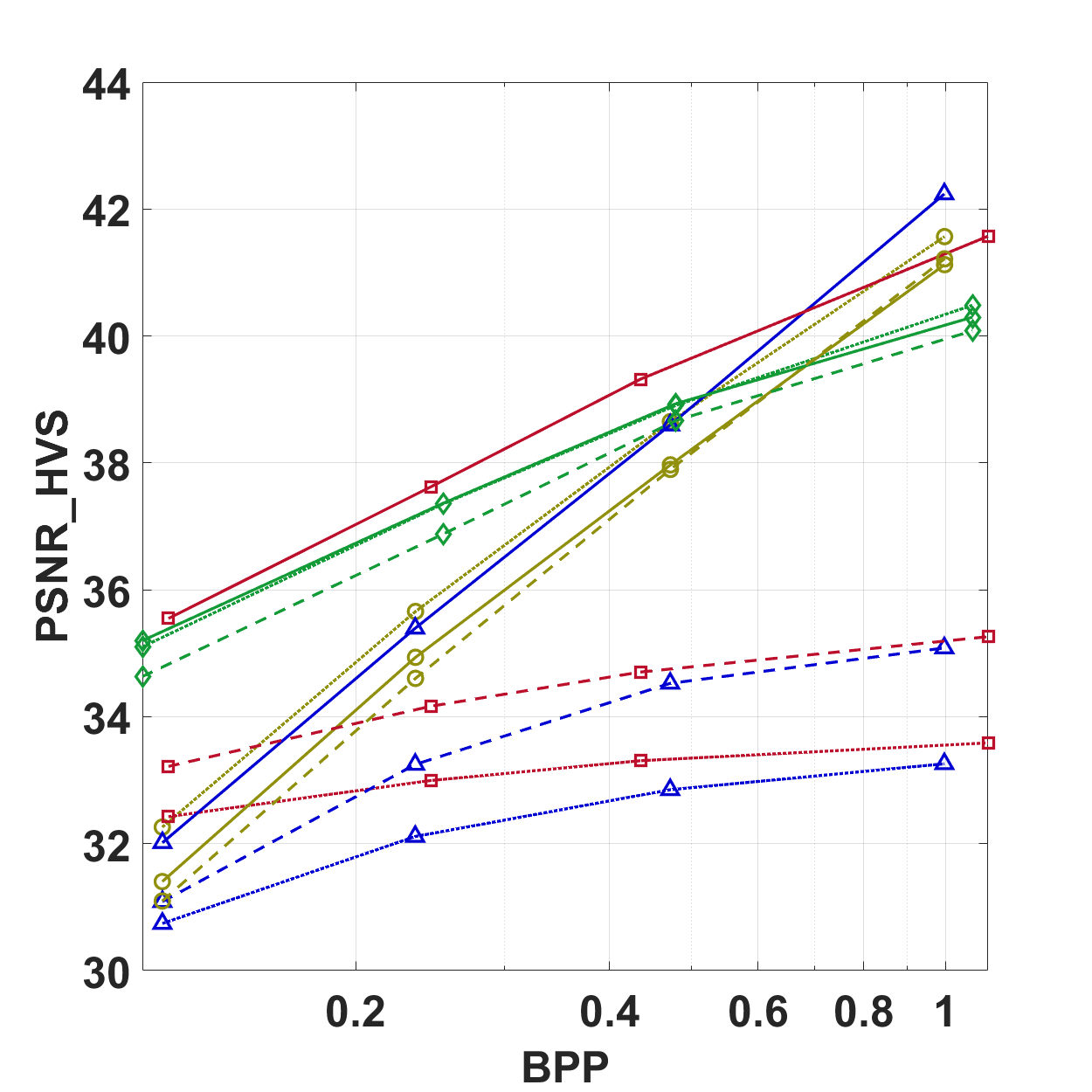}}
    \subfloat[\textit{MS-SSIM}]{\includegraphics[width=0.25\linewidth]{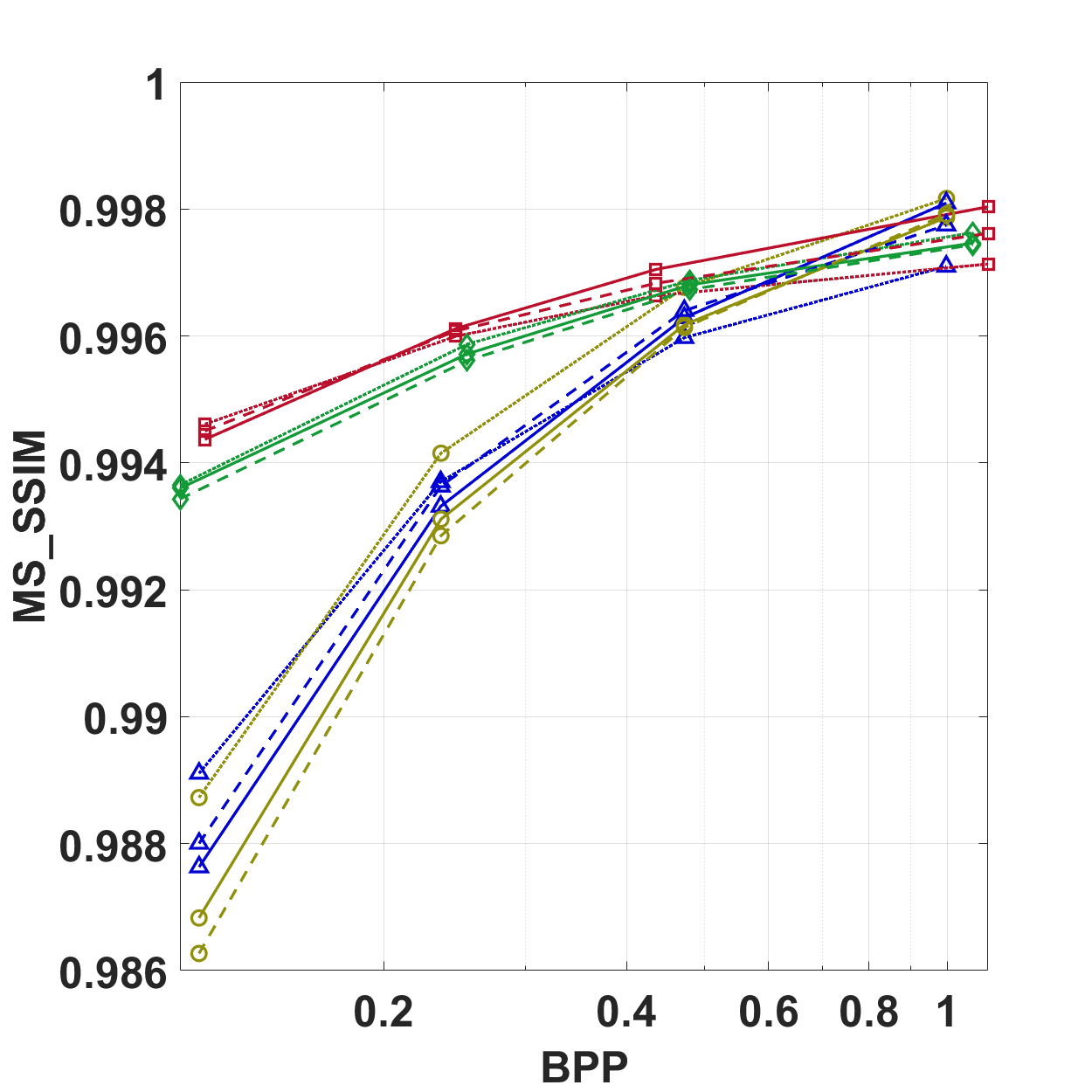}}
    \subfloat[\textit{FSIMc}]{\includegraphics[width=0.25\linewidth]{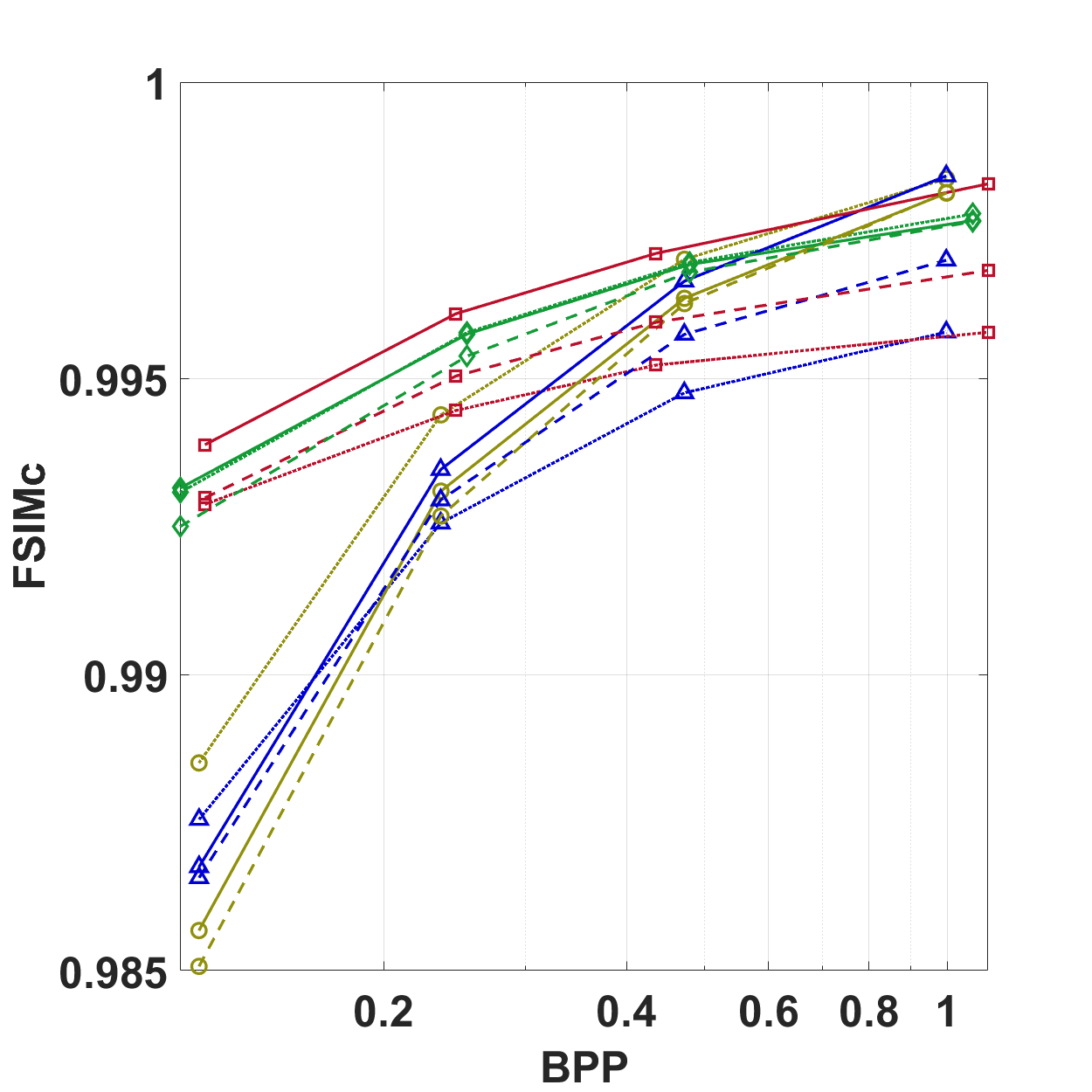}}
    \subfloat[\textit{IW-SSIM}]{\includegraphics[width=0.25\linewidth]{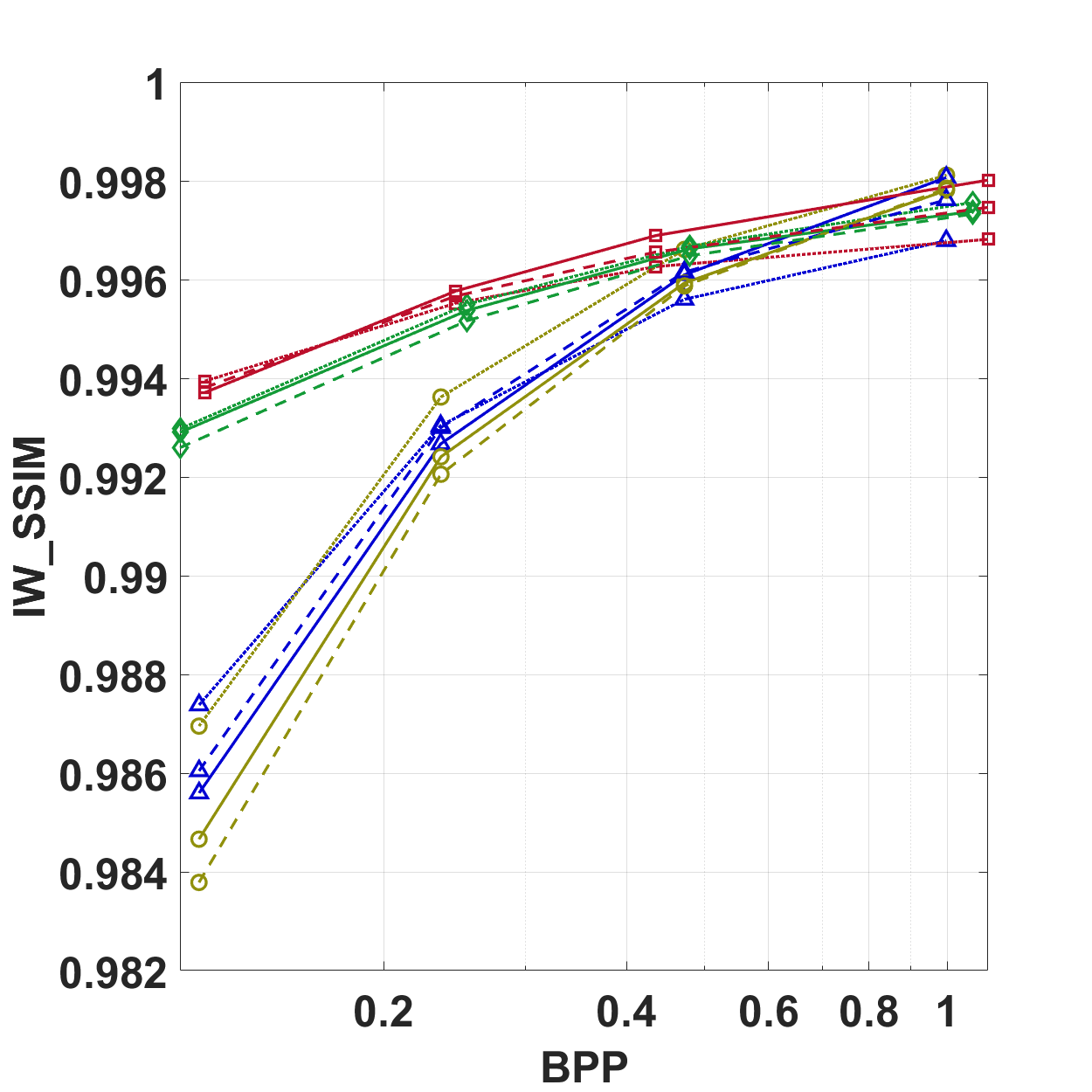}}\\
    \includegraphics[width=0.7\textwidth]{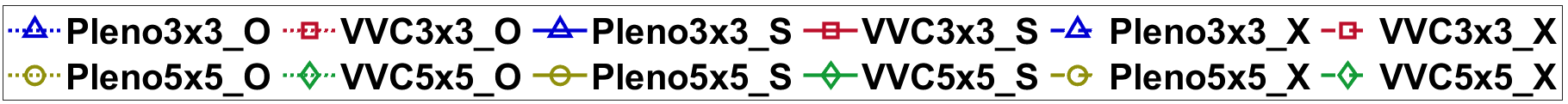}
     \caption{Objective Quality Metrics for the Bikes Light Field.}
 \label{fig:Objective_bikes}

\centering
    \subfloat[\textit{PSNR HSV}]{\includegraphics[width=0.25\linewidth]{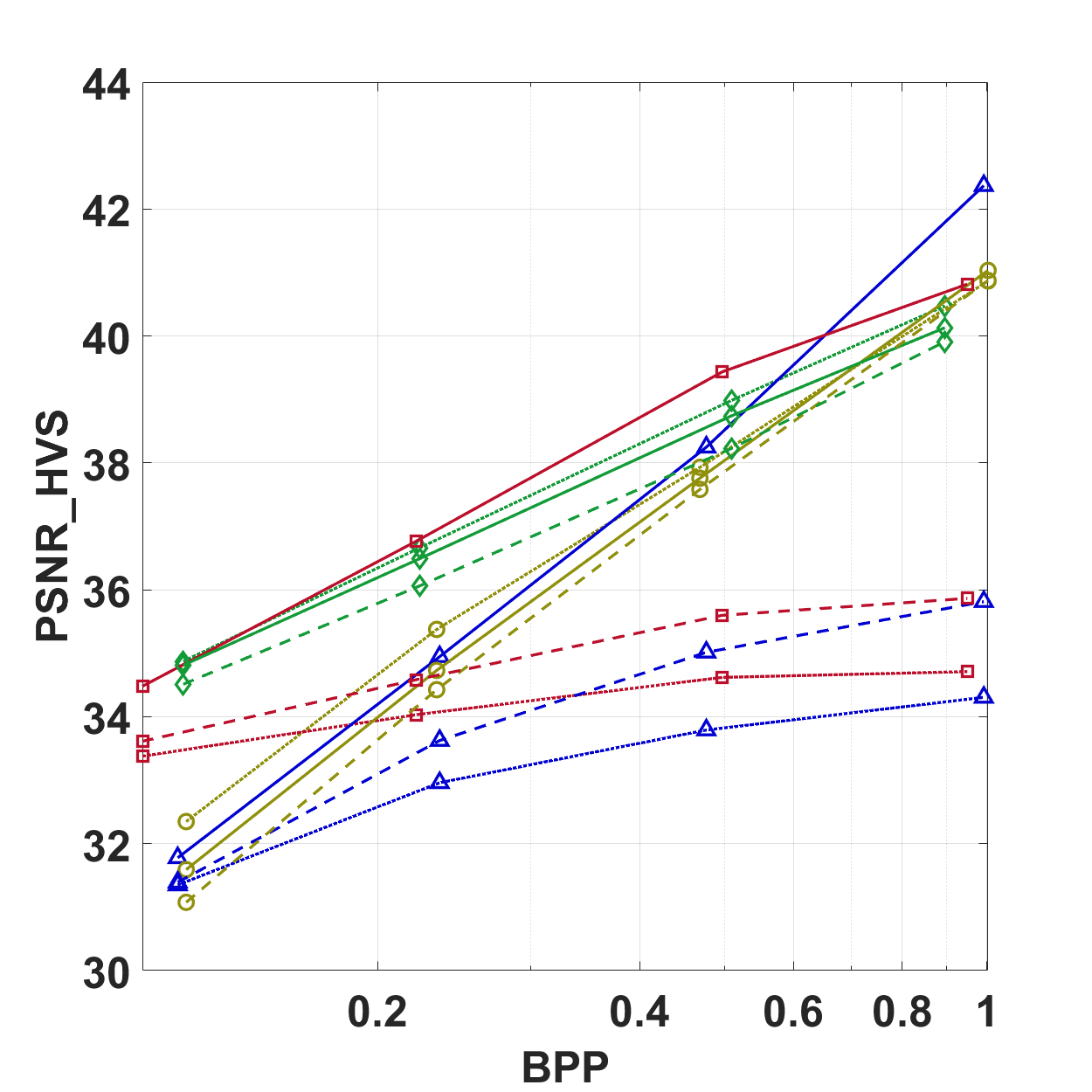}}
    \subfloat[\textit{MS-SSIM}]{\includegraphics[width=0.25\linewidth]{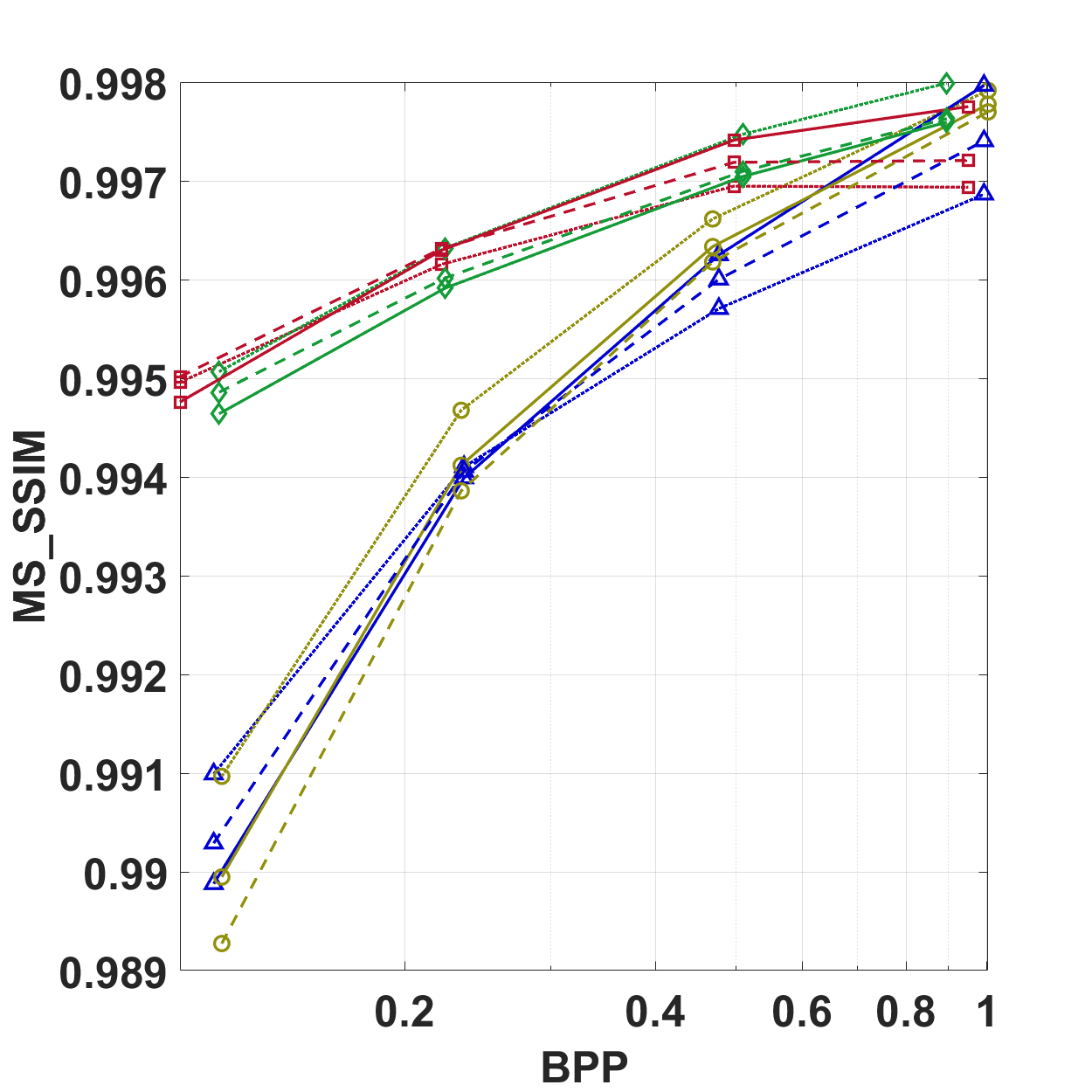}}
    \subfloat[\textit{FSIMc}]{\includegraphics[width=0.25\linewidth]{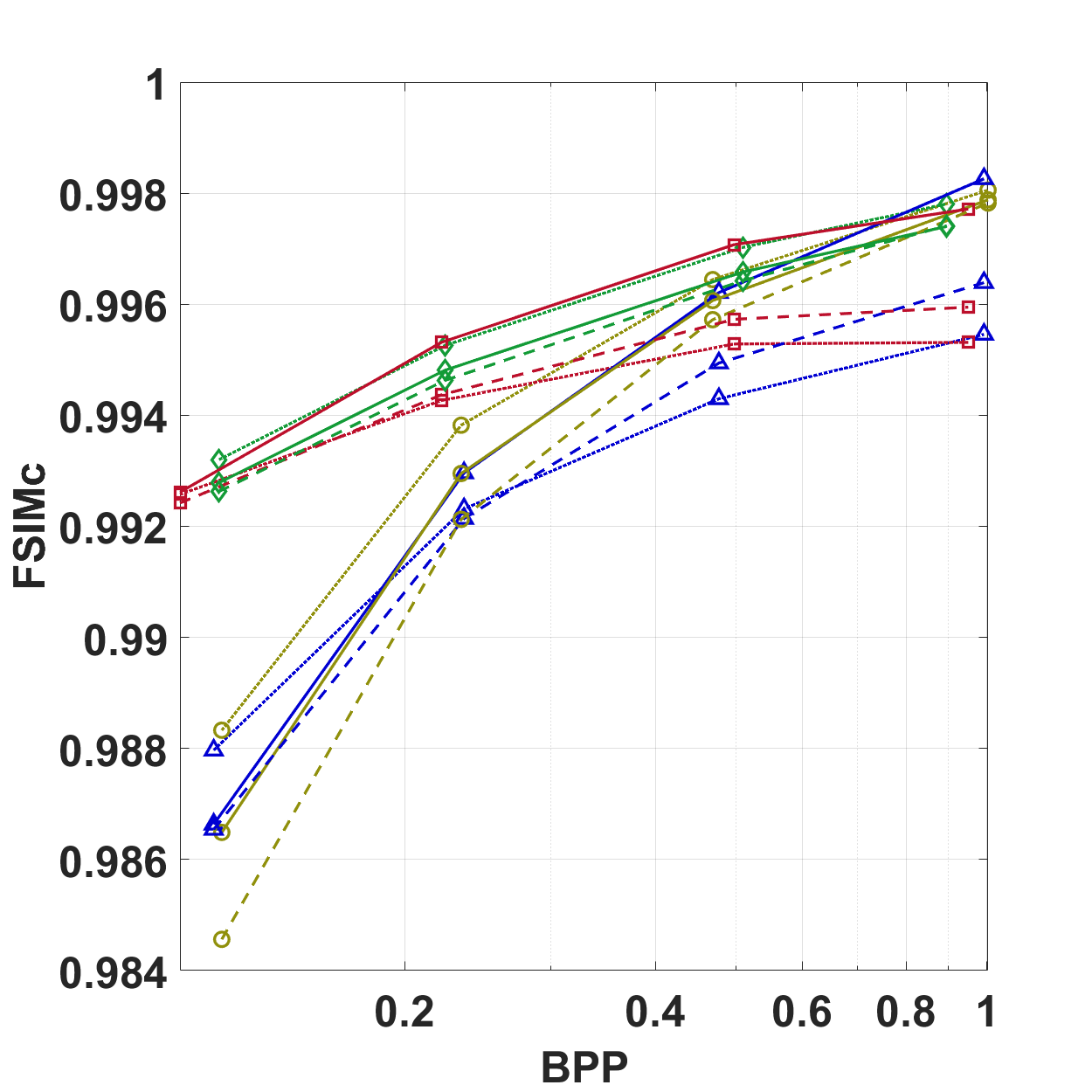}}
    \subfloat[\textit{IW-SSIM}]{\includegraphics[width=0.25\linewidth]{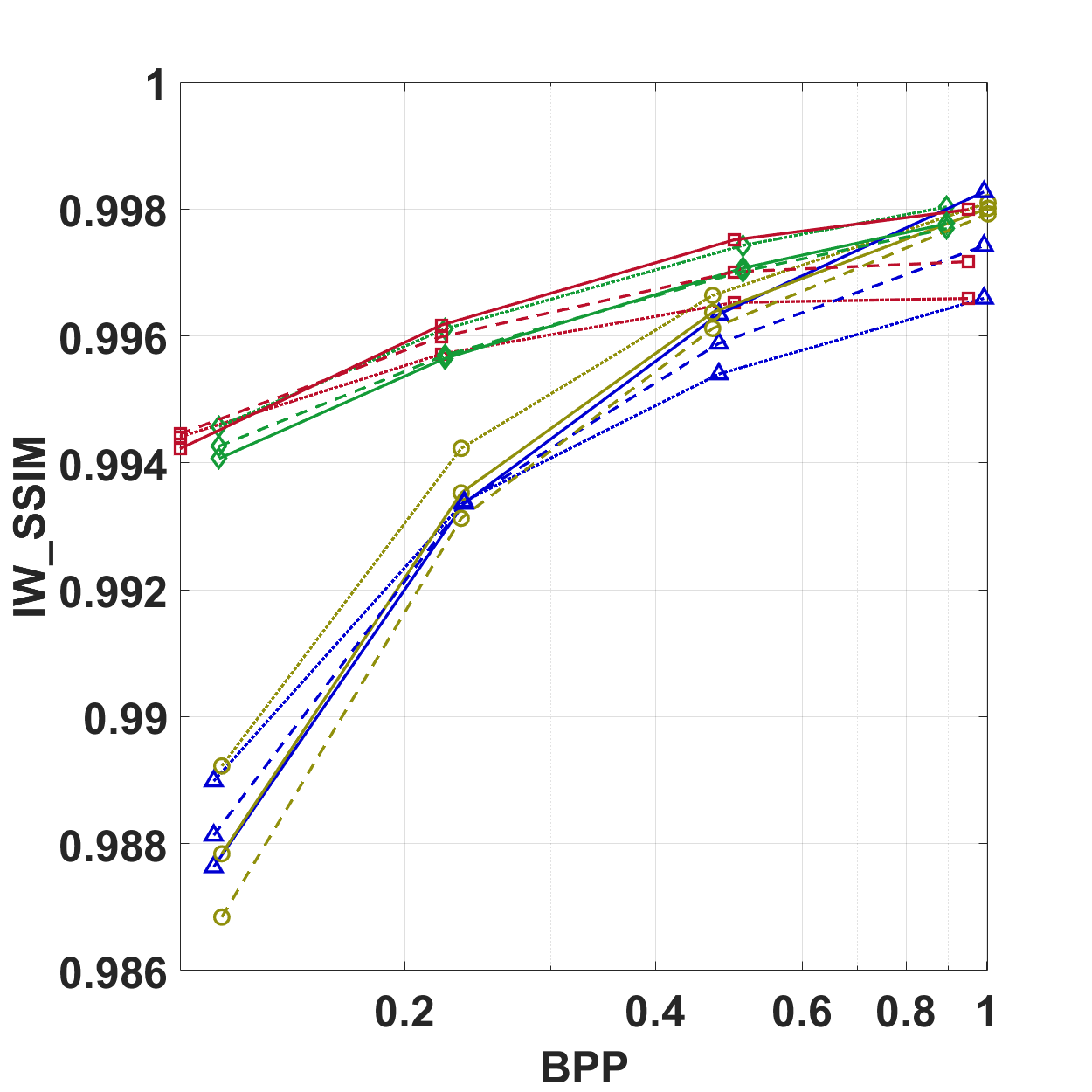}}\\
     \caption{Objective Quality Metrics for the Fountain Light Field 
     (legend in Fig. \ref{fig:Objective_bikes}).}
 \label{fig:Objective_fountain}

\centering
    \subfloat[\textit{PSNR HSV}]{\includegraphics[width=0.25\linewidth]{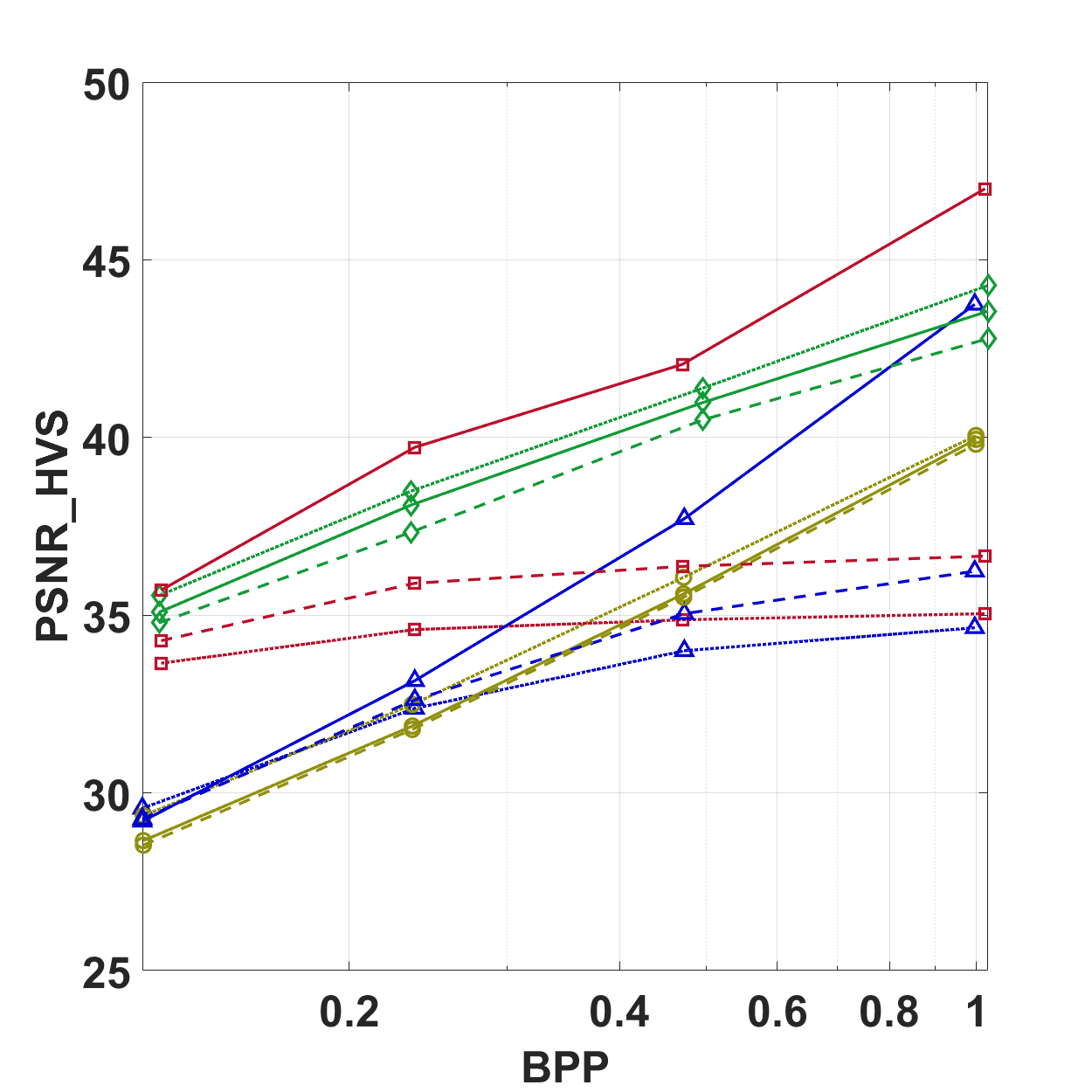}}
    \subfloat[\textit{MS-SSIM}]{\includegraphics[width=0.25\linewidth]{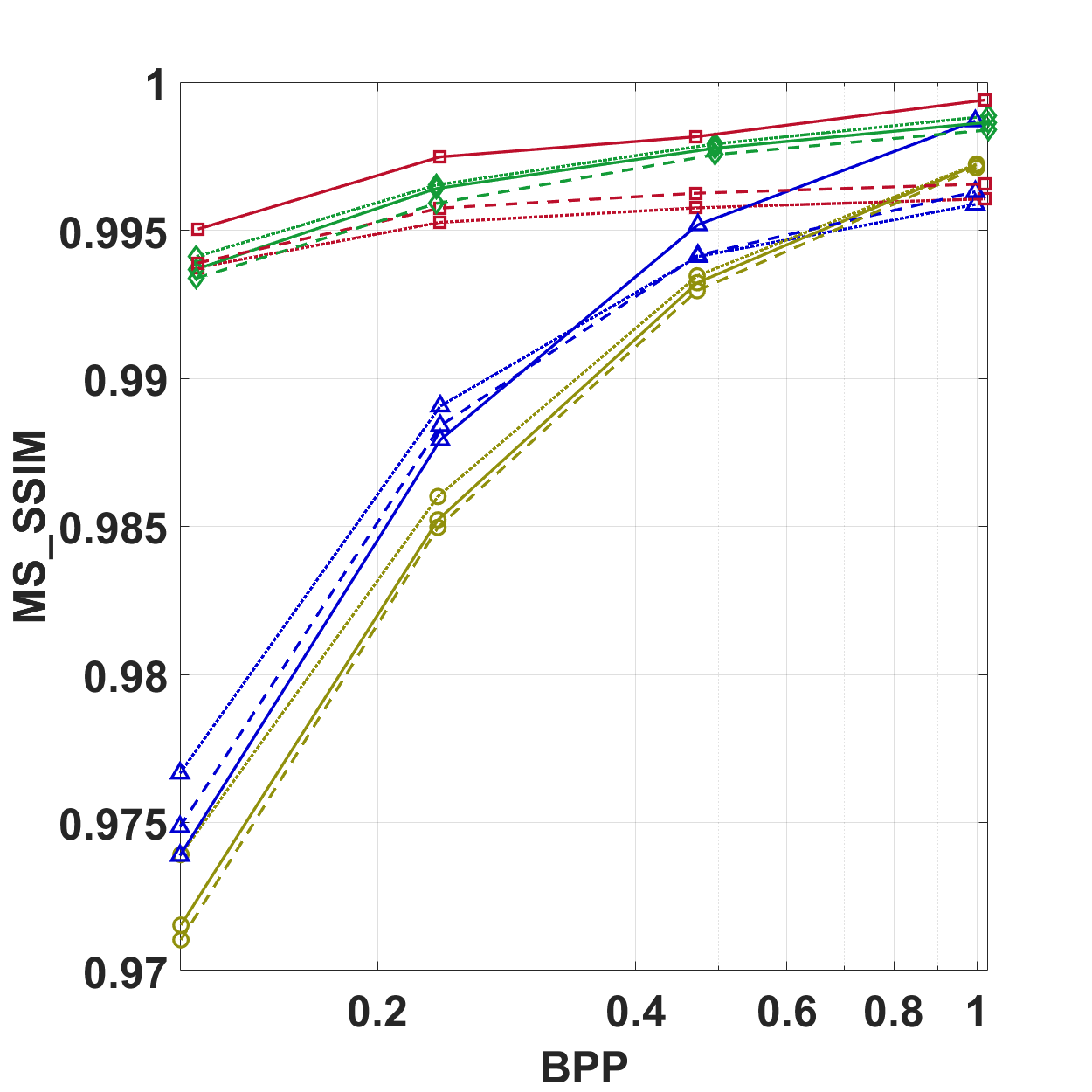}}
    \subfloat[\textit{FSIMc}]{\includegraphics[width=0.25\linewidth]{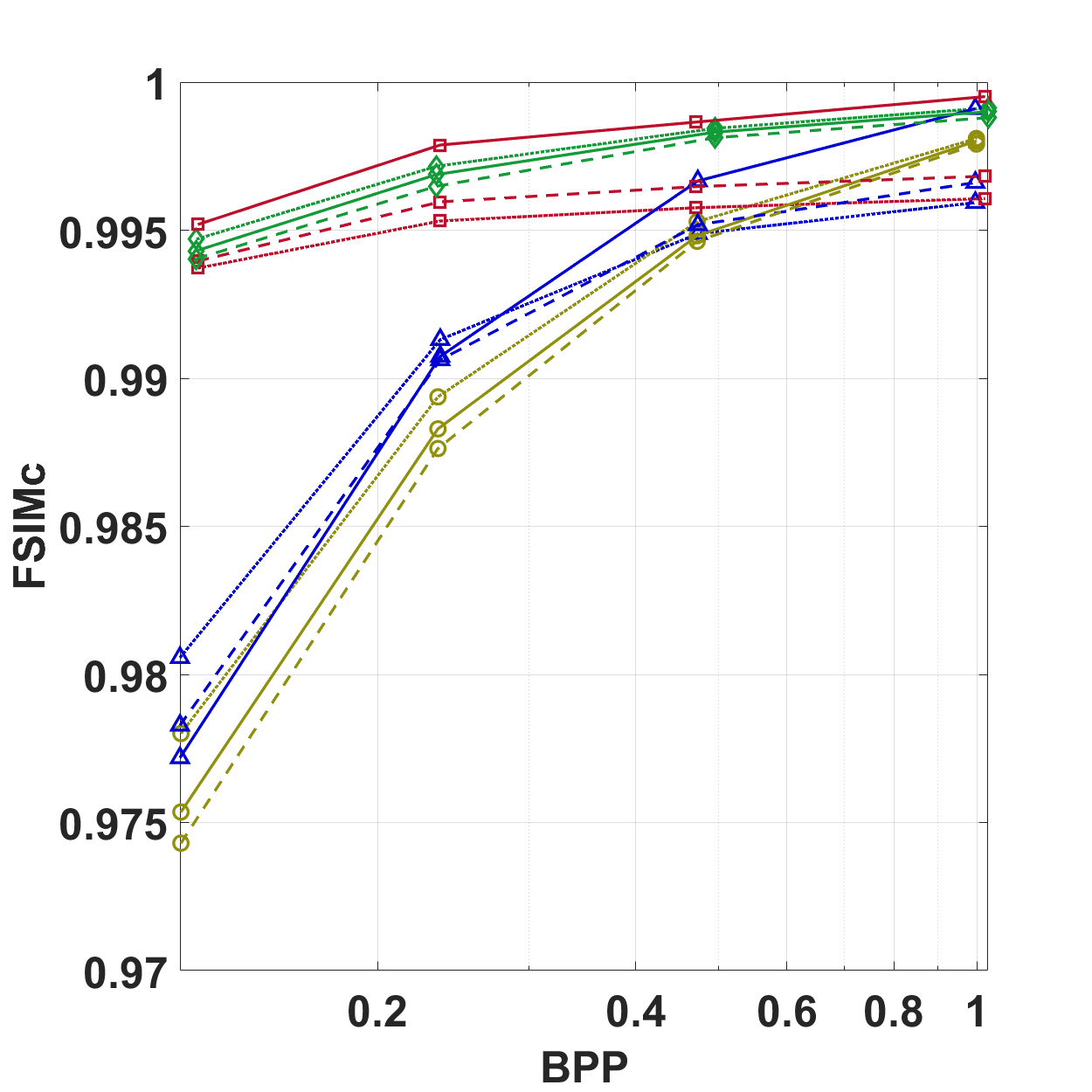}}
    \subfloat[\textit{IW-SSIM}]{\includegraphics[width=0.25\linewidth]{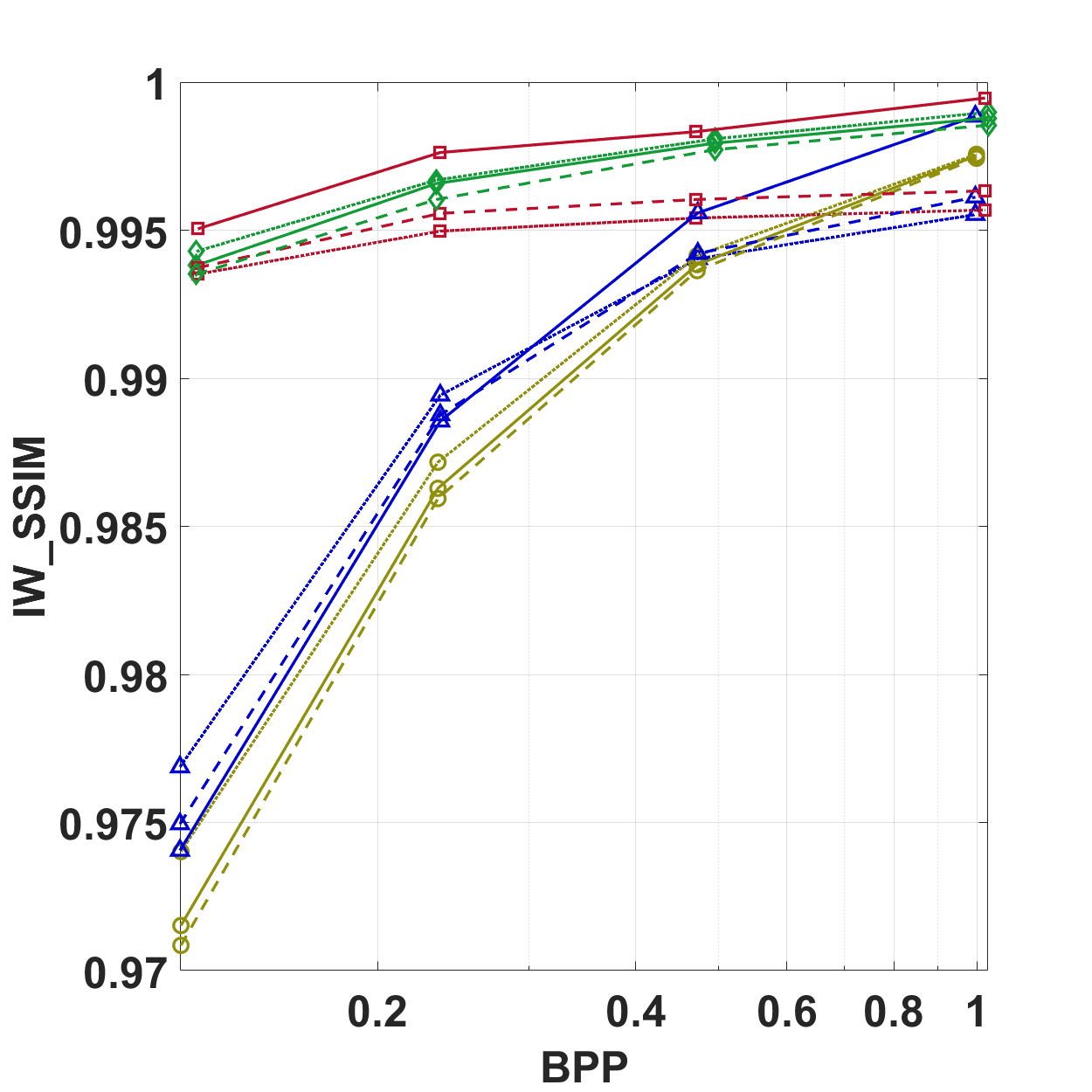}}\\
     \caption{Objective Quality Metrics for the Bicycle Light Field 
     (legend in Fig. \ref{fig:Objective_bikes}).}
 \label{fig:Objective_bicycle}

\centering
    \subfloat[\textit{PSNR HSV}]{\includegraphics[width=0.25\linewidth]{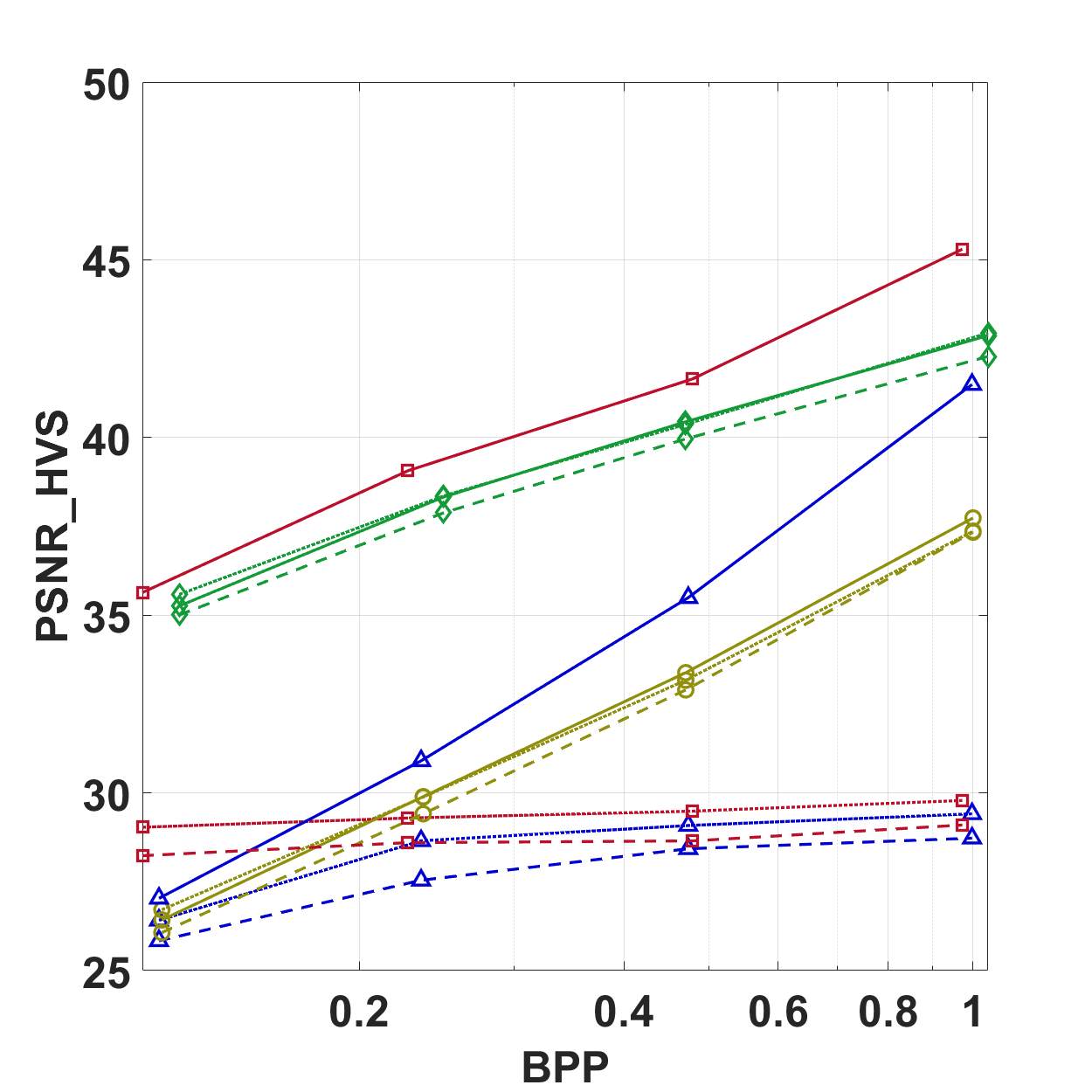}}
    \subfloat[\textit{MS-SSIM}]{\includegraphics[width=0.25\linewidth]{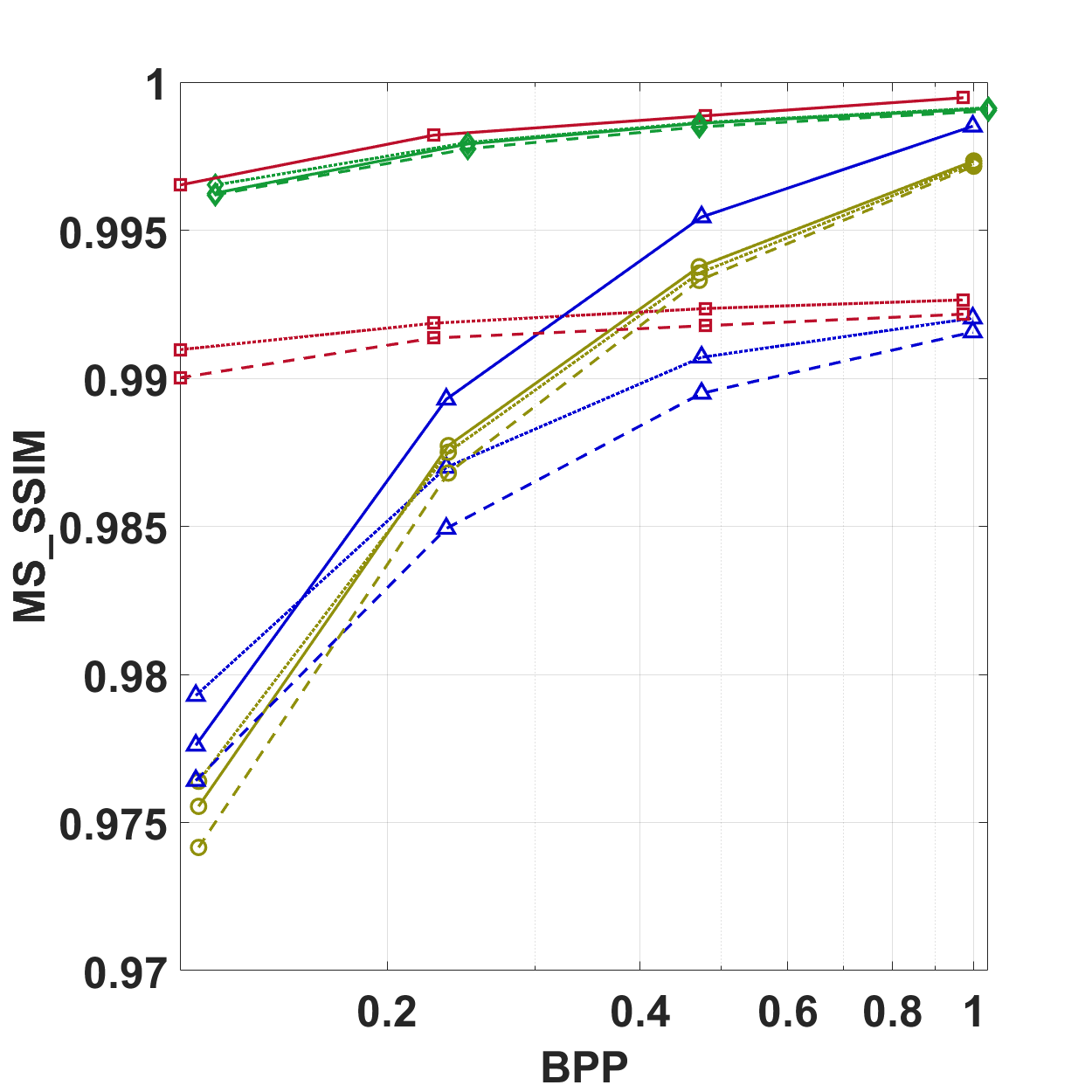}}
    \subfloat[\textit{FSIMc}]{\includegraphics[width=0.25\linewidth]{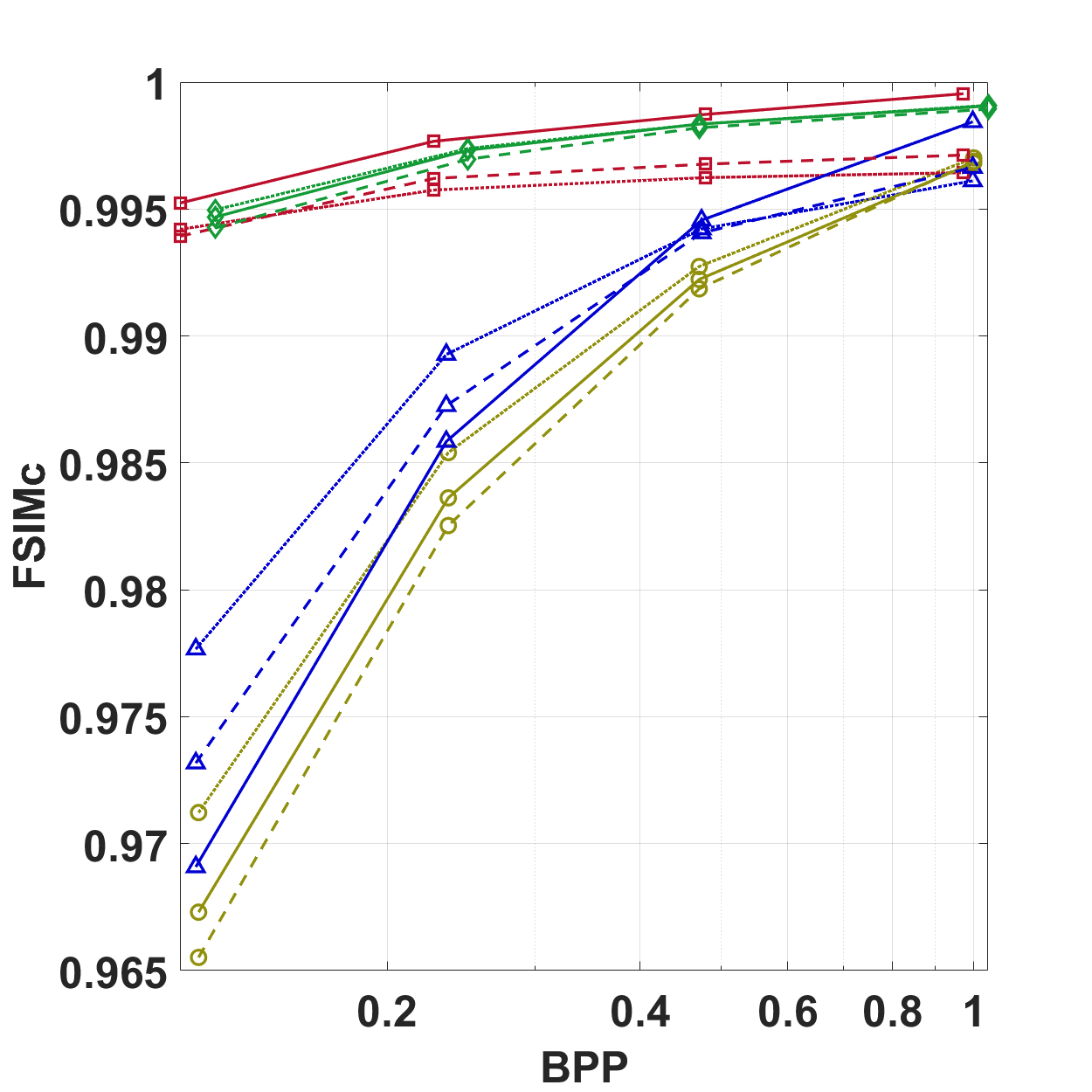}}
    \subfloat[\textit{IW-SSIM}]{\includegraphics[width=0.25\linewidth]{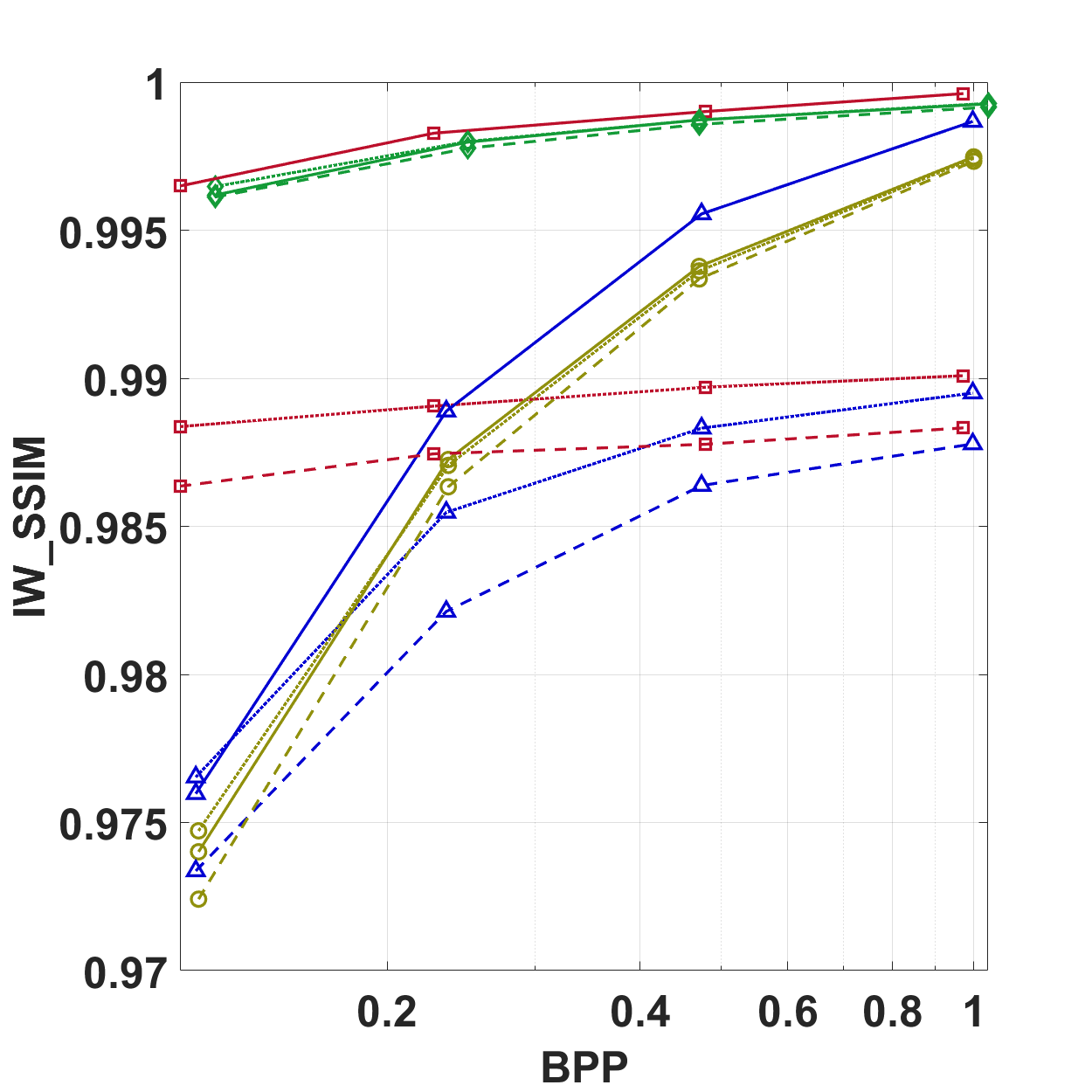}}\\
     \caption{Objective Quality Metrics for the Sideboard Light Field 
     (legend in Fig. \ref{fig:Objective_bikes}).}
 \label{fig:Objective_sideboard}

\end{figure*}

Each plot in Figs.~\ref{fig:Objective_bikes}, \ref{fig:Objective_fountain}, \ref{fig:Objective_bicycle} and \ref{fig:Objective_sideboard} corresponds to a different light field and contains 12 curves. These curves represent the four encoding configurations applied: JPEG Pleno5$\times$5, JPEG Pleno3$\times$3, VVC5$\times$5, and VVC3$\times$3. The 5$\times$5 refers to encoding the complete light field directly, while the 3$\times$3  indicates that a sparsely sampled light field was encoded and then underwent view synthesis. For each configuration, three types of views are shown: S (original compressed views), X (first-generation synthesized views), and O (second-generation synthesized views). Although the 5$\times$5 encoded light fields do not require view 
 synthesis, the same S/X/O notation is applied for easier comparison, as the corresponding views occupy the same spatial views in the light field as their 3$\times$3 counterparts. It is important to note that comparisons should only be made across the same view type, as each views represents a different angular visualization.


\begin{figure*}
\centering
    \subfloat[\textit{Fountain Reference}]{\includegraphics[width=0.40\linewidth]{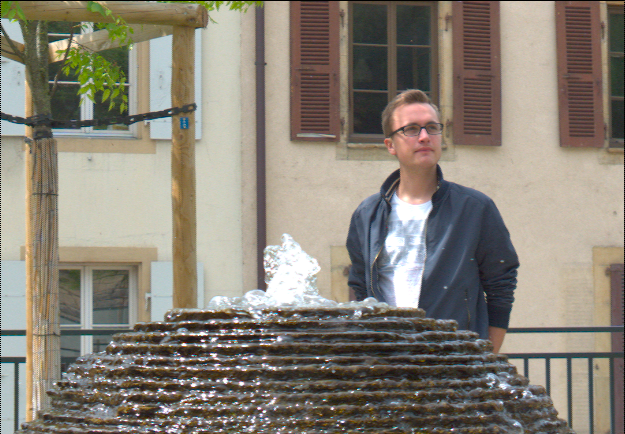}} \quad\quad
    \subfloat[\textit{Fountain with Compression Artifacts}]{\includegraphics[width=0.40\linewidth]{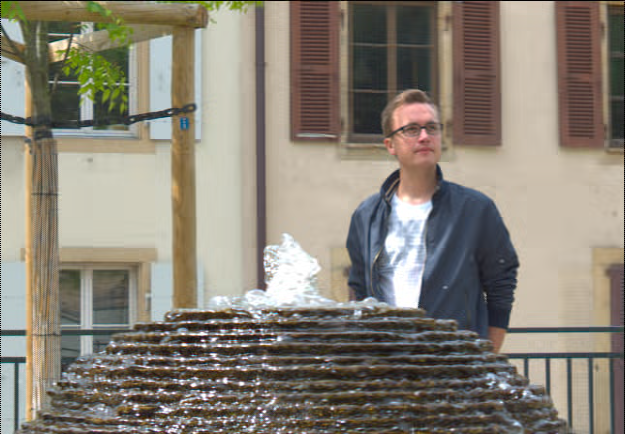}} \quad\\
    \caption{Compression artifacts in the \textit{Fountain} light field.}
\label{fig:artifacts_compression}
\end{figure*}

\begin{figure*}
\centering
    \subfloat[\textit{Sideboard Reference}]{\includegraphics[width=0.40\linewidth]{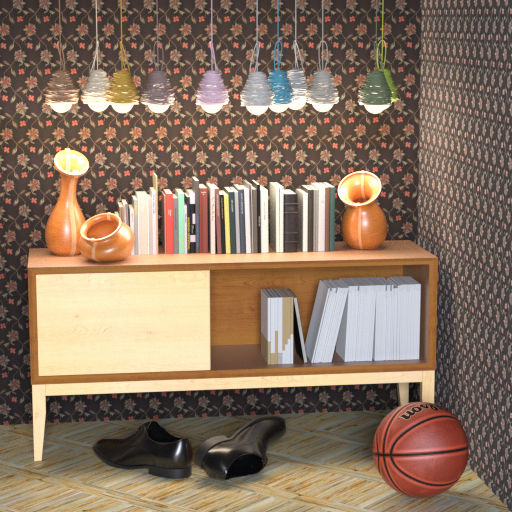}} \quad\quad
    \subfloat[\textit{Sideboard with View Synthesis Artifacts}]{\includegraphics[width=0.40\linewidth]{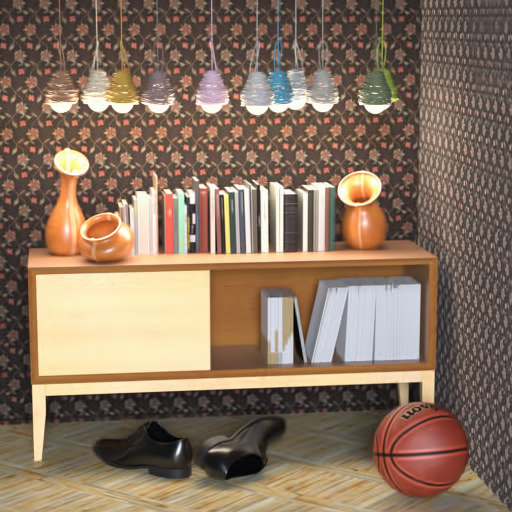}} \quad\\
    \caption{View synthesis artifacts in the \textit{Sideboard} light field.}
\label{fig:artifacts_synthesis}
\end{figure*}

\subsection{Subjective Results}
\subsubsection{Cross-Method Comparisons}
The subjective results presented in Fig.~\ref{fig:Subjective_CM_Pleno} and Fig.~\ref{fig:Subjective_CM_VVC} show a comparison between the compression methods considered in this study. One method consists on encoding the complete light field  (5$\times$5), while the the other encodes a sparsely sampled light field (3$\times$3). It then applies view synthesis to reconstruct the missing views. The 5$\times$5 method achieves a better performance. This is consistent for both codecs, across the considered light fields. By observing the different view types in the 3$\times$3 method, it can be observed that each synthesis stage decreases the perceptive quality. This is more noticeable for synthetic light fields, where the largest gap between the synthesized views and their respective 5$\times$5 counterparts is observed on the \textit{Sideboard} light field. In this light field, a very perceptible distortion caused by view synthesis is present in every 3$\times$3 view.

Compression introduces expected types of artifacts that can be observed in both the 3$\times$3 and 5$\times$5 methods, particularly at lower bitrates, as illustrated in the example of Fig.\ref{fig:artifacts_compression}. A particularly noticeable compression distortion can be observed in the water drops in front of the man's jacket in the \textit{Fountain} light field.
In contrast, view synthesis generates unique artifacts. For instance, in the \textit{Sideboard} light field, a distinctive square-shaped distortion appears consistently in the top-right corner of the synthesized views, affecting view types 
X (first stage synthesized views) and O (second stage synthesized views) across all bitrates and for both codecs, as shown in Fig.~\ref{fig:artifacts_synthesis}. This artifact is easily  observed using image flickering, but it is not very easy  to observe if two pseudo-videos are shown  side by side as  it is usual in  light field subjective quality evaluation~\cite{viola2017comparison}.


A quality stabilization can often be seen between the middle range bitrates for VVC (Fig. \ref{fig:Subjective_CM_VVC}). In the context of subjective quality evaluation, subjects tend to divide their evaluation accompanied with ``not sure scores" between stimuli that present similar distortions, resulting in quality stabilization. A slight quality decrease with the bit rate can even be observed for VCC. 

\subsubsection{Cross-Codec Comparisons}

The plots in Fig.~\ref{fig:Subjective_CC_3x3} and Fig.~\ref{fig:Subjective_CC_5x5} show the comparisons between codecs, JPEG Pleno and VVC. 
These plots highlight VVC as the best performing codec across both methods,  encoding the complete light field (5$\times$5) and encoding a sparsely sampled light field (3$\times$3) followed by views synthesis to recreate a 5$\times$5 views light field. 
All light fields quality converge with the increase of the bitrate.
VVC tends to stabilize the perceived quality for lower bit rates when compared to JPEG Pleno, starting from mid-range bitrates onward. 
These observations align with objective metrics.
Once again it was observed that in the presence of similar artifacts subjects tend to divide between the two options and it is also observed that there is a growth of the selection of the ``Not Sure" option. It is important to add that the quality levels of VVC on these bit rates almost do not differ and that although the flickering allows its observation, it becomes very difficult to understand which one has greater perceived quality.

The followed methodology   reveals also very small Confidence Intervals  which  demonstrates the reliability of  the proposed subjective evaluation model.
Furthermore, independently of the cases where an unexpected decrease of quality with bit rate happens, the computed confidence intervals still allowed to perceive a possible monotonicity of the quality.

Overall, VVC achieves a better perceived quality than JPEG Pleno, with a very significant gap at lower and medium bitrates. In terms of method, the 5$\times$5 configurations (VVC 5$\times$5 and JPEG Pleno5$\times$5) are consistently superior to their 3$\times$3 counterparts, as they avoid the view synthesis step and retain more of the original content. This results in higher perceived quality across the board. 
This also reveals that further research is needed for view synthesis methods that can be efficiently used in the context of light field coding.

\subsection{Objective results}


The objective results were obtained by computing four different quality assessment metrics, namely PSNR-HVS~\cite{PSNR}, MS-SSIM~\cite{MSSSIM}, FSIMc~\cite{FSIM} and IW-SSIM~\cite{IWSSIM}. The light fields \textit{Bikes}, \textit{Fountain\&Vincent2} and \textit{Bicycle} exhibit similar behavior, as represented in Fig.~\ref{fig:Objective_bikes} to \ref{fig:Objective_bicycle}.

Across all metrics, VVC consistently outperforms JPEG Pleno. Most results converge at higher bitrates with the exception of the PSNR-HVS metric. Regarding the method-wise performance, comparing encoding the complete light field (5$\times$5) with encoding a sparsely sampled light field and then reconstructing its missing views using view synthesis(3$\times$3), the results vary depending on the metric.

IW-SSIM and MS-SSIM show little difference when it comes to the method used. FSIMc shows to be more sensitive to view synthesis, resulting in 5$\times$5 to perform slightly better than 3$\times$3. PSNR-HVS shows the largest disparity with the 5$\times$5 method consistently outperforming the 3$\times$3 method, especially at higher bitrates. 

In the case of the \textit{Sideboard} light field (Fig.~\ref{fig:Objective_sideboard}), VVC continues to outperform JPEG Pleno across all metrics. However,  a more noticeable quality drop is observed due to view synthesis, particularly evident when analyzing the individual view types. In this case, MS-SSIM and IW-SSIM clearly favor the 5$\times$5 method, in contrast to the other light fields where both methods performed similarly. This performance drop is caused by distortions introduced during view synthesis, particularly in the upper-right corner of the synthesized views. This type of artifact appears to be specific to this light field, as it is not observed in any of the others. This finding aligns with the subjective results for \textit{Sideboard} presented in Fig.\ref{fig:Subjective_CM_Pleno}-(d) and Fig.\ref{fig:Subjective_CM_VVC}-(d), where the synthesized views consistently exhibit significantly worse performance than their 5$\times$5 counterparts, indicating that participants also noticed these distortions and were influenced by them in their evaluations.

\begin{figure*}
\centering
    \subfloat[PSNR-HVS]{\includegraphics[width=0.25\linewidth]{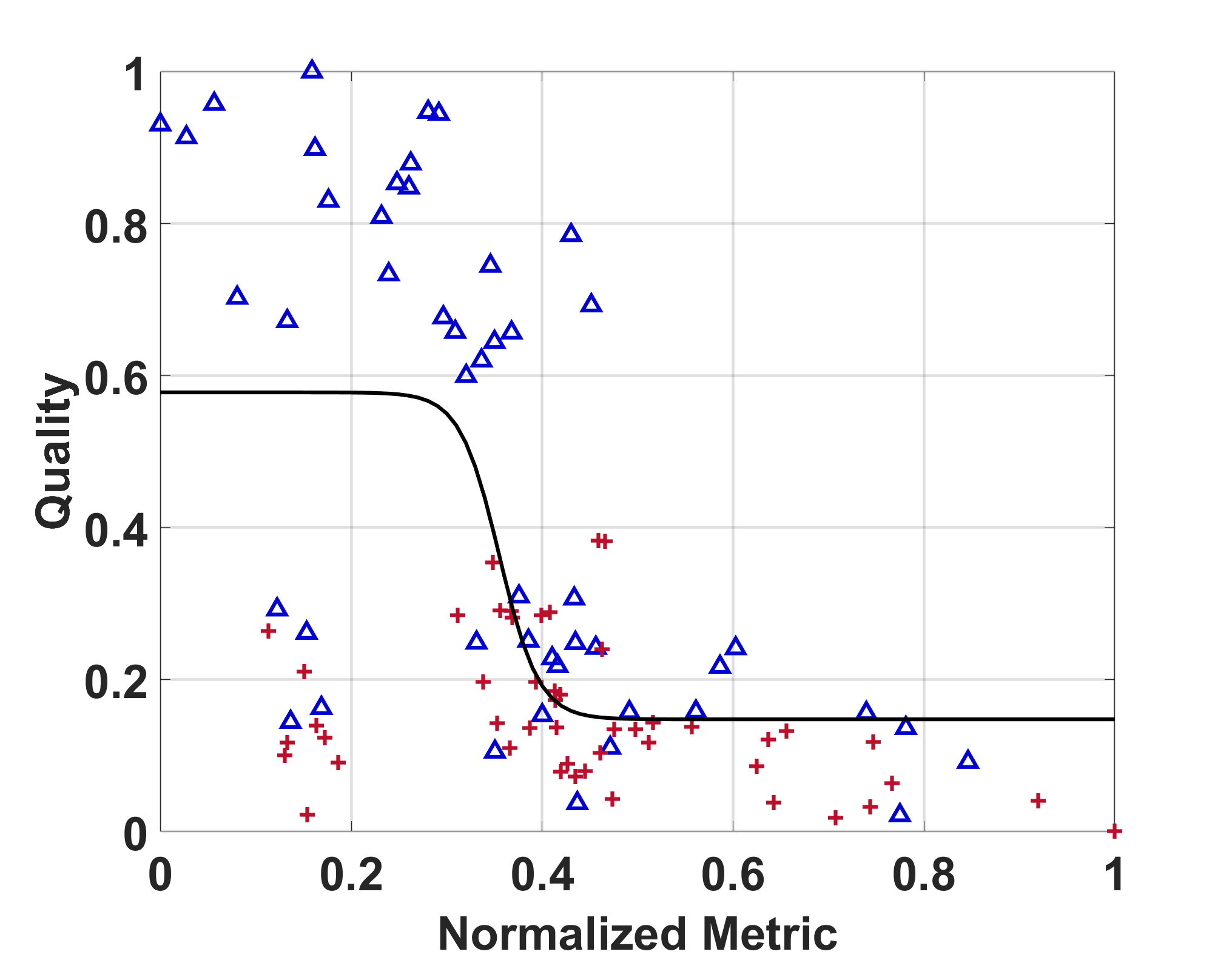}}
    \subfloat[MS-SSIM]{\includegraphics[width=0.25\linewidth]{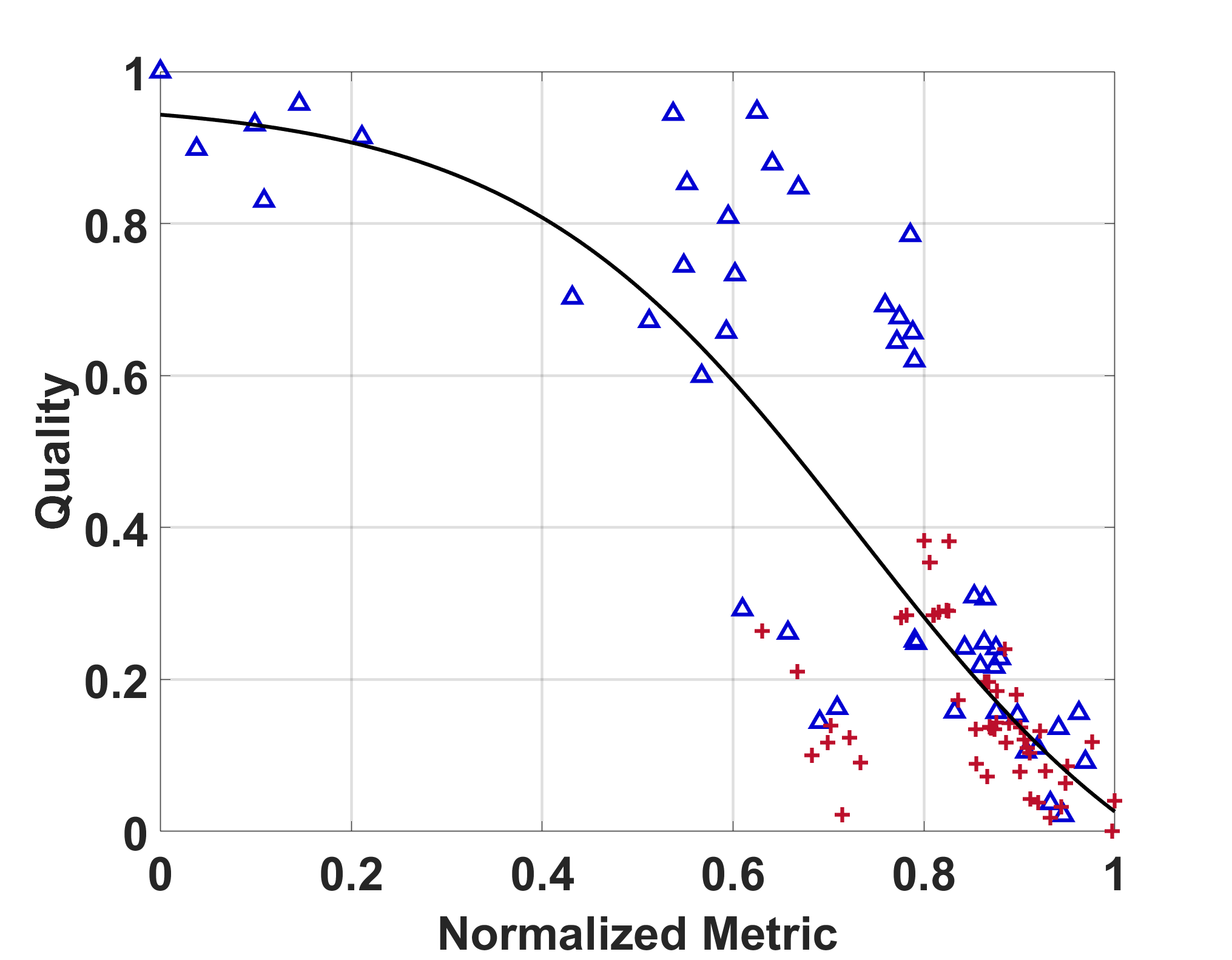}}
    \subfloat[FSIMc]{\includegraphics[width=0.25\linewidth]{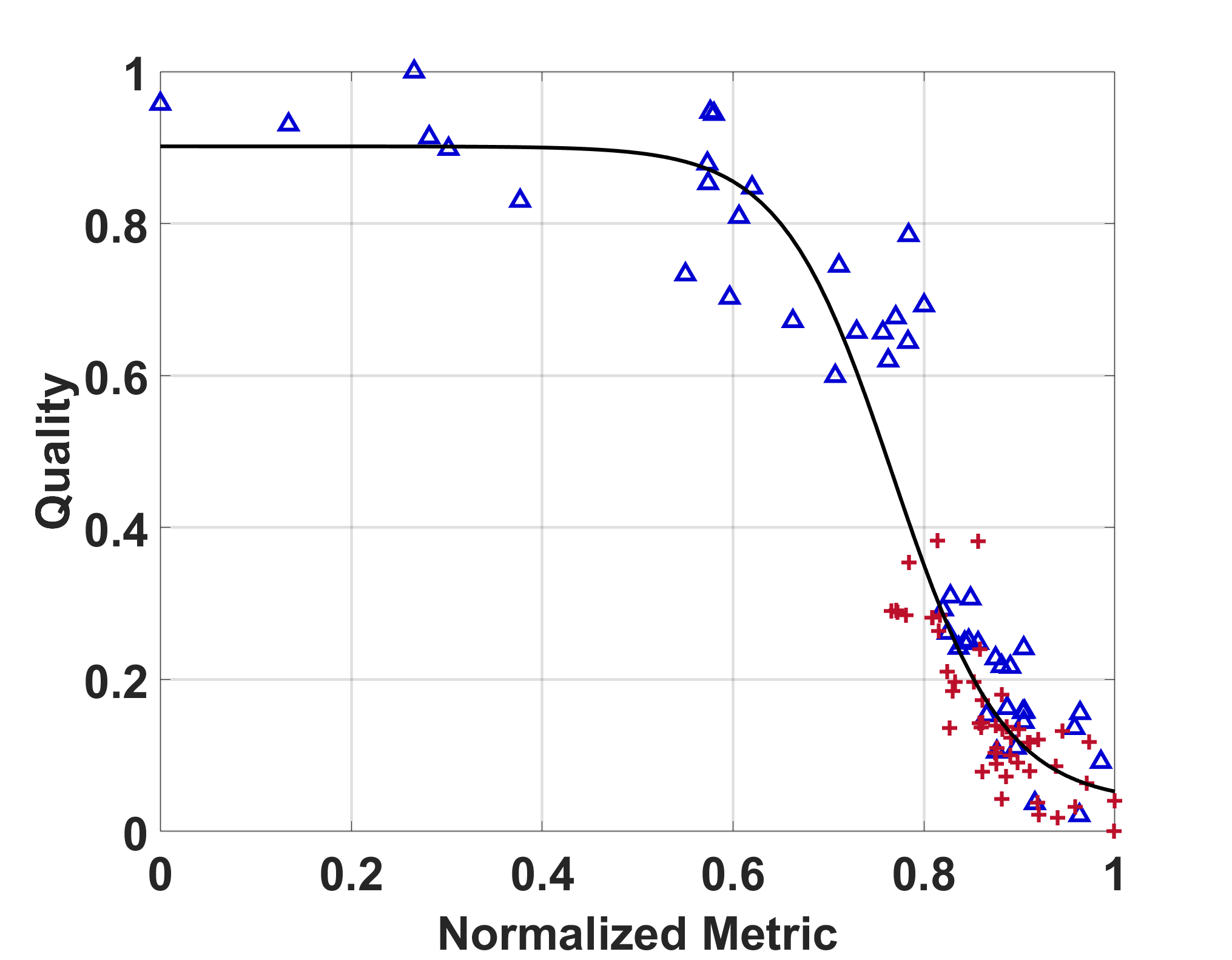}}
    \subfloat[IW-SSIM]{\includegraphics[width=0.25\linewidth]{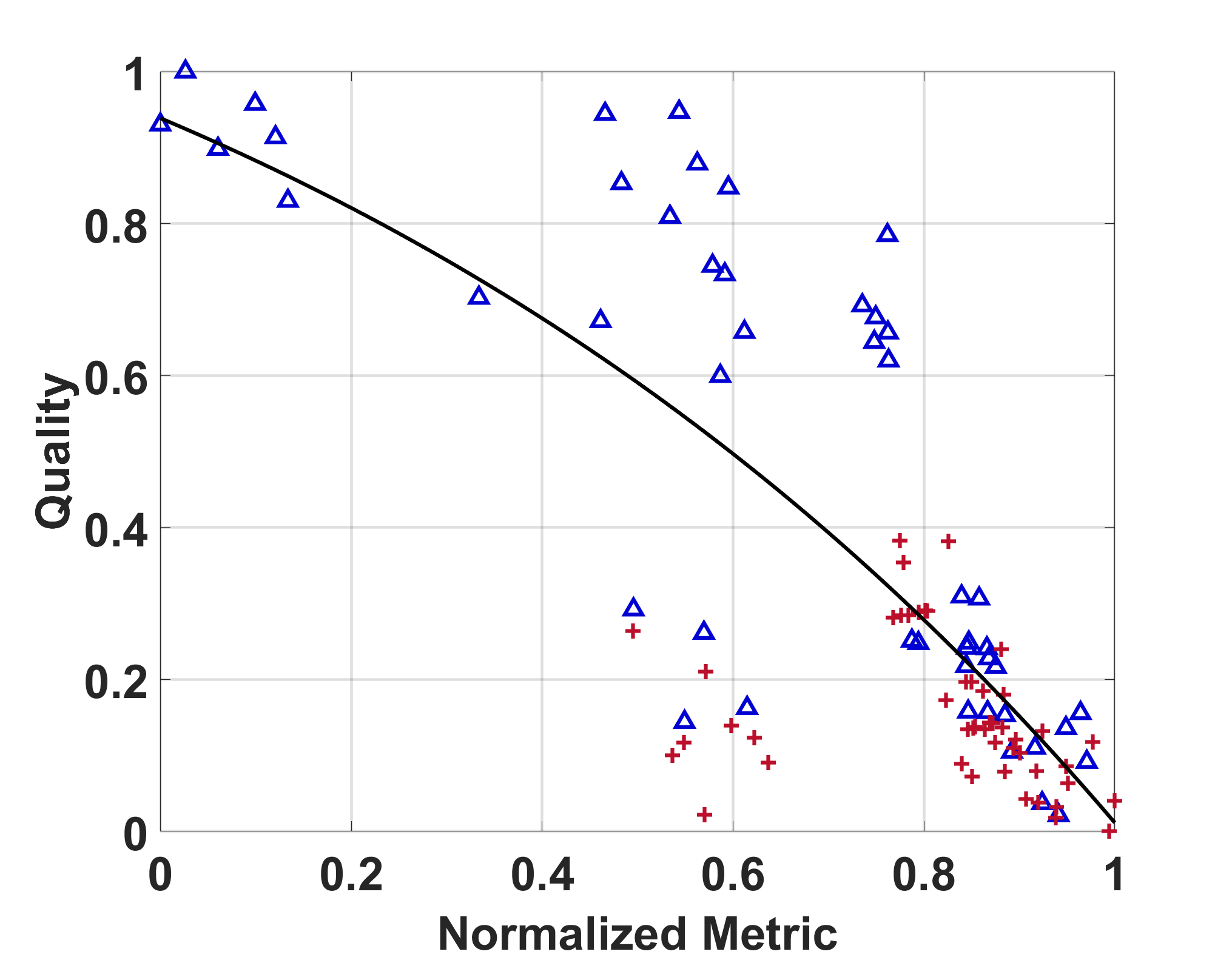}}\\
    \includegraphics[width=0.33\textwidth]{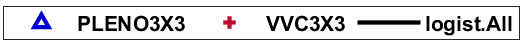}
    \caption{Logistic fitting for JPEG Pleno 3$\times$3 and VVC 3$\times$3.}
\label{fig:fitting_CC3x3}
\end{figure*}

\begin{figure*}
\centering
    \subfloat[PSNR-HVS]{\includegraphics[width=0.25\linewidth]{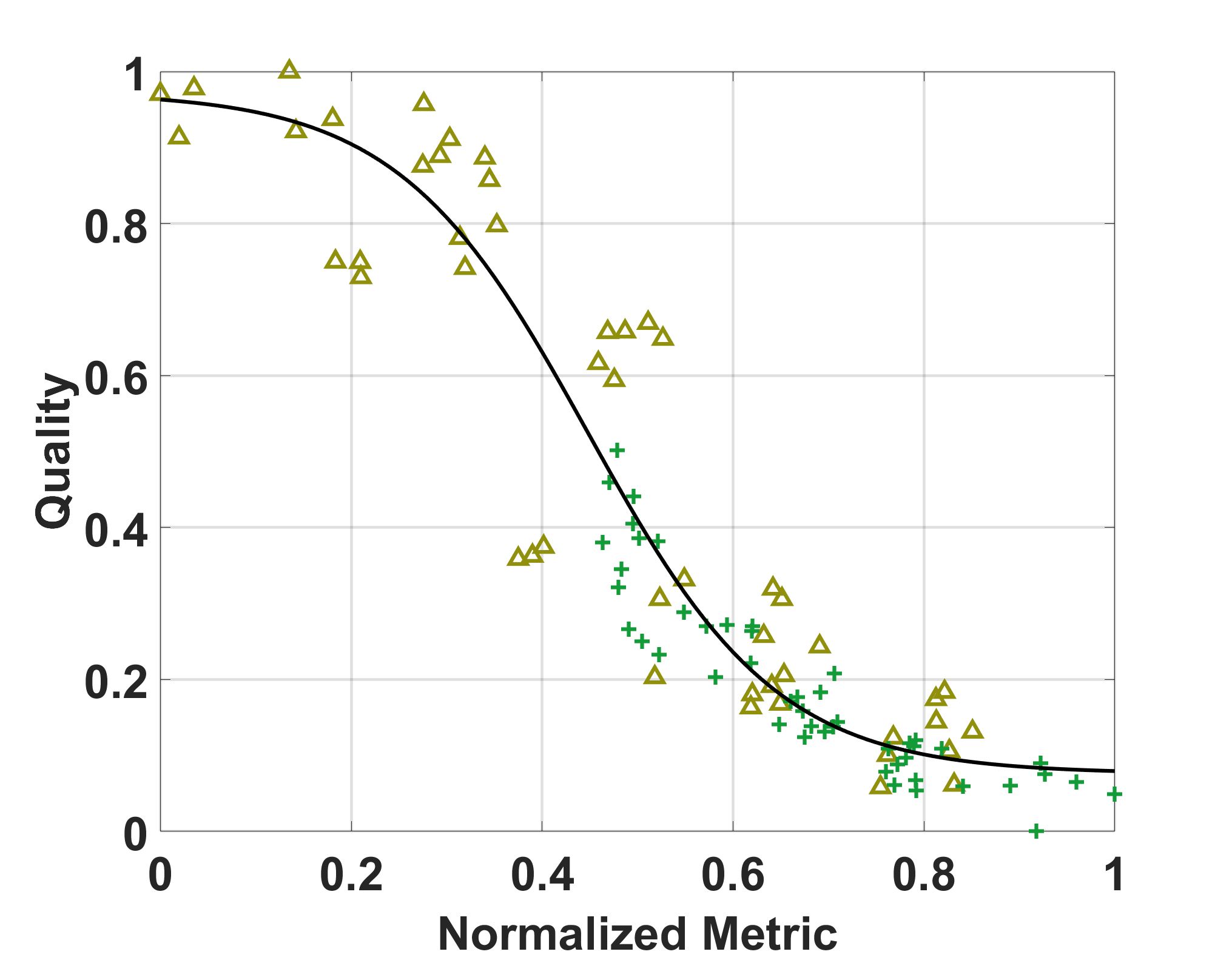}}
    \subfloat[MS-SSIM]{\includegraphics[width=0.25\linewidth]{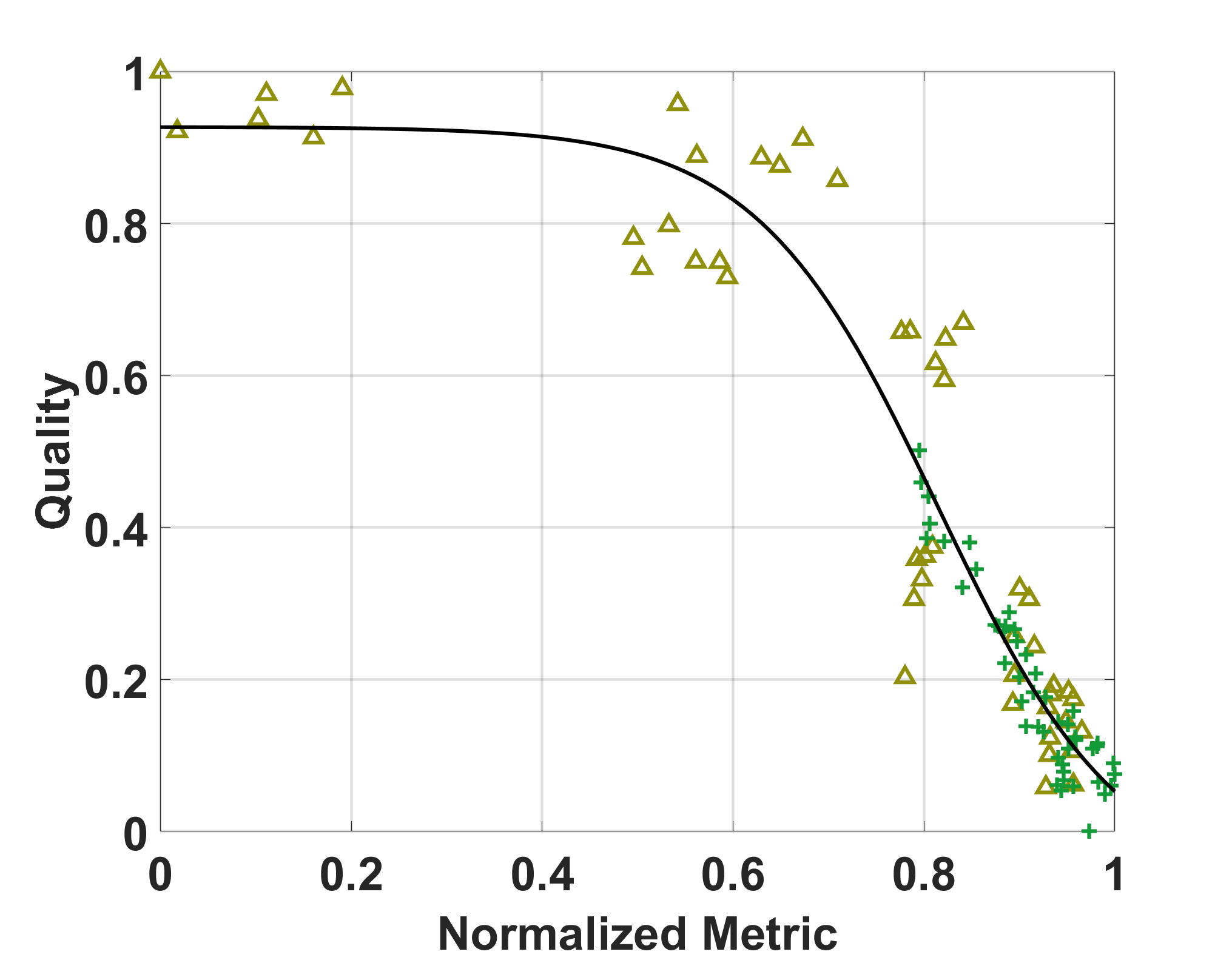}}
    \subfloat[FSIMc]{\includegraphics[width=0.25\linewidth]{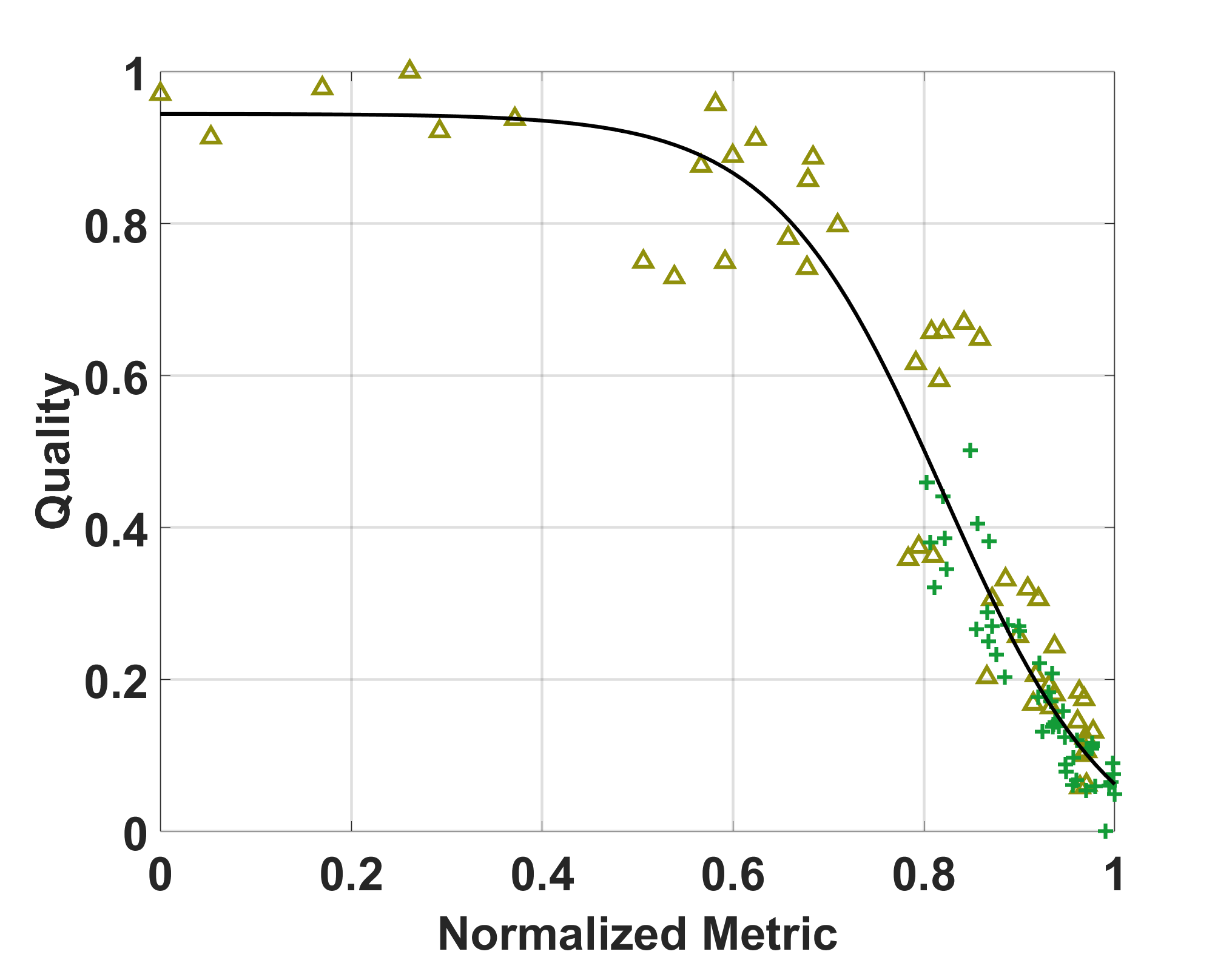}}
    \subfloat[IW-SSIM]{\includegraphics[width=0.25\linewidth]{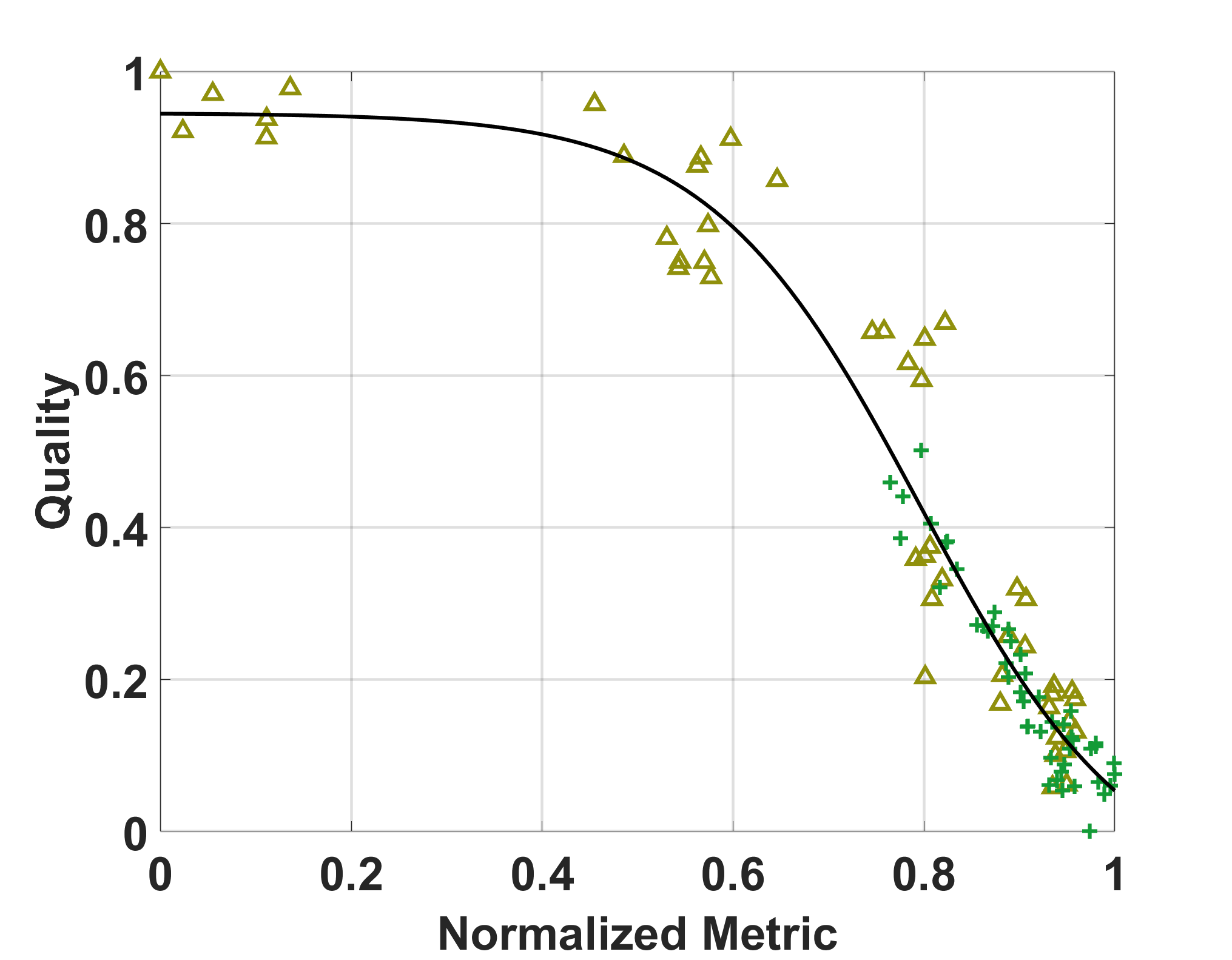}}\\
    \includegraphics[width=0.33\textwidth]{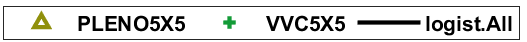}
    \caption{Logistic fitting for JPEG Pleno 5$\times$5 and VVC 5$\times$5.}
\label{fig:fitting_CC5x5}
\end{figure*}

\begin{figure*}
\centering
    \subfloat[PSNR-HVS]{\includegraphics[width=0.25\linewidth]{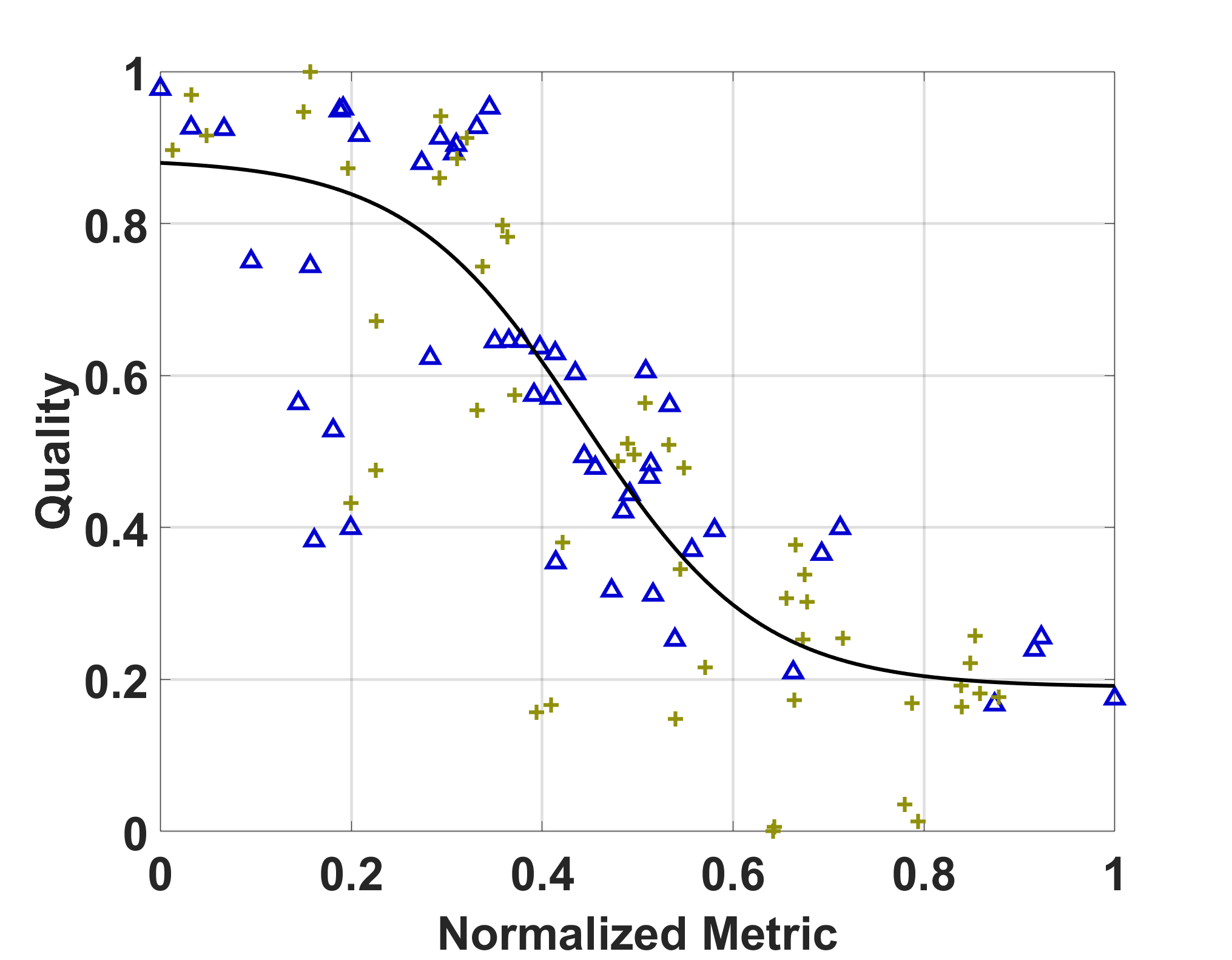}}
    \subfloat[MS-SSIM]{\includegraphics[width=0.25\linewidth]{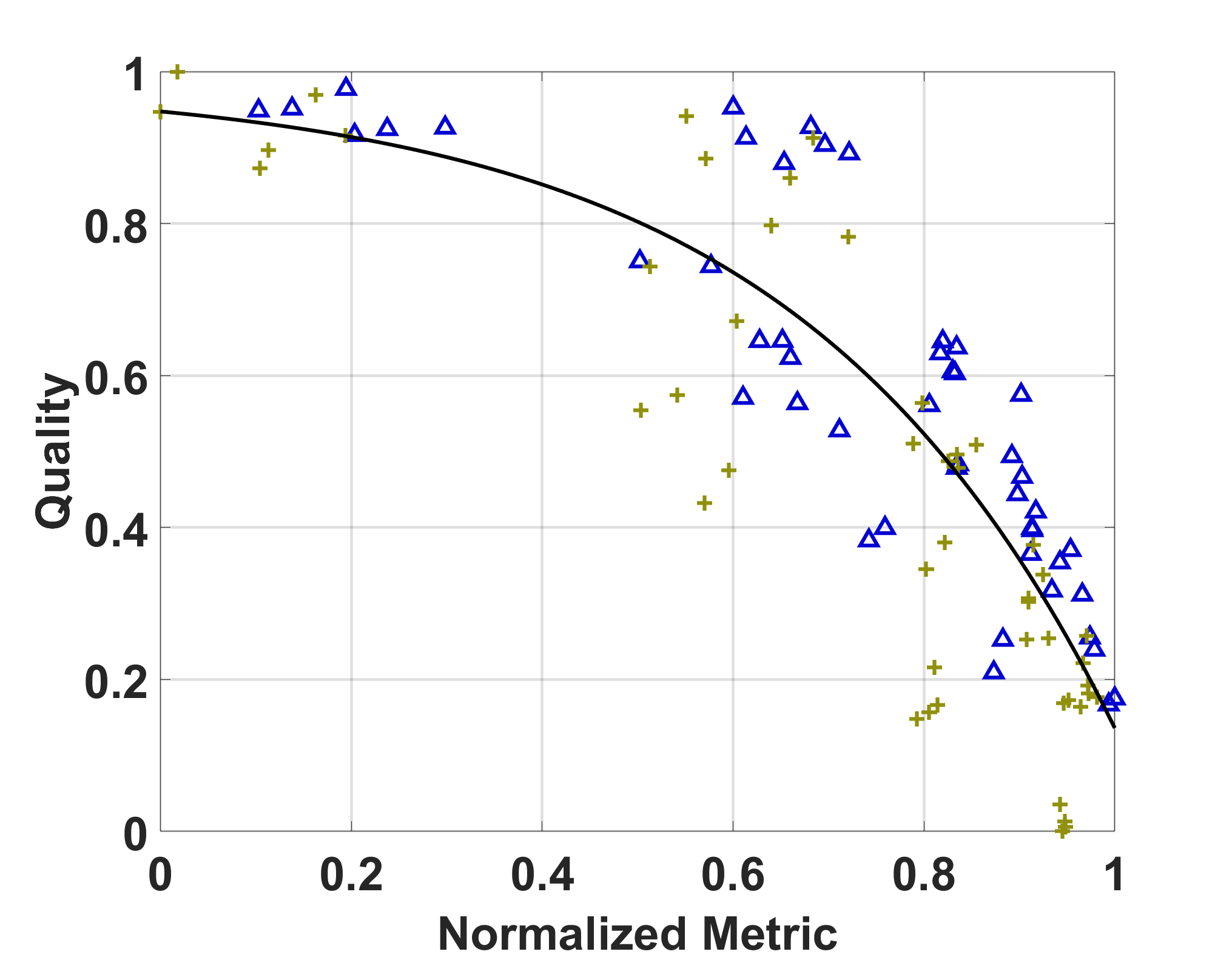}}
    \subfloat[FSIMc]{\includegraphics[width=0.25\linewidth]{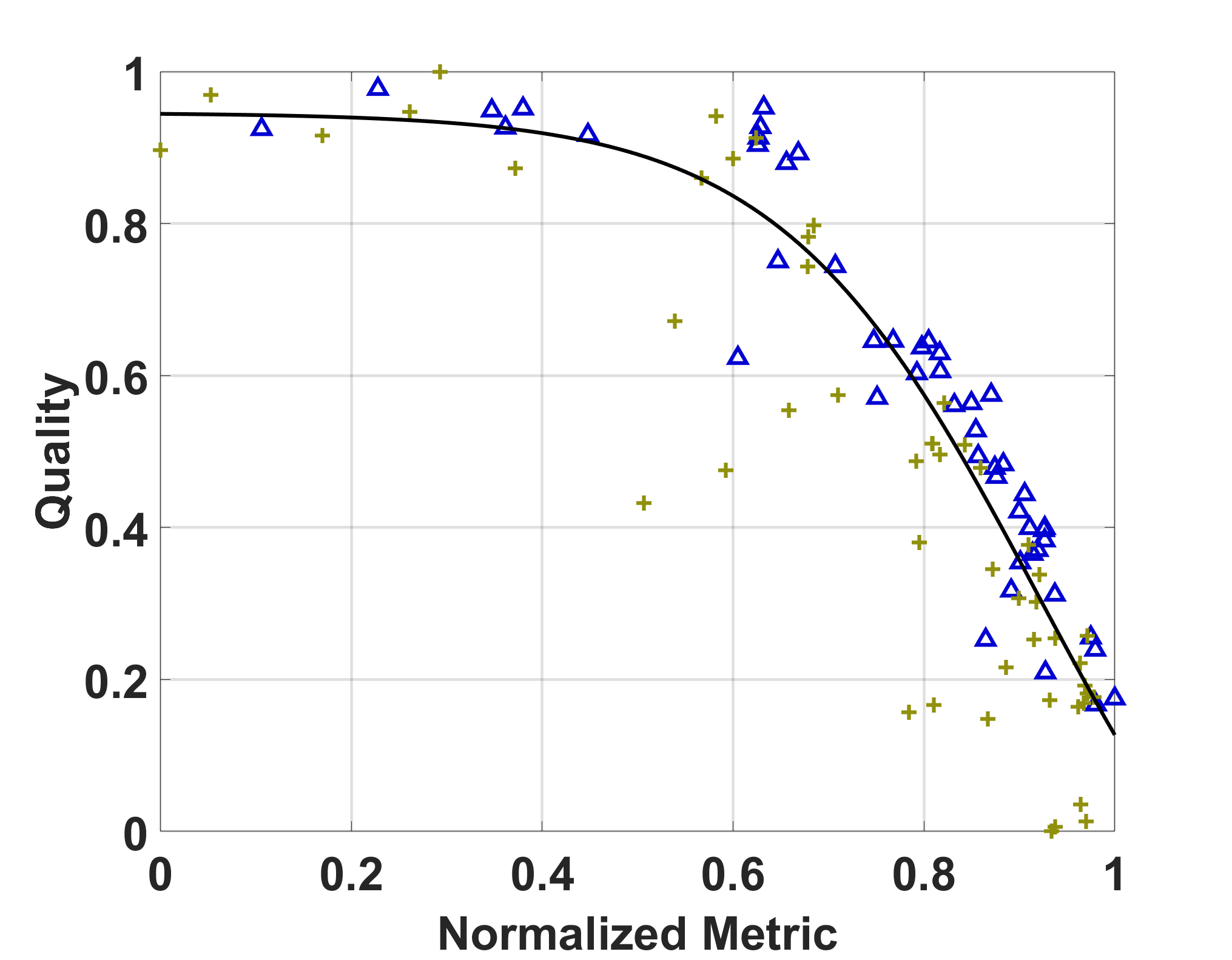}}
    \subfloat[IW-SSIM]{\includegraphics[width=0.25\linewidth]{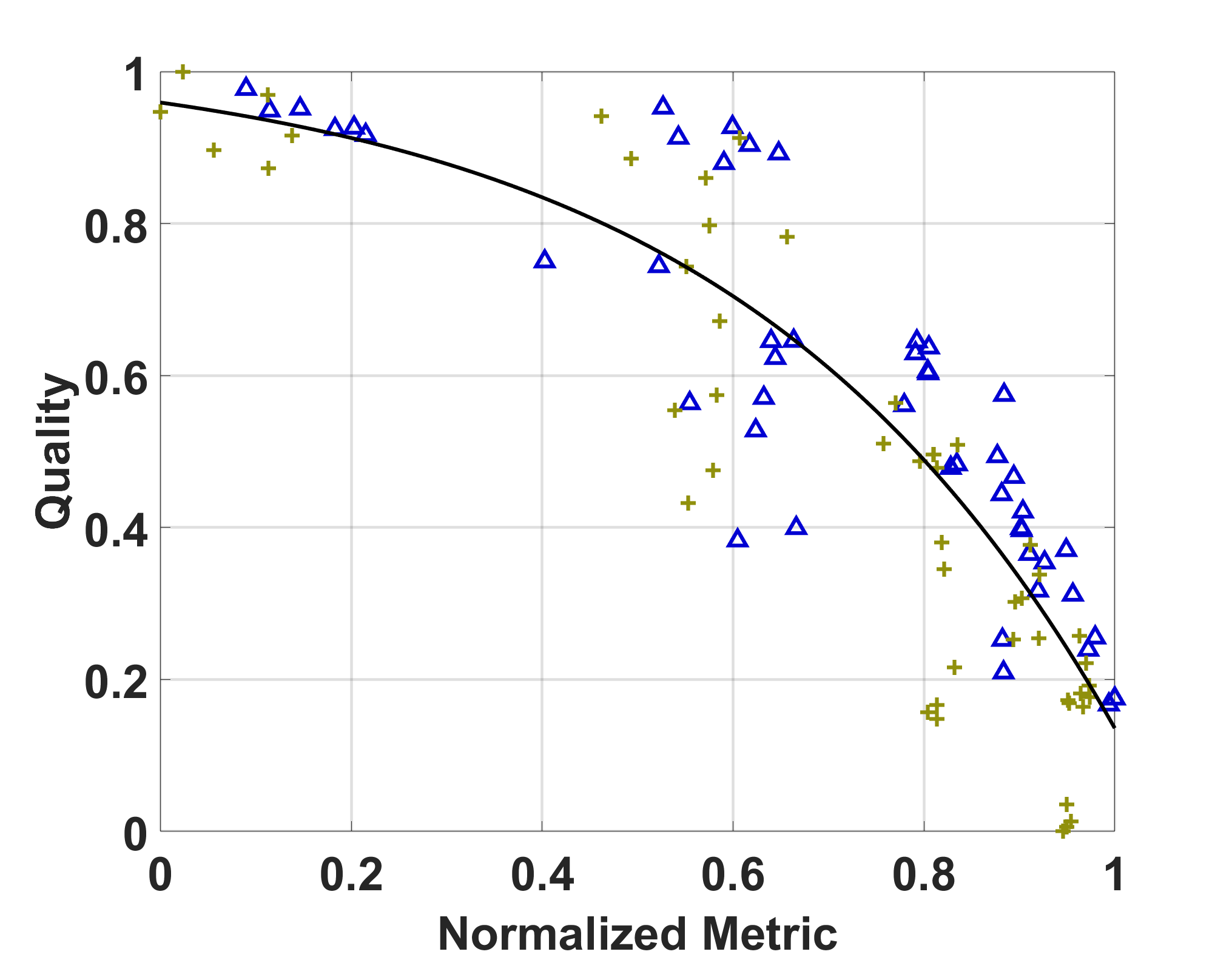}}\\
    \includegraphics[width=0.33\textwidth]{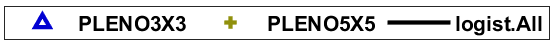}
    \caption{Logistic fitting for JPEG Pleno 5$\times$5 and JPEG Pleno 3$\times$3.}
\label{fig:fitting_CMPleno}
\end{figure*}

\begin{figure*}
\centering
    \subfloat[PSNR-HVS]{\includegraphics[width=0.25\linewidth]{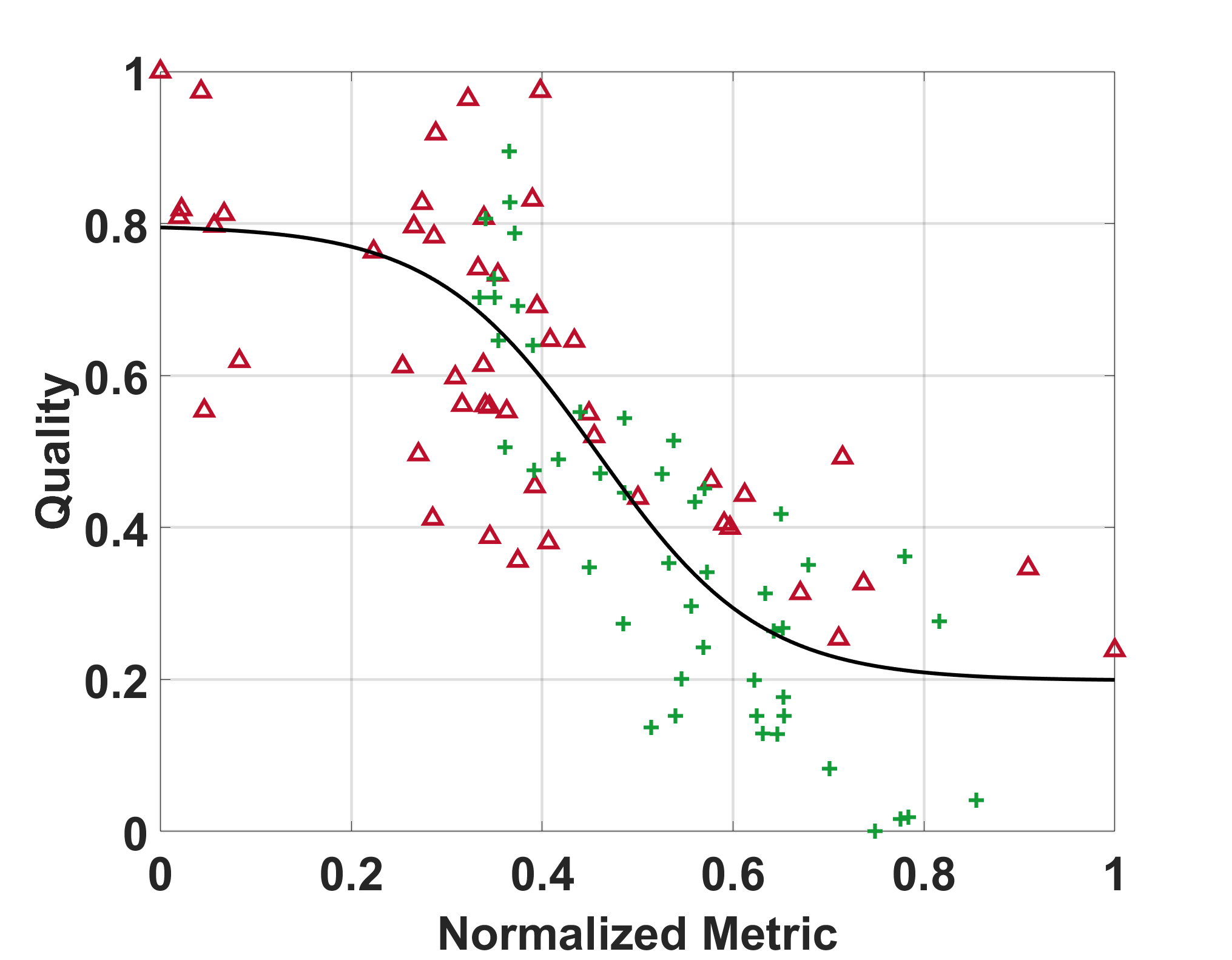}}
    \subfloat[MS-SSIM]{\includegraphics[width=0.25\linewidth]{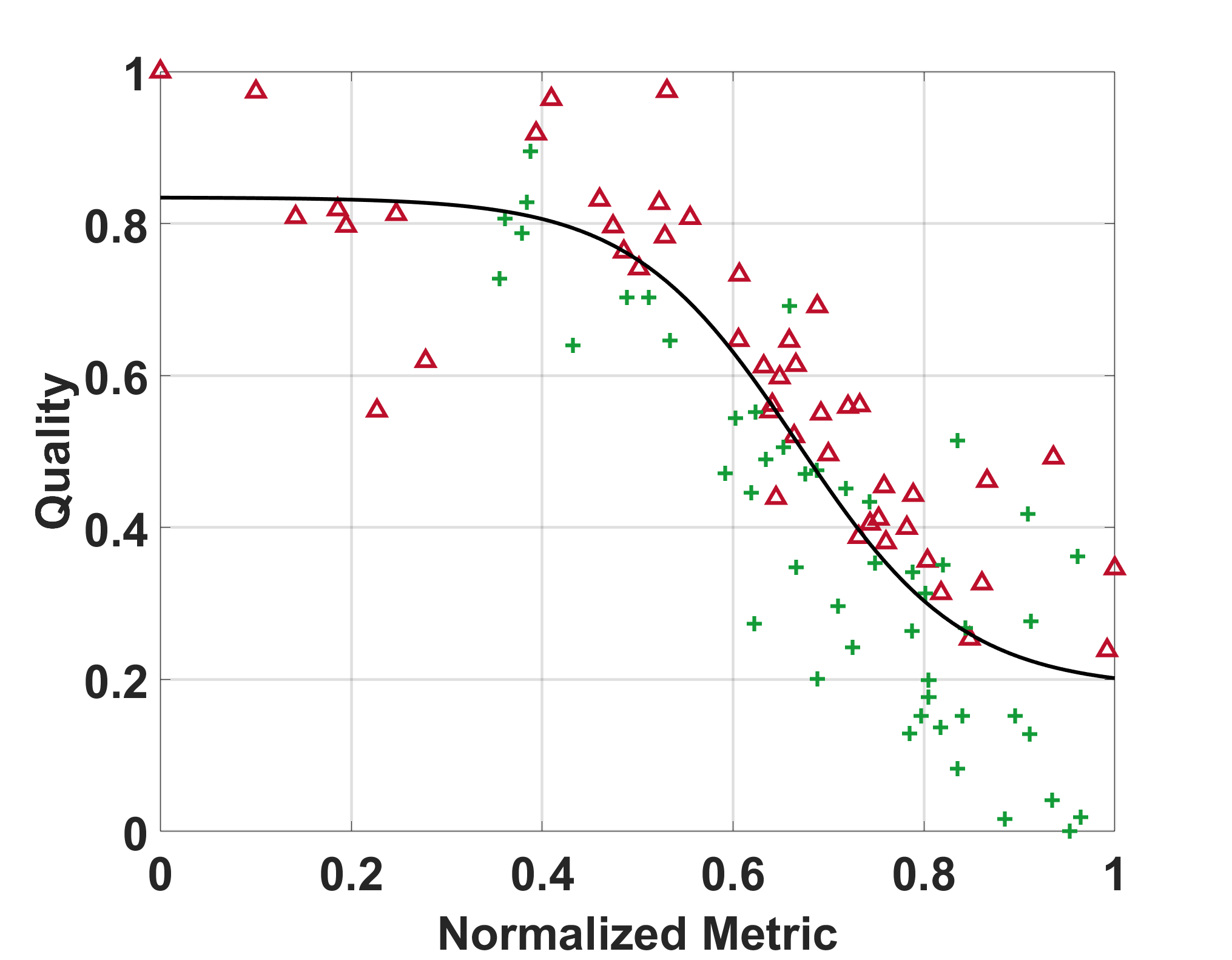}}
    \subfloat[FSIMc]{\includegraphics[width=0.25\linewidth]{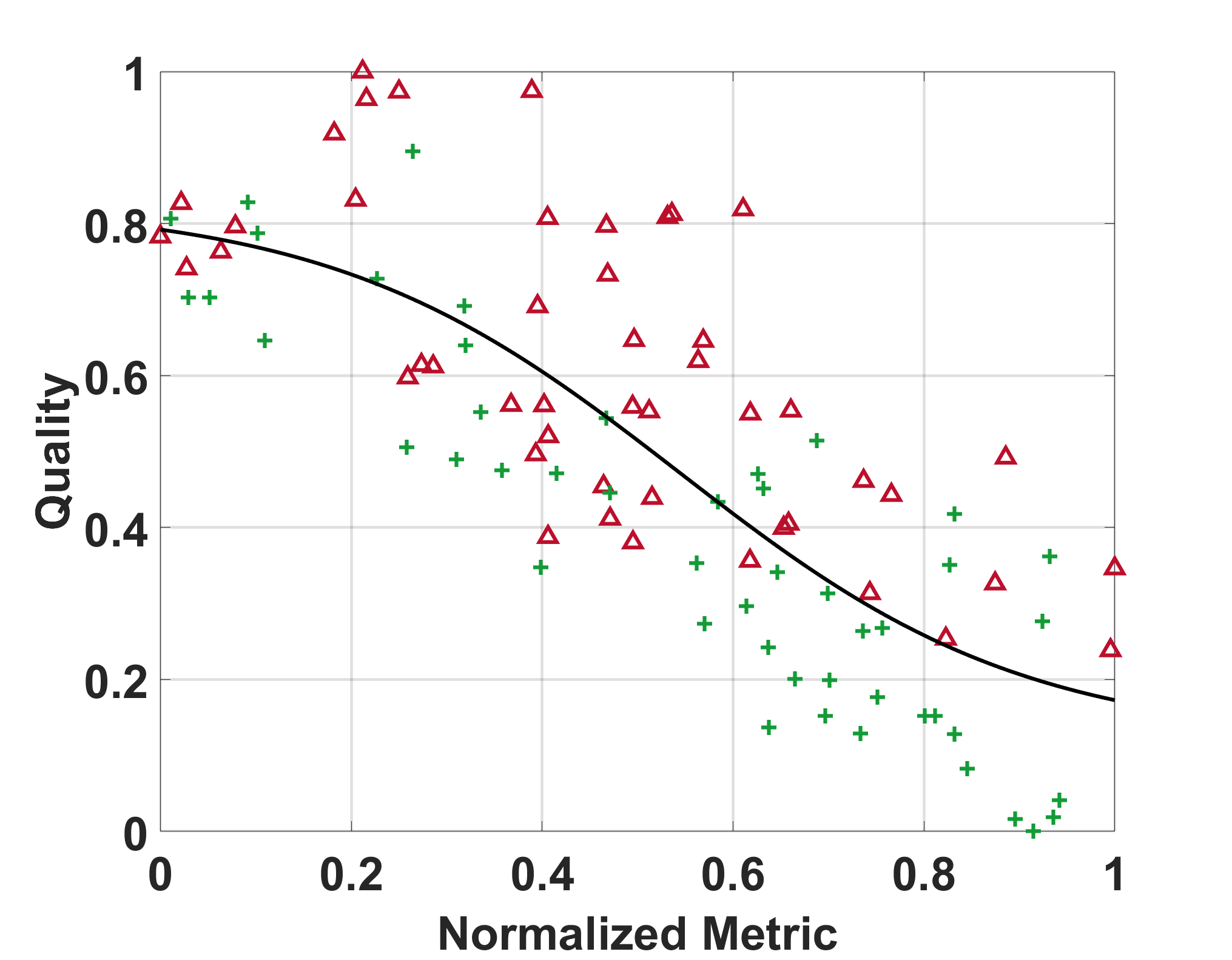}}
    \subfloat[IW-SSIM]{\includegraphics[width=0.25\linewidth]{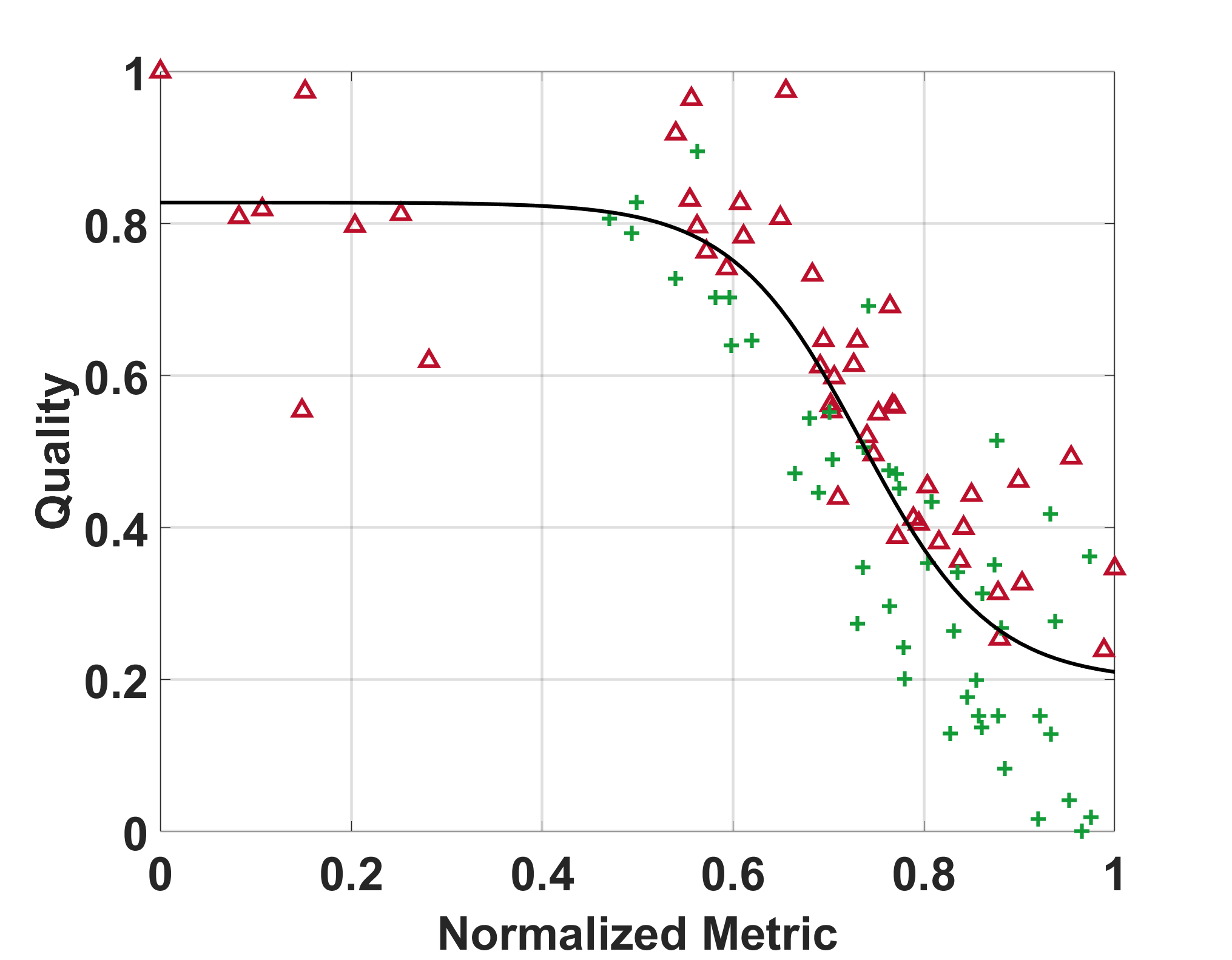}}\\
    \includegraphics[width=0.33\textwidth]{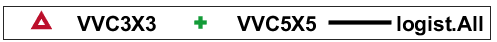}
    \caption{Logistic fitting for VVC 5$\times$5 and VVC 3$\times$3.}
\label{fig:fitting_CMVVC}
\end{figure*}

\subsection{Objective Metrics Performance}

\begin{table*}[tb]
\caption{Metrics performance.}
\centering
\begin{tabular}{|l||c|c|c|c||c|c|c|c|}
\hline
\textbf{} & \textbf{PCC} & \textbf{SROCC} & \textbf{RMSE} & \textbf{OR}  & \textbf{PCC} & \textbf{SROCC} & \textbf{RMSE} & \textbf{OR}\\ \hline
&\multicolumn{4}{|c||}{Pleno3$\times$3 vs VVC 3$\times$3} &\multicolumn{4}{c|}{Pleno5$\times$5 vs VVC 5$\times$5} \\ \hline
\textbf{PSNR-HVS} &0.586 & 0.575 & 0.236 & 0.708& 0.941 & 0.941 & 0.100 & 0.375 \\ \hline
\textbf{MS-SSIM} & 0.812 & 0.777 & 0.170 & 0.510 & 0.953 & 0.930 & 0.089 & 0.375 \\ \hline
\textbf{FSIMc} & \textbf{0.947} & \textbf{0.903} & \textbf{0.094} & \textbf{0.406} & 0.958 & \textbf{0.952} & 0.084 & 0.375 \\ \hline
\textbf{IW-SSIM} & 0.759 & 0.748 & 0.190 & 0.552& \textbf{0.967} & 0.938 & \textbf{0.074} &\textbf{ 0.302} \\ \hline
&\multicolumn{4}{|c||}{Pleno5$\times$5 vs Pleno3$\times$3} &\multicolumn{4}{c|}{VVC 5$\times$5 vs VVC 3$\times$3} \\ \hline
\textbf{PSNR-HVS} & 0.827 & 0.821 & 0.161 & 0.635 & 0.818 & 0.820 & 0.143 & 0.760\\ \hline
\textbf{MS-SSIM} & 0.855 & 0.834 & 0.147 & 0.604&0.862 & 0.855 & \textbf{0.126} & 0.656\\ \hline
\textbf{FSIMc} & \textbf{0.894} & \textbf{0.877} & \textbf{0.129} & \textbf{0.531}&0.791 & 0.804 & 0.153 & 0.750\\ \hline
\textbf{IW-SSIM} & 0.880 & 0.871 & 0.134 & 0.583 & \textbf{0.863} & \textbf{0.862} & \textbf{0.126} & \textbf{0.615} \\ \hline
\end{tabular}
\label{tab:metricPerformance}
\end{table*}


Objective quality metrics should be validated using subjective quality evaluation results as ground truth.
The statistical measures proposed in ITU-R BT.500-15~\cite{BT500} were computed, specifically the PCC, the SROCC, the Root Mean Squared Error (RMSE) and the Outlier Ratio (OR). The quality scores predicted for each of the objective metrics ($\tilde{Q}$) were computed by applying a logistic fit function to the objective scores, as it is commonly done when benchmarking objective metrics~\cite{HDRMarco}. This is computed as shown in Eq.~\ref{eq:placeholder}.
\begin{equation}
    \tilde{Q} = a + \frac{b}{1 + \exp(-c \cdot (O - d))}
    \label{eq:placeholder}
\end{equation}

Figs. \ref{fig:fitting_CC3x3} to \ref{fig:fitting_CMVVC} show the results of the logistic fitting function shown in Eq.~\ref{eq:placeholder} together with the different pairs subjective quality score and metric value. The quality and metrics scores were normalized between 0 and 1, using a min-max normalization, as recommended by ITU-R.BT500~\cite{BT500}. Those logistic functions were used for the computation of the predicted quality values from the metric values. These figures help to illustrate the results of table~\ref{tab:metricPerformance}.

Table \ref{tab:metricPerformance} shows the correlations between the predicted quality values obtained using the objective metrics regression and the subjective quality scores.
Generally, PCC and SROCC values are higher for comparisons considering the 5$\times$5 method, indicating a stronger correlation between objective metrics and subjective quality evaluations of fully compressed light fields. In contrast, the correlations are noticeably lower for 3$\times$3 comparisons, where view synthesis is employed. This trend is particularly evident in the correlation results obtained for the JPEG Pleno3$\times$3 and VVC3$\times$3 comparison. In this specific scenario, FSIMc shows the best correlation to the subjective evaluations, standing out as the best option to access objective quality when synthesized views are used.

This decrease is caused by the limitations of view synthesis, which often compromises the angular consistency, that is essential to light field data. Such disruptions introduce perceptually significant artifacts that are not adequately reflected by the tested metrics. However, the subjective evaluation methodology was defined considering the need  for an effective evaluation  that reflected the quality of the angular consistency, which is very visible in the subjective comparisons due to the flickering with the original.

A clear example of this discrepancy can be seen in the results for the \textit{Bicycle} light field. In both Fig.~\ref{fig:Subjective_CM_Pleno}-(c) and Fig.~\ref{fig:Subjective_CM_VVC}-(c), the subjective evaluations reveal that the perceived quality of the synthesized views, namely JPEG Pleno3$\times$3$\_$X, JPEG Pleno3$\times$3$\_$O, VVC3$\times$3$\_$X and VVC3$\times$3$\_$O, is significantly lower than that of their 5$\times$5 counterparts. This suggests that view synthesis introduces perceptually noticeable distortions. However, as shown in Fig.~\ref{fig:Objective_bicycle}, this degradation is not appropriately captured by perceptual metrics such as MS-SSIM, FSIMc, and IW-SSIM. 
In some cases, these metrics even assign higher quality scores to the synthesized views than to the original coded ones (of view type ``S"), further emphasizing the disconnection between the metric predictions and the perceived quality in scenarios involving view synthesis.

It can be observed in Fig. \ref{fig:fitting_CC3x3} that for the comparison between JPEG Pleno 3$\times$3 and VVC 3$\times$3, the results for PSNR-HVS and IW-SSIM tend to be quite far from the logistic curve. This tendency is less present in MS-SSIM. The scores for FSIMc are the ones that are closer to the logistic curve.

For the comparison between JPEG Pleno 5$\times$5 and VVC 5$\times$5 (Fig. \ref{fig:fitting_CC5x5}), it can be observed that all the metrics present similar results, quite close to the 
fitting curve.

The comparison between JPEG Pleno  5$\times$5 and 3$\times$3 and VVC 5$\times$5 and 3$\times$3 (Figs. \ref{fig:fitting_CMPleno} and \ref{fig:fitting_CMVVC}) show a very similar behavior. 

Additionally, the RMSE and OR results support this analysis, with the JPEG Pleno5$\times$5 and VVC5$\times$5 comparisons exhibiting the lowest error and outlier ratio, further reinforcing the higher reliability of objective metrics in the absence of view synthesis.

These findings suggest that current objective metrics may  be inappropriate for evaluating light fields that have synthesized views, which are often used in the literature for the evaluation of the performance of the view synthesis algorithms.
Furthermore,  most of the works on  light field coding also rely on the PSNR or SSIM/MS-SSIM metrics, which limits their validity.

\subsection{Compression Times}

The compression times reported in Table \ref{CompressionTimes} were measured on a system running Ubuntu 22.04.5 LTS with an AMD Ryzen 7 2700X Eight-Core Processor and 32 GB of RAM.  They represent the average compression time 
for each codec and bitrate, calculated for the four testing light fields. 
The 3$\times$3 sparsely sampled light fields (Pleno 3$\times$3, VVC 3$\times$3) exhibit significantly lower encoding times compared to their fully compressed 5$\times$5 counterparts, for the same bitrates. Specifically, Pleno achieves a 30\% increase in speed, VVC Random Access  by 38.3\%.

\begin{table}[tbh]
\caption{Compression times (in seconds) across different light fields for JPEG Pleno and VVC.}
\resizebox{\linewidth}{!}{%
\begin{tabular}{|c||c|c|c|c|}
\hline
\textbf{\makecell[c]{Target \\ Bitrates}}       & \textbf{Pleno5x5} & \textbf{Pleno3x3} & \textbf{VVC5x5} & \textbf{VVC3x3} \\ 
\hline\hline
\textbf{1.003} & 12,20 & 8,31  & 3209,7   & 1833,3   \\ \hline
\textbf{0.472}      & 8,69  & 6,00  & 2033,6   & 1225,411 \\ \hline
\textbf{0.236}    & 6,748 & 4,113 & 1281,618 & 794,346  \\ \hline
\textbf{0.118}       & 5,452 & 3,041 & 790,970  & 493,899  \\ \hline
\end{tabular}%
}
\label{CompressionTimes}
\end{table}

\section{Conclusions}\label{sec:conclusions}

This work studied the effect of view synthesis on light field compression, with focus on how it affects visual quality perception through subjective testing. 
The results reveal that view synthesis negatively impacts perceptual quality, with synthesized views consistently rated lower than their directly encoded counterparts. From these subjective quality evaluation results, it is implied that the view synthesis using the selected algorithm, compromises the angular consistency that is an inherent aspect of light field data. The resulting angular incoherence manifests as visible flicker, particularly at lower bitrates.

This effect was most pronounced in the synthetic light fields \textit{Sideboard} and \textit{Bicycle}, where artifacts introduced by view synthesis were both noticeable and disruptive. 

A critical insight from the correlation analysis is that objective quality metrics often fail to capture the distortions introduced by view synthesis. Metrics like MS-SSIM, IW-SSIM, and FSIMc sometimes assigned higher scores to synthesized views than to fully encoded ones, despite clear subjective preferences for the latter. This misalignment led to lower correlation values,particularly in 3$\times$3 configurations, as measured by PCC, SROCC, RMSE, and OR, and highlights the inadequacy of current metrics in reflecting synthesis-induced artifacts. However, FSIMc stood out as the best metric obtaining the best correlation values.

The subjective evaluation followed in the JPEG AIC-3 methodology, which proved effective at detecting subtle differences between high-fidelity views that would otherwise be indistinguishable using previous methods, such as the original-coded side-by-side approach. The use of coded/reference flicker helped reveal inconsistencies. However, the increase in ``Not Sure” responses at higher bitrates suggests a limitation of the approach when quality differences become minimal.

When comparing codecs, VVC consistently outperformed JPEG Pleno in both subjective and objective evaluations, particularly at low to medium-high bitrates. This performance gap becomes less apparent at higher bitrates, where both codecs tend to converge or stabilize in the subjective results.

Future research aims to focus on the following directions:

\begin{itemize}   
    
    \item \textbf{Development of new quality models:} Explore quality metrics that are sensitive to angular inconsistencies and synthesis-induced artifacts, in order to better align with subjective perception.
    
    \item \textbf{Advancement in view synthesis techniques:} Explore machine learning–based view synthesis approaches tailored for light fields, which better preserve angular consistency and minimize perceptual artifacts due to their data-specific training.

\end{itemize}

\bibliographystyle{IEEEtran}
\bibliography{refs}

\begin{IEEEbiography}
[{\includegraphics[width=1in,height=1.25in,clip,keepaspectratio]{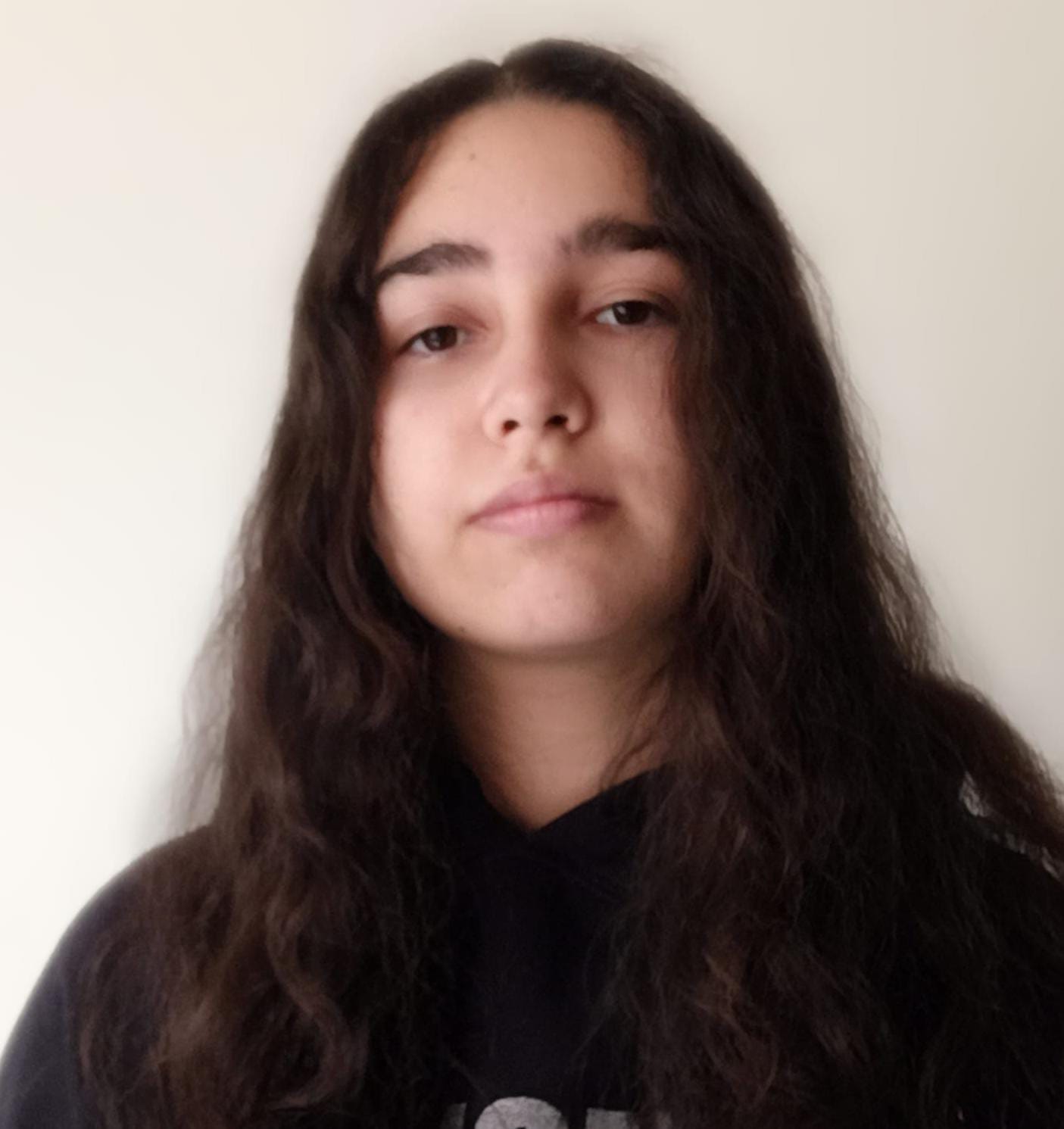}}]{Daniela Saraiva}  (IEEE Student Member) is a PhD student at Universidade da Beira Interior (UBI), Covilhã. Completed a bachelor's degree in Electrical and Computer Engineering at UBI in 2022. Completed a master’s degree in Electrical and Computer Engineering at the UBI in 2024.
\end{IEEEbiography}
\vspace{-1.5cm}
\begin{IEEEbiography}
[{\includegraphics[width=1in,height=1.25in,clip,keepaspectratio]{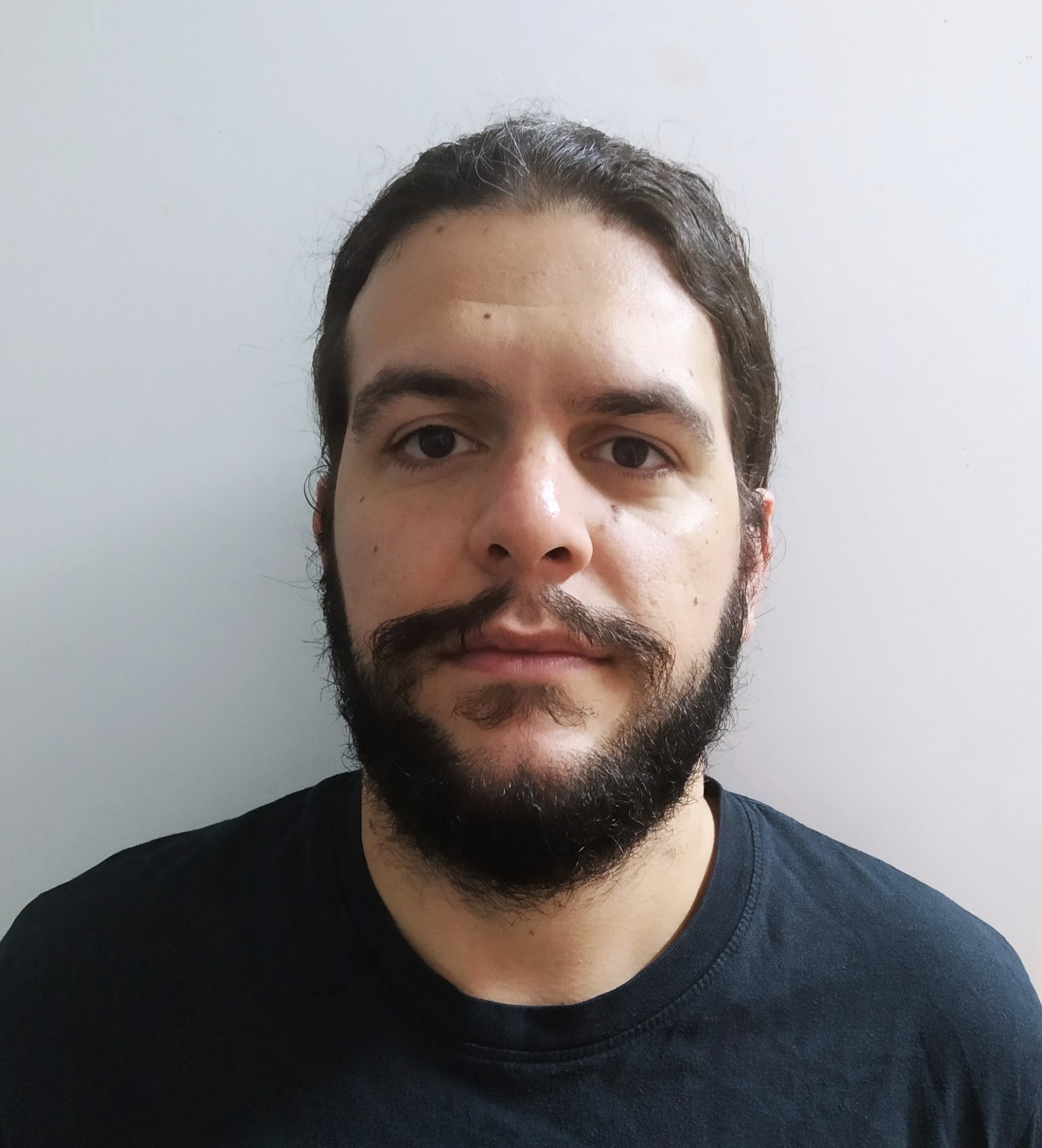}}]{João Prazeres}  (IEEE Student Member) is a PhD candidate from Universidade da Beira Interior (UBI), Covilhã. He graduated in Electrical and computer engineering in Universidade da Beira Interior in 2018 and received his master degree in 2020. 
He has been deeply involved in the JPEG PLENO Point Cloud Coding activity.
Recently he received a best paper award in 3D Imaging and Applications of the Electronic Imaging Symposium 2022. 
\end{IEEEbiography}
\vspace{-1.5cm}
\begin{IEEEbiography}
[{\includegraphics[width=1in,height=1.25in,clip,keepaspectratio]{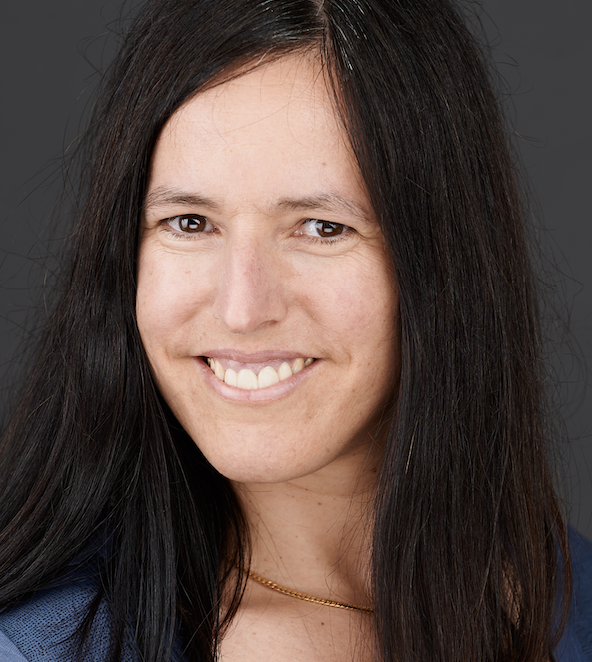}}]{Manuela Pereira}
received the 5-year B. S. degree in Mathematics and Computer Science in 1994 and the M. Sc. degree in Computational Mathematics in 1999, both from the University of Minho, Portugal. She received the Ph. D. degree in Signal and Image Processing in 2004 from the University of Nice Sophia Antipolis, France. She is an Associate Professor in the Computer Science Department 
of the University of Beira Interior, Portugal. Her main research interests include: Image and Video Coding; Multimedia technologies standardization; 
Signal Processing for Telecommunications; Information theory; Real-time video streaming; 3D and 4D Imaging; Medical Imaging.
\end{IEEEbiography}
\vspace{-1.5cm}
\begin{IEEEbiography}
[{\includegraphics[width=1in,height=1.25in,clip,keepaspectratio]{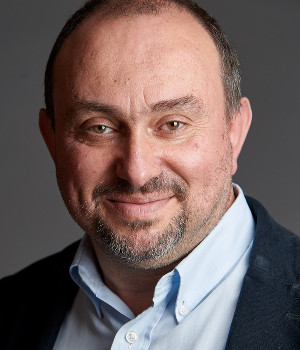}}]{António M.G Pinheiro}
(M'99, SM'15)
Is an Associate Professor at UBI (Universidade da Beira Interior), and a researcher at IT (Instituto de Telecomunicações), Portugal. He received the "Licenciatura" in Electrical and Computer Engineering from IST, Lisbon in 1988 and the PhD in Electronic Systems Engineering from University of Essex, UK in 2002. 
He is a Portuguese delegate to ISO/IEC JTC1/SC29 and the Communication Subgroup chair of JPEG. 
He was the PC co-chair of QoMEX 2015, special session co-chair of QoMEX 2016, 
and organizer of the tutorial in ACM Multimedia 2021 ”Plenoptic Quality Assessment: The JPEG Pleno Experience”.
He is Associate editor of IEEE Trans. on Multimedia and a senior member of IEEE. \end{IEEEbiography}

\end{document}